\documentclass[apj,twocolumn]{aastex63}

\usepackage[T1]{fontenc}
\usepackage{ae,aecompl}

\usepackage{graphicx}   % Including figure files
\usepackage{amsmath}    % Advanced maths commands
\usepackage{amssymb}    % Extra maths symbols
\usepackage{url}

\usepackage[normalem]{ulem}
\usepackage{color}
\usepackage{bm}
\usepackage{booktabs} % added by WHB: makes nicer looking tables

\def\mlt/{\ifmmode \alpha_{\text{MLT}}\else $\alpha_{\text{MLT}}$\fi}

\begin{document}

%\title{Standing on the shoulders of giants: A new distance to $\alpha$ Orionis supported by evolutionary, asteroseismic, and hydrodynamical simulations with MESA }
\title{Standing on the shoulders of giants: New mass and distance estimates for Betelgeuse through combined evolutionary, asteroseismic, and hydrodynamical simulations with MESA}

\shorttitle{Uncovering $\alpha$ Orionis}

\author{Meridith Joyce}
\affiliation{Research School of Astronomy and Astrophysics, Australian National University, Canberra, ACT 2611, Australia}
\affiliation{ARC Centre of Excellence for All Sky Astrophysics in 3 Dimensions (ASTRO 3D), Australia}
% \affiliation{South African Astronomical Observatory, Observatory Road, Cape Town, South Africa}

\author{Shing-Chi Leung\thanks{Email address: scleung@caltech.edu}}
%\author{Shing-Chi Leung\thanks{Email address: scleung@caltech.edu}}
\affiliation{TAPIR, Walter Burke Institute for Theoretical Physics, 
Mailcode 350-17, Caltech, Pasadena, CA 91125, USA} 
 
\author{L\'aszl\'o Moln\'ar}
\affiliation{Konkoly Observatory, Research Centre for Astronomy and Earth Sciences, Konkoly-Thege \'ut 15-17, H-1121 Budapest, Hungary}
\affiliation{MTA CSFK Lend\"ulet Near-Field Cosmology Research Group, Konkoly-Thege \'ut 15-17, H-1121 Budapest, Hungary}
\affiliation{ELTE E\"otv\"os Lor\'and University, Institute of Physics, Budapest, 1117, P\'azm\'any P\'eter s\'et\'any 1/A }
 
\author{Michael Ireland}
\affiliation{Research School of Astronomy and Astrophysics, Australian National University, Canberra, ACT 2611, Australia}

\author{Chiaki Kobayashi}
\affiliation{Centre for Astrophysics Research, Department of Physics, Astronomy and Mathematics, University of Hertfordshire, College Lane, Hatfield AL10 9AB, UK}
\affiliation{Kavli Institute for the Physics and 
Mathematics of the Universe (WPI),The University 
of Tokyo Institutes for Advanced Study, The 
University of Tokyo, Kashiwa, Chiba 277-8583, Japan}
\affiliation{ARC Centre of Excellence for All Sky Astrophysics in 3 Dimensions (ASTRO 3D), Australia}

\author{Ken'ichi Nomoto\thanks{Email address: nomoto@astron.s.u-tokyo.ac.jp}}
\affiliation{Kavli Institute for the Physics and 
Mathematics of the Universe (WPI),The University 
of Tokyo Institutes for Advanced Study, The 
University of Tokyo, Kashiwa, Chiba 277-8583, Japan}

\shortauthors{Joyce et al.}

\correspondingauthor{Meridith Joyce}
\email{meridith.joyce@anu.edu.au}

\label{firstpage}

\date{Accepted XXX. Received YYY; in original form ZZZ}

\begin{abstract}
We conduct a rigorous examination of the nearby red supergiant Betelgeuse by drawing on the synthesis of new observational data and three different modeling techniques. Our observational results include the release of new, processed photometric measurements collected with the space-based SMEI instrument prior to Betelgeuse's recent, unprecedented dimming event. We detect the first radial overtone in the photometric data and report a period of $185\pm13.5$\,d.

Our theoretical predictions include self-consistent results from multi-timescale evolutionary, oscillatory, and hydrodynamic simulations conducted with the Modules for Experiments in Stellar Astrophysics (MESA) software suite. Significant outcomes of our modeling efforts include a precise prediction for the star's radius:
$764^{+116}_{-62}\,R_{\odot}$. In concert with additional constraints, this allows us to derive a new, independent distance estimate of $168^ {+27}_{-15}$\,pc and a parallax of $\pi=5.95^{+0.58}_{-0.85}$\,mas, in good agreement with \textit{Hipparcos} but less so with recent radio measurements. 

Seismic results from both perturbed hydrostatic and evolving hydrodynamic simulations constrain the period and driving mechanisms of Betelgeuse's dominant periodicities in new ways. Our analyses converge to the conclusion that Betelgeuse's $\approx 400$ day period is the result of pulsation in the fundamental mode, driven by the $\kappa$-mechanism. Grid-based hydrodynamic modeling reveals that the behavior of the oscillating envelope is mass-dependent, and likewise suggests that the non-linear pulsation excitation time could serve as a mass constraint. 

Our results place $\alpha$ Ori definitively in the early core helium-burning phase of the red supergiant branch. We report a present-day mass of $16.5$--$19 ~M_{\odot}$---slightly lower than typical literature values. 

\end{abstract}


\section{Introduction}
Since November of 2019, the red supergiant $\alpha$~Orionis---popularly known as Betelgeuse---has experienced an unprecedented brightness drop of nearly $2$ magnitudes in the V band. The severity of this decrease and the deviation from its typical pattern of variability
have sparked much public speculation about the physics responsible and its likelihood of undergoing a cataclysmic event. 

To investigate these questions first requires an understanding of the short-timescale behavior of variable red giants. Such stars are known to exhibit a complex spectrum of variability, where cyclic variations with different driving mechanisms occur over a range of timescales. Though we can explain and fully capture some pulsation physics in 1D stellar models (e.g., pressure and gravity modes; see review by \citealt{ConnyReview}), 
other mechanisms are not well understood \citep{LSP-wood2004,LSP-nicholls2009}. In this latter class fall many of the variations we observe on human timescales, as such behavior is, with rare exception, too rapid to be explained by classical stellar evolution \citep{TUMi}. Modeling such processes may require 3 dimensions, time-dependent convection, or otherwise more sophisticated physical formalisms that are beyond the scope of typical 1D stellar evolution programs.
Nevertheless, 1D stellar models are among the most powerful devices for gaining insight on the sub-surface physics responsible for observed changes in real stars  (\citealt{YY, Piet04, VDB2006, BaSTI, GARSTEC, Dotter, GYRE, MESAIV} and others). When conducted on a range of timescales, their calculations can be exploited to great effect.

In red supergiants, the $\kappa$-mechanism drives radial pulsations in the hydrogen ionization zone, and simulations show the emergence of periods and growth rates of the dominant fundamental pulsation mode---typically on the order of years---both in linear and non-linear models, as shown in e.g.\ \citet{LiGong1994}, \citet{Heger1997},  \citet{YoonCantiello2010}, and \citet{MESAII}.  
In addition to these, previous modeling work on $\alpha$ Ori and similar red supergiants (RSGs) includes \citet{Dolan2016}, \citet{BetelgeuseProjectI}, \citet{BetelgeuseProjectII}, and  \citet{Goldberg2020}.

In both \citet{YoonCantiello2010} and \citet{MESAII}, models of rotating and non-rotating RSGs with approximately solar metallicity and initial masses of $25 M_{\odot}$ were found to exhibit pulsations on the order $1$--$8$ years.  Obtaining frequencies of this magnitude required lowering the evolutionary timestep to a fraction of a year during helium burning. The limiting factor on these calculations was the emergence of supersonic radial velocities in the envelope (see Section 6.6 in \citealt{MESAII} for more details on their example).

A rigorous estimation of the model-derived fundamental parameters of $\alpha$ Ori was undertaken by \citet{Dolan2016}. In particular, their models find a best estimate of $20^{+5}_{-3} M_{\odot}$ for the progenitor mass. They also attempt to model the pulsation properties of $\alpha$ Ori, but find they were unable to reproduce the fundamental mode (FM) and first overtone (O1) frequencies with adiabatic models alone. They suggest that interplay between non-adiabatic pulsations and convection could be responsible for some variability, noting that 3D simulations of similar red supergiants show the development of large-scale granular convection that can itself drive pulsation (\citealt{Xiong1998, Jacobs1998, freytag2002, Chiavassa2011, FreytagChiavassa2013,   Dolan2016}).

Recent 1D modeling efforts in ``The Betelgeuse Project'' series and other works suggest that a past merger may be required to explain the present-day surface rotation of $\alpha$ Ori, which is more rapid than standard stellar evolutionary calculations including rotation can reproduce (\citealt{BetelgeuseProjectI, BetelgeuseProjectII, Chatzopoulos2020}). The \citet{BetelgeuseProjectII} study also examines the star seismically, but the authors are primarily focused on rapid waves in the convection zone that might precede a cataclysmic event.
This concept was also addressed in depth by \citet{Goldberg2020}, who modeled the observable features of supernova events as a function of the point of onset during the stellar pulsation.

In this paper, we use a range of tools to investigate the variability of $\alpha$ Orionis.
We use the Modules for Experiments in Stellar Astrophysics (MESA, \citealt{MESAI, MESAII, MESAIII, MESAIV, MESAV}) stellar evolution software suite to generate both classical evolutionary tracks and short timescale, hydrodynamic simulations of stars. We likewise use the GYRE pulsation program to construct complementary predictions of the pressure mode ($p$-mode) oscillations in models of red giants \citep{GYRE}.

This paper proceeds as follows: In Section \ref{section:constraints}, we discuss the current knowledge of $\alpha$ Ori's classical constraints, including pulsation periods, evolutionary stage, radius, temperature, and distance.
We present a lightcurve highlighting $\alpha$ Ori's recent behavior, constructed from data collected from the American Association of Variable Star Observers (AAVSO) and newly processed space-based photometry from the Solar Mass Ejection Imager instrument.
In Sections \ref{section:evolution}, \ref{section:seismic}, and \ref{section:hydro}, we discuss our evolutionary, seismic, and hydrodynamic models, respectively. Section \ref{section:conclusions} concludes our analysis and presents best estimates of its fundamental parameters based on detailed photometric analysis and comprehensive, multi-timescale simulation.

%%%%%%%%%%%%%%%%%%%%%%%%%%%%%%%%%%%%%%%%%%%%%%%%%%%%%%%%%%%%%%%%%%%%%%%%%%%%%%%%%%%%%%%%%%%%%%%%%%%%%%%%%%%%%%%%%%%%%%%%%%%%%%%%%%%%%%%
\section{Observational Constraints}
\label{section:constraints}
\begin{table} 
\centering 
\caption{Processed SMEI photometry of $\alpha$~Ori. Observations were corrected for systematics and averaged into 1-day bins. Errors calculated as simple shot noise. $V$ mag is the same light curve, scaled to existing $V$-band data. The full data set is available in the online journal}
\begin{tabular}{ccccc}  
\hline\hline
BJD-- & SMEI & SMEI  & $V$ & $V$  \\
2400000 (d) & mag & error & mag & error\\
\hline
52677.959995 & 0.3759 & 0.0037 & 0.5168 & 0.0037\\
52678.983194 & 0.3849 & 0.0039 & 0.5330 & 0.0039\\
52680.041678 & 0.3717 & 0.0043 & 0.5094 & 0.0043\\
52680.959028 & 0.3801 & 0.0039 & 0.5244 & 0.0039\\
52681.911649 & 0.3869 & 0.0035 & 0.5365 & 0.0035\\
52682.934838 & 0.3908 & 0.0036 & 0.5436 & 0.0036\\
52683.887465 & 0.3991 & 0.0035 & 0.5586 & 0.0035\\
52684.875377 & 0.4032 & 0.0035 & 0.5662 & 0.0035\\
52685.863269 & 0.4030 & 0.0035 & 0.5657 & 0.0035\\
52686.851169 & 0.4068 & 0.0035 & 0.5727 & 0.0035\\
52687.415625 & 0.4177 & 0.0093 & 0.5929 & 0.0094\\
52689.038675 & 0.4142 & 0.0042 & 0.5863 & 0.0042\\
\multicolumn{5}{l}{\dots}\\
 \hline
\end{tabular}
\label{table:phot}
\end{table}

$\alpha$~Ori is well studied interferometrically; together with R~Dor and IRC~10216, it is among the stars with the largest angular diameters ever measured (\citealt{Bedding1997,Menten2012,Stewart2016}). In their Table~3, \citet{Dolan2016} provide a clear summary of previous measurements.

The earliest interferometric measurement from \citet{Michelson21} resulted in an
angular diameter of 47$\pm$5\,mas at visible wavelengths, assuming a uniformly illuminated disk model. In recent years, it was realized that there were elevated layers of molecules and dust above the photosphere \citep[e.g.,][]{perrin2004}, complicating the interpretation of diameter measurements. In the context of the recent dimming event of $\alpha$~Ori, \citet{Haubois19} solved for the photospheric diameter, dust shell diameter and optical depth. They found a Uniform Disk diameter of 44.0$\pm$0.5\,mas in their 1.04\,$\mu$m, bandpass, which had relatively little influence from molecular bands. This would be equivalent to a limb-darkened diameter of 46.0$\pm$0.6\,mas using a linear limb darkening coefficient of $\sim$0.5 \citep{Claret11,HanburyBrown74}. However, the relatively simple dust model consisting of of a 64.7\,mas diameter thin shell scattering 4.4~\% of the light means that the statistical error from that work is not fully representative of the model uncertainty.
%}

We adopt as observational reference the diameter from the recent work of \citet{Montarges14} of 42.28$\pm 0.43$\,mas for the limb-darkened (i.e., physical photospheric) diameter. Those authors resolved the photosphere significantly past the first null in the visibility curve, so were insensitive to low optical depth shells, unlike many of the other measurements. Additionally, the relatively high spectral resolution observations in the K band, away from main molecular absorption features, mean that this measurement is relatively unaffected by apparent molecular shells.

Radius estimates are further complicated by uncertain parallax measurements, which are made difficult by variability and known $>$2\,au-scale asymmetries on the surface of the star both at optical and radio wavelengths \citep{Young2000,Kervella2018}. The revised \textit{Hipparcos} astrometric solution gave an optical-only distance of $153^{+22}_{-17}$~pc \citep{vanLeeuwen2007}. Combination of the \textit{Hipparcos} data with radio observations captured by the Very Large Array (VLA) extended that distance out to $197\pm45$~pc, which was also used by \citet{Dolan2016}.

The revised \textit{Hipparcos}-only value is inconsistent at the 1.7$\sigma$ level with the most recent radio measurement of $222^{+48}_{-34}$~pc \citep{Harper2017}, which took into account both VLA and Atacama Large Millimeter Array (ALMA) observations but which was also significantly affected by ``cosmic noise''.\footnote{``Cosmic noise'' is an umbrella term used to describe the elevated dispersion of the residuals of the astrometric solution compared to the formal errors. It can include various physical effects such as source size, unresolved companions, unresolved properties of stars in the stellar models used for fitting, variability of the stellar parameters, and instrumental effects such as excess noise due to saturation.}
The star is well beyond the established brightness limit of \textit{Gaia}, and data enabling a future parallax measurement were not collected in the first years of the mission. A parallax estimate of Betelgeuse is therefore not included in \textit{Gaia} Data Release~2 \citep{Gaia2016,Sahlmann2018}. Given the very long time-baselines needed to overcome the effects of photospheric motions and variability, there is unlikely to be a reliable direct parallax measurement of Betelgeuse with $<10$\% uncertainty in the near term.

%%%%%%%%%%%%%%%%%%%%%%%%%%%%%%%%%%%%%%%%%%%%%%%%%%%%%

\begin{table*} 
\centering 
\caption{Observational best values, estimated ranges, and model-derived constraints (where indicated) for $\alpha$ Ori. 
The temperature constraints reflect the spectroscopically derived temperature from $\alpha$ Ori at its brightness minimum, which is not necessarily reflective of its mean temperature. However, even \citet{Levesque2020}'s $100$ K error bars accounting for decadal variations are more restrictive than the theoretical uncertainty imposed by modeled variations in \mlt/.
Though we quote a ``best'' radius and reference a wide range of values, in practice we do not impose any constraints on the radius when modeling. The range of possible radii derived from the models is smaller than the uncertainties reported by many observers.
We quote the initial and present-day mass ranges preferred by our seismic models. Masses considered in the initial grid range from 10 to 26 $M_{\odot}$.
%FM pulsation period. 
}
\begin{tabular}{l c  l  l }  
\hline\hline
Property				&  Value                & Source  &Comment\\ \hline
$T_{\text{eff}}$ 	& $3600 \pm 25$	K&   \citet{Levesque2020} & range extended by $\sigma_{\text{theory}}$ to $\pm200$ K \\ \hline
Angular Diameter & 42.28$\pm$0.43       & \citet{Montarges14} & Limb-darkened \\ \hline %Was 41.8$\pm$0.6 mas
Radius upper limit 	  & $ \sim 1100 R_{\odot}$ 	&   \citet{Dolan2016} & data collated from many sources  \\ \hline
Radius lower limit 	  & $ 500 R_{\odot}$ 	&   \citet{Dolan2016} & data collated from many sources   \\ \hline
Distance  	        & $ 197\pm45 $	pc &   \citet{harper2008}  & parallax data adopted by \citet{Dolan2016} \\ \hline
Period of variability    & $ 388 \pm 30$ days	&   \citet{Kiss2006} &  dominant, higher frequency; likely FM  \\ \hline
Period of variability    & $2050 \pm 460$ days	&   \citet{Kiss2006} &  lower frequency; likely LSP \\ \hline \hline
Period of variability (FM)    & {$416 \pm 24$ days}	&   this work & SMEI+\textit{V} data; mode ID from GYRE  \\ \hline
Period of variability (O1)    & {$185 \pm 14$ days}	&   this work & SMEI+\textit{V} data; mode ID from GYRE  \\ \hline
Period ratio O1/FM    & {$0.445 \pm 0.041$}	&   this work & SMEI+\textit{V} data  \\ \hline
Radius 	   & {$ 764 +116, -62 R_{\odot}$}	&  this work  & {$3\sigma$} range; seismic analysis \\ \hline
Initial Mass 	   & {$18$--$21 M_{\odot}$}	&  this work  & median range; seismic analysis \\ \hline
Present-day mass  & {$16.5$--$19 M_{\odot}$}	&  this work  & median range; seismic analysis \\ \hline
Distance 	   & {$ 168.1 +27.5, -14.9$ pc}	&  this work  & {$3\sigma$} range; seismic analysis \\ \hline
Parallax 	   & {$5.95 +0.58, -0.85$ mas}	&  this work  & {$3\sigma$} range; seismic analysis \\ \hline
 \hline
\end{tabular}
\label{table:obs}
\end{table*}

%%%%%%%%%%%%%%%%%%%%%%%%%%%%%%%%%%%%%%%%%%%%%%%%%%%%%

Estimates of Betelgeuse's mass are derived from models and range from roughly $15$--$25 M_{\odot}$, with previous modeling work suggesting that $\alpha$ Ori is in the midst of its core helium-burning giant branch phase \citep{Neilson2011,Dolan2016, BetelgeuseProjectI, BetelgeuseProjectII}.
However, while \citet{Dolan2016} state that its mass loss rate---the primary piece of evidence supporting the claim that it is on the red supergiant branch (RSB)---is ``consistent with having recently begun core helium burning,'' they also note that a previous interaction of Betelgeuse with a binary companion could account for similar mass loss rates without necessitating that Betelgeuse currently exist on the RSB. Since nearly half of $\sim20 M_{\odot}$ stars have a companion close enough to induce mass loss, this is, in fact, ambiguous \citep{deMink2014}. 
It is demonstrated by \citet{BetelgeuseProjectI} that rotating models of $\alpha$ Ori do not produce reasonable evolutionary predictions (a finding consistent with our present work), but they do not draw any specific conclusion about whether the star is core helium burning. 

As it is impossible to measure either mass or evolutionary status directly, and the evidence regarding its phase is not definitive, we do not assume a particular evolutionary phase a priori in our models. Instead, we consider the relative probabilities that $\alpha$ Ori is in a particular evolutionary stage based on
(1) the masses of tracks that match the other observational constraints and (2) the duration of the possible evolutionary stages. 

The first-order, theoretical constraints on its mass and age are provided by the linear pulsation calculations, which rule out any model in an evolutionary stage earlier than the RSB. 
From an observational perspective, we note that Betelgeuse is far in the foreground of the known $<$10\,Myr age young associations in Orion \citep{Grossschedl2018}, and it is not known to have kinematics consistent with ejection. In particular, its radial velocity of +21.9$\pm$0.5\,km\,s$^{-1}$ is consistent with the $\sim$+23\,km\,s$^{-1}$ of typical high mass stars in the Orion OB1 association \citep{Morrell91,Famaey2005}, but would differ by $\sim$20\,km\,s$^{-1}$ if it had travelled 200\,pc in $\sim$20\,Myr. The $(U,V,W)$ space motion of Betelgeuse is $(-22,-10,12)$ km\,s$^{-1}$ with respect to the Sun, which is $(-11,2,19)$ km\,s$^{-1}$ with respect to the local standard of rest \citep{Famaey2005,Schoenrich2010}. The high $W$ velocity in particular is of note, as it is discrepant at $3\sigma$ from the kinematics of the young disk \citep{Robin2003}. If this high $W$ velocity were due to ejection from a young association lying on the Galactic disk, now falling back through the disk due to vertical epicyclic motion, this would imply an origin of $\sim$50\,Myr ago.
With these proper motion estimates in mind, we are left with a few possible scenarios of varying likelihood: (1) Betelgeuse was formed very recently in a region where there is no star formation; (2) it is $\gtrsim$50 Myr old, or (3) it underwent some kind of binary interaction that propelled its trajectory.
Scenario (1) is not reasonable, and scenario (2) would be consistent with a mass below $10M_{\odot}$---a possibility that is readily ruled out by our other constraints. We are thus left with the third scenario, which is likewise supported by observations of Betelgeuse's present-day surface rotation and the inability of 1D, rotating models to reproduce it (see subsequent discussion).

%%%%%%%%%%%%%%%%%%%%%%%%%%%%%%%%%%%%%%%%%%%%%
We construct an age-prior function that performs a Monte Carlo interpolation over a grid of stellar tracks with masses ranging from $16$--$26 M_{\odot}$ (other parameters fixed; $\mlt/=2.1$) and a power law IMF. For two sets of realizations, we adopt a minimum age constraint of 8 Myr and permit radii between $600$ and $900 R_{\odot}$. In the first statistical run, masses are heavily skewed towards the head of the distribution, peaking at $16 M_{\odot}$, and the bulk of the trials fall from $16$--$18 M_{\odot}$. This indicates that the lower-mass regime is strongly statistically preferred, which is consistent with our understanding of the prevalence of high-mass stars in general. In the second statistical run, the distribution peaks a bit higher, at $18M_{\odot}$, and tapers off rapidly beyond $17.5$ and $19.5$ in either direction. The number of trials that do not fall somewhere on the core helium-burning RSG is negligible regardless of mass, though this is even more strongly the case for trials with masses between $17$ and $19M_{\odot}$. 

%}

%%%%%%%%%%%%%%%%%%%%%%%%%%%%%%%%%%%%%%%%%%%%%%
As we will conduct estimates of the stellar mass, and many other parameters, in several ways throughout this analysis, we treat the above statistical experiment strictly as sufficient evidence to assume that Betelgeuse is core helium burning in subsequent modeling.

Recent spectral analysis of Betelgeuse presents an effective temperature of $3600 \pm 25$ K (e.g., \citealt{Levesque2020, Guinan2020}). \citet{Levesque2020} write that the difference between the spectroscopically-derived temperature measured in 2004--5 and that measured during Betelgeuse's brightness minimum in 2020 is at most a decrease of $100$K, and at minimum, negligible. We note, however, that there is some disagreement on the reliability of the method by which \citet{Levesque2020} derive their temperature, with, e.g., \citet{IrelandScholzWood2008} and \citet{Davies2013} suggesting it may be underestimated.
We discuss this in more detail in Section \ref{section:evolution}.
Adopting the results of \citet{Levesque2020} essentially rules out convective turnover as an explanation for its recent dimming, but surface temperature is less informative on other oscillation driving mechanisms.

Critically, the brightness of Betelgeuse varies in a systematic way on at least two different timescales, and these periodicities were measured with good precision by \citet{Kiss2006} (and later corroborated by \citealt{Chatys2019}). 
The shorter occurs with a period of $\sim388$ days and the longer with a period of $\sim5.6$ years ($2050$ d). 
The period--luminosity relation depicted in Figure 6 of \citet{Kiss2006}
provides some evidence that the $388$ d brightness variation is caused by $p$-mode pulsation in the fundamental mode. This is likewise supported by various observational and theoretical considerations, including the position of the star on the $\log P$--$M_\mathrm{K}$ diagram, where the absolute K brightness provides the observational proxy for the stellar luminosity. T
%hese periods were later corroborated by \cite{Chatys2019}.
%
\citet{Kiss2006} also found that the shorter periods fit the theoretical calculations of \citet{guo2002}, forming an extension to sequences B and C of the supergiant variables observed in the Magellanic Clouds that correspond to the FM and the O1 pulsation modes, respectively \citep{wood1999,KissBedding2003,soszynski2007}. This also suggests that these variations correspond to $p$-mode pulsation. 

The longer, $\sim2050$ d periodicity likely falls in a class of signal known as ``Long Secondary Periods,'' or LSPs. These have been observed in multiple semiregular and red supergiant variables, but the cause of the LSP is still debated \citep{Wood2000,Chatys2019}.
Proposed mechanisms include rotational modulation caused by spots or a nearby companion followed by a dust cloud, among other possibilities \citep{Wood2000,soszynski-LSP-2014}.
Such signals were observed in the LMC supergiant population as ``sequence D,'' and the long periods found by \citet{Kiss2006} extend that sequence to higher luminosities \citep{derekas2006}. 
Among other things, rotational modulation was proposed as a possible mechanism for the LSP \citep{Percy-Deibert-2016}. However, the rotational period of $\alpha$~Ori has recently been estimated at $P_{\rm rot} = 31 \pm 8$~yr, which is considerably longer than the LSP of the star \citep{Kervella2018}.
Models in this work shed more light on the questions of mode classification and driving mechanism.
% !!! 

\begin{figure*} 
\centering
\includegraphics[width=\linewidth]{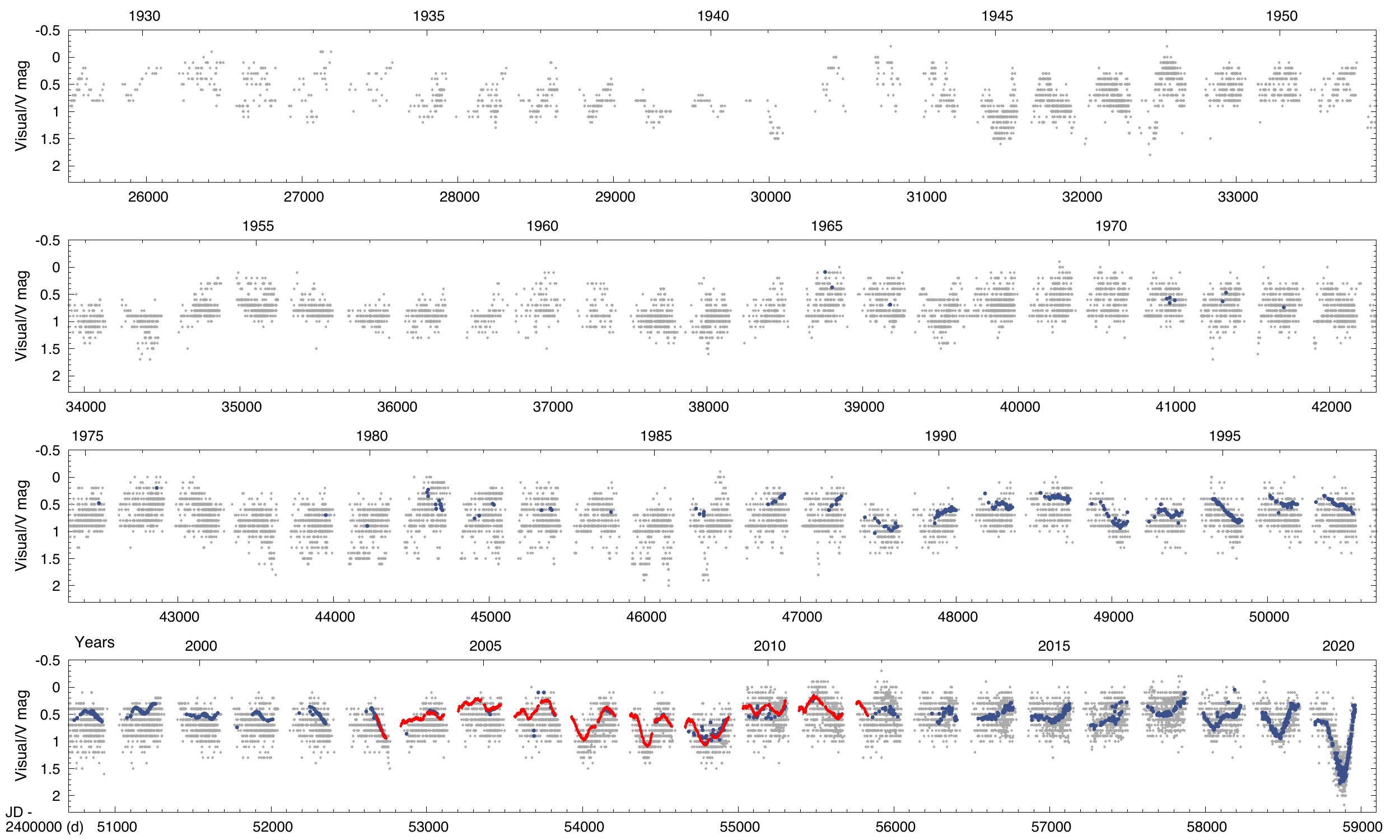}
\caption{Lightcurve of $\alpha$ Ori assembled from publicly available data compiled by the AAVSO, from 1928 to present, and from the SMEI observations. Horizontal axes are marked in both years (UPPER) and JD ${}+2400000$ (LOWER). Grey points are visual estimates, blue are $V$-band photometry, from AAVSO. Red points are the SMEI data. }
\label{LongitudinalB}
\end{figure*}

\subsection{Photometric Observations}
Both the $\sim 400$ day and 5.6 yr (2050 d) periods are visible in Figure \ref{LongitudinalB}, which shows the longitudinal brightness variations of $\alpha$ Ori over the last $90$ years. These visual brightness estimates were collected in large part by amateur observers and archived by the American Association of Variable Star Observers (AAVSO).

Examining Figure \ref{LongitudinalB} more closely, we see that the amplitude of the brightness drops corresponding to the $\sim 400$ day pulsation period are about $0.3$--$0.5$ mag in the $V$ band. The difference between these and the 1 mag drop in 2019--20 is clear. We do note, however, that Betelgeuse has undergone other periods of drastic dimming a few times over the last 100 years. Dimming events of comparable magnitude are visible in Figure \ref{LongitudinalB}, for instance, in the mid to late 1980s and arguably in the early 1950s. An argument could be made for the existence of a $35$--$40$ year dimming cycle, particularly if we take into account that the sensitivity of instruments has improved considerably in the last few decades.
We note that this 3--4 decade variation is of the same order as the suggested rotational period. While this could potentially be a manifestation of rotational modulation, confirmation will require ongoing observation.

\begin{figure} 
\centering
\includegraphics[width=\linewidth]{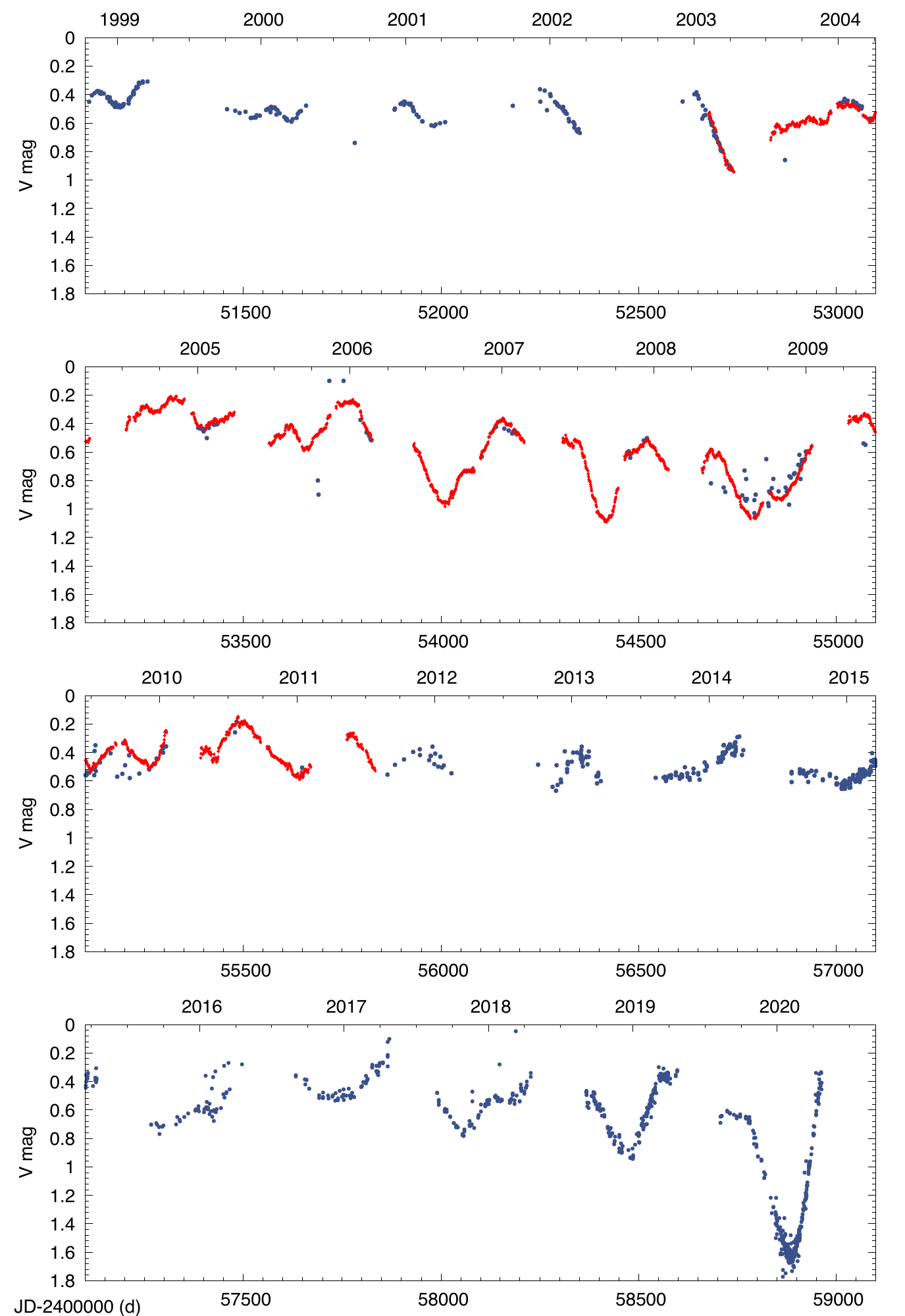}
\caption{Detailed plot of the recent photometric data. Blue: AAVSO $V$-band photometry. Red: rectified and scaled SMEI photometry.}
\label{newBphot}
\end{figure}

The low amplitude and scarcity of adequate comparison stars make visual estimates less sensitive to smaller changes from one pulsation cycle to another. Digital photometric observations exist for the last three decades, but both the quality and quantity varied over time. Most of the publicly available data have been archived by the AAVSO and provide good coverage from the mid-1980s to the early 2000s and from 2010 onward. To fill in the gap, we supplement the AAVSO data set with the observations taken with the SMEI (Solar Mass Ejection Imager) instrument aboard the Coriolis satellite \citep{jackson2004}.

\subsubsection{SMEI photometry of Betelgeuse}
SMEI was designed to follow Coronal Mass Ejections (CMEs) from the Sun, and in order to do this, stellar signals must be removed from its images. About 6000 stellar sources plus the brightest Solar System objects were catalogued and then subtracted from the images. It was soon realized, however, that the source subtraction procedure used by the mission can be processed into time series photometry of the brightest stars in the sky, essentially turning SMEI into one of the early space photometry missions \citep{buffington2007,hick2007}. SMEI observed $\alpha$~Ori from early 2003 to late 2011 with a cadence of 104 mins. Each year, data collection was split between the three cameras whose outputs needed to be rectified. Yearly systematics arise from the changing thermal conditions in each of the cameras \citep{tarrant2008}. Slow degradation of the camera sensitivity is also apparent in the data. 
%this is pretty cool actually

We could not remove the annually repeating instrumental signals directly, as the timescale is on the same order as the variation of $\alpha$~Ori. Therefore, we relied on the ensemble photometry of neighboring stars to derive common instrumental characteristics. We inspected ten nearby bright stars and selected $\gamma$~Ori, $\varepsilon$~Ori and 32~Ori to generate a template for the instrumental signals. We calculated a smoothed systematics curve by calculating the medians of the combined relative intensity data of these three stars in 4-day windows placed around every time stamp of $\alpha$~Ori, and for each camera separately. 

The rectified SMEI light curve of $\alpha$~Ori is the result of scaling the raw data with the systematics curve and then transforming it to magnitudes using $m_{\rm SMEI} = 10.0$~mag as the magnitude zero point. However, the passband of SMEI is not the $V$ band, therefore requiring that we scale and shift the light curve to match the AAVSO data. To compute the appropriate scaling, we determined the brightness difference for six other stars with M1-2 spectral class in the SMEI catalog to be $m_{V} - m_{\rm SMEI} \approx 0.15$~mag. We found that we needed to stretch the amplitude by a factor of 1.8 to match the $V$ data points. We then averaged the raw photometry points into 1-day bins.  While the shape of the variation could also be passband-dependent to some extent, the scarcity of overlapping $V$ data prohibited us from performing a more detailed comparison. The final light curve is plotted in Fig.~\ref{newBphot}, along with the AAVSO $V$-band data. The SMEI data
%fill the gap in publicly available photometric data during the late 2000s and 
confirm that the star did not dim excessively during the LSP minimum occurring in 2007--08.

A sample table of the processed and binned SMEI photometry and the scaled $V$-band values can be found in Table~\ref{table:phot}. Here, we provide formal errors calculated as the shot noise from the number of electron counts.

The photometric light curve reveals a richer set of features than the visual light curve.
The SMEI observations, in particular, show both the slow variation from the LSP along with additional smaller, more rapid variations. 
The SMEI data also put the severity of the recent dimming event in perspective: the brightness of the star did not drop below 1.1 mag in the $V$ band during the last 40 years, whereas the dip commencing in November of 2019 dimmed the star to 1.6 mag in that band. The light curve also highlights some smaller variations on the order of a few hundredths of a magnitude on timescales of days to weeks. Similar variations are present in the SMEI light curves of other nearby stars as well, so we do not consider these to be an intrinsic feature of $\alpha$~Ori. 
%}  

\begin{figure} 
\centering
\includegraphics[width=\columnwidth]{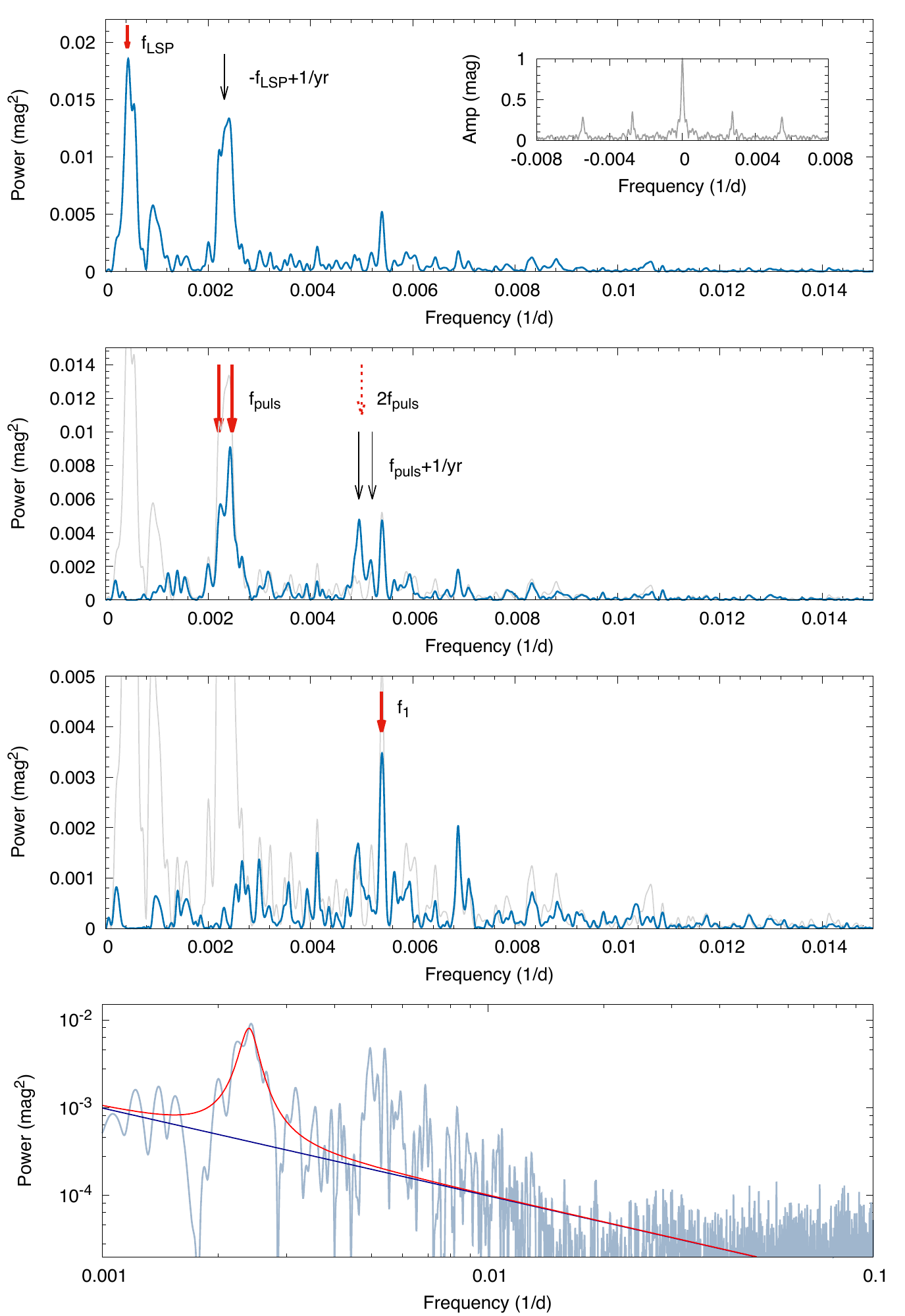}
\caption{Top: power spectrum of the photometric observations. The strongest LSP frequency and the position of a yearly alias are indicated. The insert shows the spectral window function of the data with the prominent yearly aliases. Second plot: spectrum after we removed the LSP signal. Pulsation peaks and aliases indicated. Third plot: residual spectrum after both the LSP and the pulsation frequency removed: $f_1$ marks the significant peak that remained. Bottom plot: power spectrum with the LSP removed from the data, in log space: we fitted the granulation noise component with a $1/f$ function and the pulsation frequency region with a Lorentzian profile. }
\label{fig:powerspectra}
\end{figure}

\subsubsection{Frequency analysis of observations}
\label{sec:freqs}
We analyzed the frequency spectrum of the photometric light curve with Period04 \citep{period04}. We are able to identify the LSP and pulsation frequency regions easily, as shown in Fig.~\ref{fig:powerspectra}, but the identification of individual frequency components intrinsic to the star
was hindered by the presence of yearly aliases. Most notably, the $-f_{\rm LSP}+1/yr$ component coincides with the pulsation frequency region. As the FM pulsation mode itself is only slightly longer than one year, its harmonics and/or overtones could coincide with yearly aliases.

We first apply a pre-whitening procedure to the data with LSP components. Figure \ref{fig:powerspectra} shows that the LSP is not strictly cyclic and that $\alpha$ Ori hovered in a bright state throughout the 2010s. We test combinations of multiple harmonics and subharmonics of the main $f_{\rm LSP} = 0.000423\,{\rm d}^{-1}$ frequency ($P_{\rm LSP} = 2365 \pm 10$~d), which is 15\% longer than that determined by \citet{Kiss2006}, but still within their uncertainty range. We use the 0.5 and $2.5\,f_{LSP}$ components for the final fit, which successfully reproduces the deep LSP minima in 1985/1989 and in 2001/2006--7. 
Non-sinusoidal features in LSP light curves are common for smaller red giants in the Magellanic Clouds: one half of the LSP cycle shows an eclipse-like dip, and the other half resembles a plateau. 
The model proposed by \citet{soszynski-LSP-2014} to reproduce this shape assumes a nearby orbiting companion and associated dense cloud. Currently, there is no indication of a companion orbiting $\alpha$~Ori, but some observations suggest that the recent dimming was likely caused by excess dust\footnote{We note that this view is not held uniformly, however; e.g. \citealt{Dharmawardena2020} do not agree)}. \citep{Levesque2020, Safonov2020}.

We detect two significant frequency components ($f_{\rm puls1} = 0.002469$, $f_{\rm puls2} = 0.002213\,{\rm d}^{-1}$) at the pulsation frequency peak, in agreement with the expected short lifetime of the mode.
We likewise detect the first harmonic ($2\,f_{\rm puls}$) of the stronger pulsation component. 

Since the pulsation signal is non-coherent, we fit it with a Lorentzian profile as in \citet{Kiss2006}, but in combination with a $1/f$ component to account for the red noise component of the convective motions (bottom panel of Fig.~\ref{fig:powerspectra}). 
We calculate a pulsation frequency of $f_{\rm puls} = 0.00240\pm0.00014\,{\rm d}^{-1}$ from the peak of the profile, corresponding to a period of $P_{\rm puls} = 416\pm24\,{\rm d}$. We can also use the the full width at half maximum ($\Gamma$) of the profile to estimate a mode lifetime of $\tau = 1/\pi\Gamma = 1174 \,{\rm d}$, or $\approx 3$ pulsation cycles. The mode lifetime matches the value calculated by \citet{Kiss2006}; the pulsation period, however, is 7.2\% longer, though still within their uncertainty range. We note that \citet{dupree1987} determined a similar, 420 d period, but this was based on only three years of photometric observations. 

Apparent changes to the period likely arise from (1) the non-coherent nature and short lifetime of the mode and (2) interference with photometric variations caused by convective motions and the evolution of hot spots. 
%Although the visual data are less accurate than the photometry, the visual data average the pulsation over a considerably longer time. 
Differences of up to 15\% among apparent periods calculated from shorter and longer data sets have been found for other RSGs as well \citep{Chatys2019}.
Presently, we report a new period for the photometric data covering only the last three decades; disentangling the temporal evolution of the pulsation is beyond the scope of this work. 
%We therefore prefer to use the $388\pm30$~d period determined by \citet{Kiss2006} as reference in this work.

\subsubsection{Detection of the first overtone}
After pre-whitening the data using these frequencies, one significant peak remained at $f_1 = 0.005392\pm0.000002\,{\rm d}^{-1}$ ($P_1 = 185.5\pm0.1$\,d)\footnote{Uncertainties for $f_1$ were calculated with the assumption of a single coherent Fourier component: more data will be needed to assess the validity of this assumption.}.
Neither this component nor the harmonic was described by \citet{Kiss2006}, nor is it present in the power spectrum of the complete visual light curve. However, $f_1$ can be identified in some segments. This peak could suggest the presence of the first overtone with 
a period ratio of $P_1/P_0 = 0.445\pm0.014$ (using the Lorentzian fit to $P_0$) in $\alpha$ Ori. Overall, we expect the $P_1/P_0$ period ratio to be $\sim 0.5$ for red giant and supergiant variables, though model predictions typically focus on lower mass ranges \citep{fox-wood-1982,kiss1999}. We discuss period ratios derived from our own models in Section \ref{sec:o1}. 

Multi-periodicity is not uncommon among red giants and supergiants. \citet{kiss1999} detected more than one periodicity in 60\% of 93 well-observed, semiregular variable stars. Although some of these were a combination of pulsation and LSP signals, 30\% of the stars clearly pulsated in the fundamental mode and first overtone simultaneously. We therefore consider it likely that $\alpha$~Ori also pulsates in more than one mode, though we cannot conclusively exclude the possibility that the $f_1$ signal detected in our analysis corresponds to a yearly alias or a harmonic of the non-coherent pulsation signal that the photometric data do not resolve properly.

It would be informative to collect photometric observations of $\alpha$ Ori throughout the year for as long as possible in order to minimize the gaps in the data and diminish such aliasing in the frequency domain.  

\subsection{Timing of minima}
A standard means of identifying deviations from an assumed periodic signal is the O--C method.\footnote{O--C refers to the observed minus calculated method, where we measure the time differences between observed events (e.g., cycle minima or maxima) and a periodic ephemeris.} Here we attempted to identify and time the various larger and smaller minima in the light curve. The light curve data appear to alternate between two states: one defined by deep minima exceeding 0.5 mag (e.g., at JD 49800, 52750, 54000, 54400, 58500 and the dip itself at 58800), the other by shallower and more frequent meandering (e.g., around JD 51500, 53200, 55000 to 57000). 
However, the annual gaps make it difficult to identify enough minima accurately, and it is thus possible that we simply miss one type during certain intervals. Time spans between consecutive shallow minima can be as short as 60--100 days---much shorter than the FM pulsation period. We see no indication of discrete frequency components corresponding to these intervals in the power spectrum of the star, which suggests they are not high-degree pulsation modes.
The timescales and low amplitudes, however, do match the convective turnover times of giant convection cells:
our photometric results agree with predictions of timescales from 3D radiative hydrodynamic models and the time-resolved results of spectropolarimetric observations of the surface of the star \citep{freytag2002,Lopez2018}. 
We therefore attribute these short-timescale, stochastic variations to the photometric effects of convective cell turnover on the surface of $\alpha$~Ori.

The critical observational features of Betelgeuse are summarized in Table \ref{table:obs}.
%
%% second dust cloud comment
We add a final note that observations released since the presentation of the lightcurve in this work show a new brightness minimum occurring less than half a pulsation period later, but at a regular depth of $V \sim 1.0$ mag \citep{atel-stereo,atel-ground2020}. These observations are consistent with behavior seen in our longitudinal data, particularly those periods in which the short-lifetime mode shifts in phase abruptly. It is likewise consistent with the photometric effects of changing convective cells. Our results thus demonstrate that these two phenomena are sufficient to explain the subsequent, short-term semiregular variability of $\alpha$~Ori without invoking explanations such as multiple, opaque dust clouds orbiting the star.

%\section{Modeling}
\section{Classical Evolutionary Models}
\label{section:evolution}

Having carefully collated the set of observational criteria described in Table \ref{table:obs}, we proceed in modeling the system. Our numerical efforts include three types of simulation: (1) classical evolutionary tracks; (2) linear pulsation models; and (3) short-timescale, 1D, implicit hydrodynamical evolution. We discuss results from each in this order.

We compute evolutionary tracks for stellar models with initial masses of $10$--$26~M_{\odot,i}$. Calculations are carried out from the pre-main sequence to the termination of the helium-burning giant branch, with the terminating condition set by the amount of helium remaining in the core of the star ($M$($^4$He) $\sim10^{-8} ~M_{\odot}$).
Models in an evolutionary phase more advanced than core helium burning are less favored probabilistically, as the star will spend considerably less time in such phases. As shown above, they are also unlikely to be consistent with the existing array of observational constraints, especially since these constraints prove to be discriminating even within the set of strictly core helium-burning models. As such, we do not consider post-core helium-burning models in further detail.

\begin{figure} 
\centering
\includegraphics[width=\linewidth]{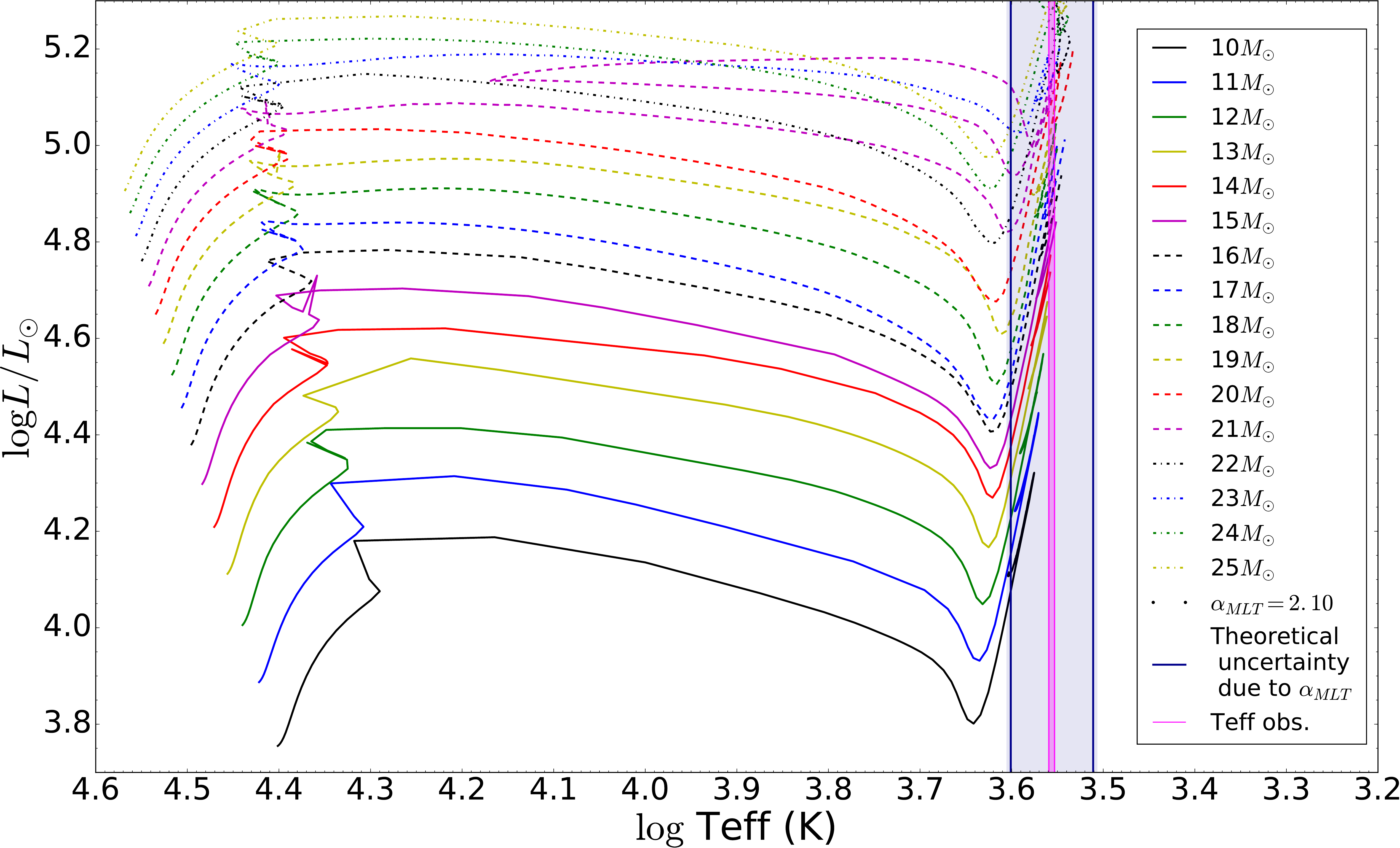}
\includegraphics[width=\linewidth]{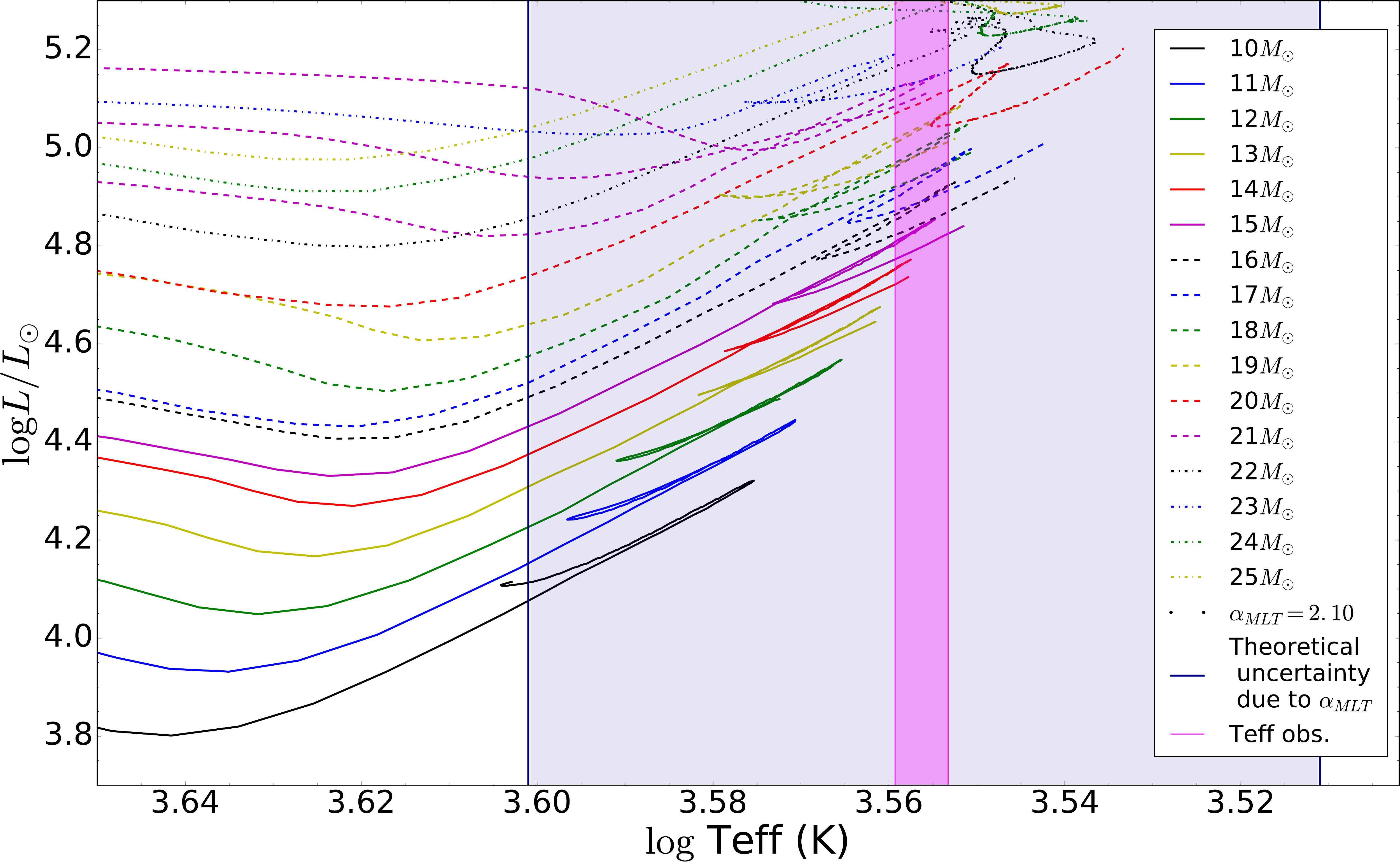}
\caption{
\\(UPPER) A set of classical evolutionary tracks for $10$--$25\, M_{\odot}$ computed with MESA. Initial mass per track as indicated. All models shown are computed until the end of helium burning and shown from ZAMS. All tracks adopt a mixing length of $\mlt/=2.1$.
(LOWER) Same as above, but rescaled to highlight the evolutionary region consistent with the temperature constraints provided by \citet{Levesque2020}, shown in pink. The extended temperature regime permitted by the mixing length degeneracy is shown in blue.
}
\label{fig:setoftracks}
\end{figure}

Figure \ref{fig:setoftracks} shows a set of evolutionary tracks evolved from the zero-age main sequence (ZAMS) to the end of core helium burning. Masses indicated refer to the initial 
%(i.e. progenitor) 
mass. 
In subsequent discussion, we refer to models by their initial masses, though typically the mass of the star will be between $2$--$3 ~M_{\odot}$ smaller at the termination of its evolution (and onset of its hydrodynamic evolution) due to mass loss during its prior stages.
%, primarily during the red giant phase. 

Our initial grid of models does not invoke rotation and has fixed, solar metallicity represented by a heavy metal fraction of $Z_{\text{in}}=0.02$. We consider multiple values of the convective mixing length \mlt/, ranging from $1.8$ to $2.5$.
As massive stars are quite sensitive to the prescriptions used for convective boundaries and convective overshoot, we adopt convective overshoot settings of $f_{\text{ovs}} = 0.010 H_p$\footnote{Multiples of the pressure scale height, $d\ln P/d\ln T$.} surrounding hydrogen- and helium-burning zones \citep{Herwig2000, MESAIV}. 
We set convective boundaries according to the Schwarzschild criterion, and we do not include semi-convection\footnote{only available when the Ledoux criterion (i.e. composition gradient) is used to set convective boundaries}. We use the Cox prescription of the mixing length formalism \citep{Cox1968}. We do not use the MLT$++$ option in the calculation of our evolutionary models. Our surface boundary conditions are set by a simple photosphere model, whose implementation is described in \citet{MESAI}.
%Add discussion of choices for semi-convection, MLT$++$, surface boundary conditions

%
We account for mass loss in the evolutionary calculations via MESA's implementation of the ``Dutch'' wind schemes, a composite of prescriptions summarized in \citet{Reimers75, deJager88, Bloecker95} and \citet{vanLoon2005}.
We model the low-temperature mass loss via the prescription of \citet{deJager88}, adopting a wind coefficient of $\eta=0.8$ as default.

We test a range of $\eta$ values and find that while the choice of $\eta$ does impact the terminal mass of the evolutionary model, our results are predominantly sensitive to the radius. The relationship between an evolutionary model's terminal radius and its input controls---mass, metallicity, mixing length, convective overshoot, mass loss coefficient, etc.---is complex, and we do not gain much insight on this interplay by varying $\eta$. 
We do not use mass loss or rotation during the hydrodynamic evolution itself, as the impact of these processes on a timescale of several decades is negligible.

A critical component of our classical modeling objective involves reproducing the recently observed temperature of $\alpha$ Ori. However, given limited a priori information on the star's mass and evolutionary phase, there is a strong degeneracy between choice of \mlt/ and predicted temperature. While this issue emerges even for well-constrained systems \citep{Joyce2018a, Joyce2018b}, the magnitude of the degeneracy is exacerbated as observational constraints loosen and the structural complexity of the stellar models increases.
Hence, even if we assume that the atmospheric models used by \citet{Levesque2020} can determine the temperature corresponding to the observed line profile with high accuracy and precision, the underlying evolutionary models themselves may shift by about $\pm200$~K. This introduces, at minimum, the same uncertainty on the evolutionary stage at which the star crosses the observed temperature.
It is likewise prudent to extend the uncertainties on our temperature measurements anyway, as RSG temperatures are notoriously difficult to infer. In particular, \citet{Davies2013} note that an SED fitting approach would give a higher temperature for Betelgeuse than the one we adopt here, and these authors raise concerns about the validity of the TiO-band fitting method use by \citet{Levesque2020}. 
Likewise, \citet{IrelandScholzWood2008} find that fits based on TiO and made under the assumption of LTE can give temperatures that are much too low; based on these findings, Betelgeuse could easily have an effective temperature that is hotter by 200 K.
A looser interpretation of the temperature constraints is implemented in practice by extending the observational boundaries by a theoretical uncertainty of appropriate order. This is done by measuring the shift in temperature a track of fixed mass undergoes when its mixing length is adjusted to extremal values. In the case of our grid, this shift is calculated for $\mlt/=1.8$ vs $\mlt/=2.5$ and corresponds to a shift in modeled temperature, in the relevant part of the HR diagram, of roughly 0.1 dex for a track with middling mass $17 ~M_{\odot}$. 
We likewise note that the mixing length is not the only parameter that contributes to uncertainties in the derivation of RSG temperatures; others include convective overshoot and semi-convection, both of which also affect the appropriate choice of mixing length itself. We do not account for variations in either in this study, but refer to \citet{Chun2018} for a detailed analysis of the impact of theoretical assumptions on RSG temperatures.

We choose to account, approximately, for variations among temperature estimates caused by differences in observational inference methods and theoretical parameter choices as follows: 
We expand the observational temperature constraints by an amount equal to the shift in modeled temperature for two otherwise identical stellar tracks that adopt minimal and maximal values, respectively, of \mlt/. This adjusted temperature boundary is indicated by the blue strip in Figure \ref{fig:setoftracks}. The effective temperature constraints of \citet{Levesque2020} alone are shown in the much more restrictive pink band. 
The observational constraints on luminosity are not strong and do not themselves rule out any of the models shown in Figure \ref{fig:setoftracks}.

Attempts to reproduce Betelgeuse's present-day rotation of $\sim5$~km\,s$^{-1}$ ($v\sin i = 5.47 \pm 0.25$~km\,s$^{-1}$, \citealt{uitenbroek1998,Kervella2018}) with single-star evolutionary models are unsuccessful.
To this end, we compute tracks that use an initial surface rotation of up to $\Omega=0.65 \Omega_{\text{crit}}$, or roughly $200$ km/s on the ZAMS, in accordance with \citet{BetelgeuseProjectI}. In cases where the models do not fail outright, the results are not consistent with even the most generous interpretation of the observational constraints. Among tracks that converge, even those with the highest values for $\Omega_i$ still fail to predict a present-day rotation rate in the vicinity of the observed value.

In particular, tracks with initial rotations approaching breakup velocity ($ \Omega /\Omega_{\text{crit}}\sim 0.7$) fail to intersect the (extended) effective temperature regime with large enough present-day surface rotations. The highest values attained by our grid only just reach 1 km/s, and these correspond to models with initial masses as low as 6--10 $M_{\odot}$. Such low-mass models are easily ruled out by other constraints, especially period.

Our results from this exercise are thus similar to those of \citet{BetelgeuseProjectI}, who find that ``models at the tip of the RSB typically rotate at only $\sim0.1$ km/s, independent of any reasonable choice of initial rotation.'' Though \citet{BetelgeuseProjectI} are able to create rotating models consistent with $3 \sigma$ uncertainties on their observational constraints at the time, our constraints prioritize the fundamental mode frequency and include a much tighter range on effective temperature. 
More sophisticated modeling of the rotational aspects of $\alpha$ Ori's evolution are beyond the scope of this paper.

The terminal models from the evolutionary run provide both the structural input for calculations with the linear pulsation program (next section) and the initial conditions for the hydrodynamic study (Section \ref{section:hydro}).

\section{Seismic Models}
\label{section:seismic}
\begin{figure} 
\centering
\includegraphics[width=\columnwidth]{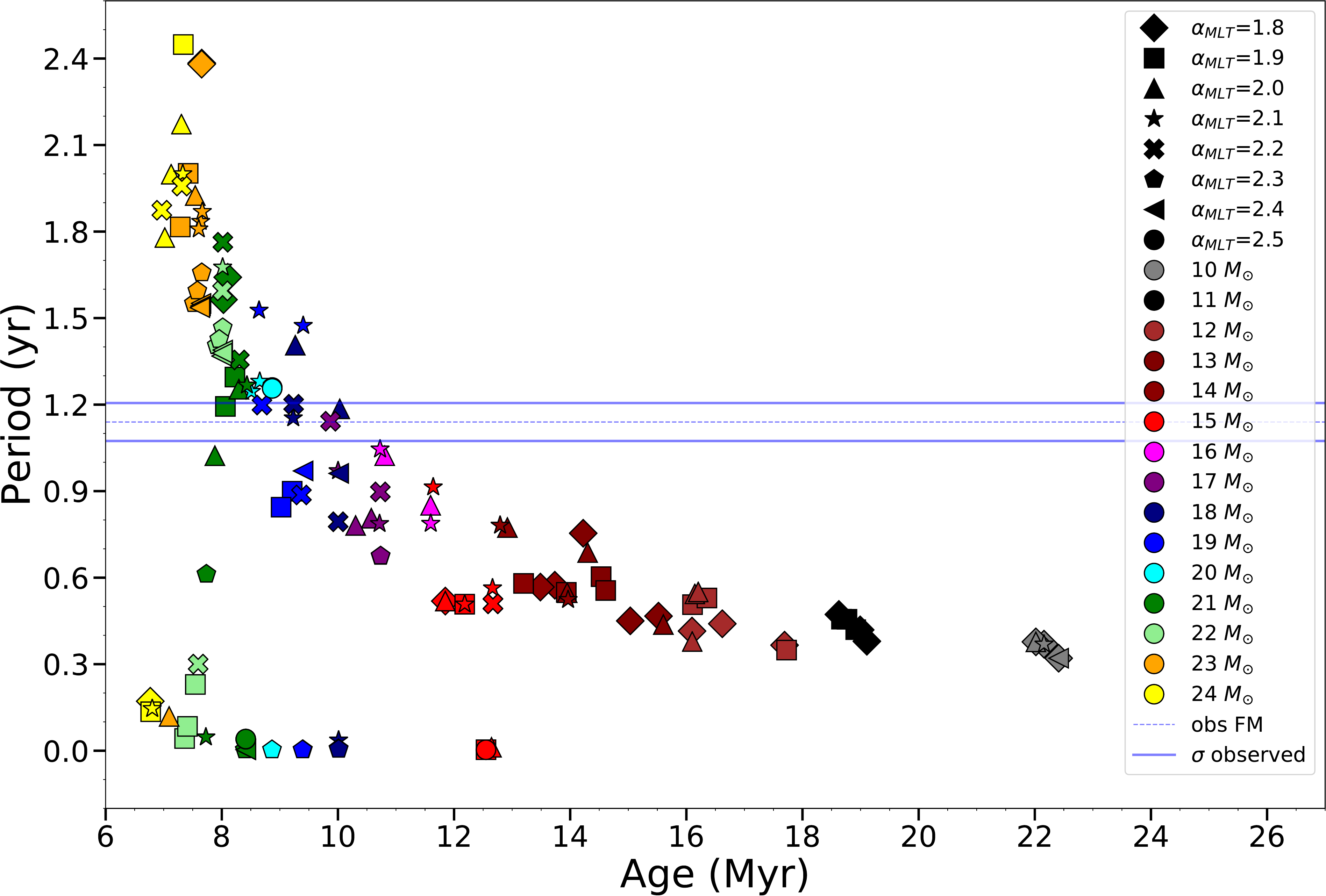}
\includegraphics[width=\columnwidth]{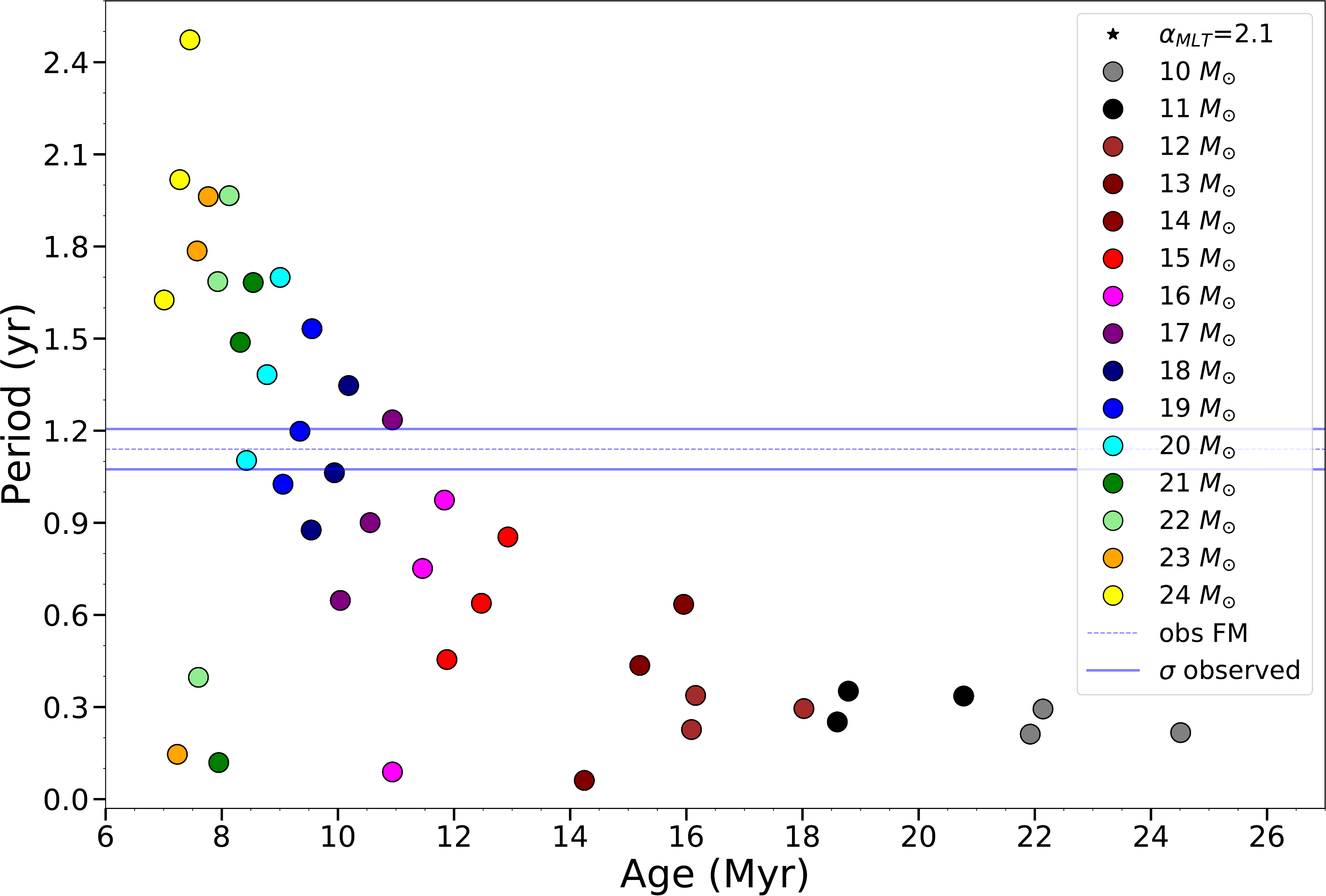}
\caption{ 
Adiabatic $p$-modes are calculated with GYRE for all relevant evolutionary tracks. Periods, in days, of all models consistent with the observed temperature constraints are shown, coded by color for mass and by marker style for mixing length, as indicated.
(UPPER) Masses range from $10$--$24\, M_{\odot}$ at a resolution of $1.0 M_{\odot}$ and mixing lengths range from $1.8$ to $2.5$ at a resolution of $0.1$. 
(LOWER) Masses range from $10$--$24\, M_{\odot}$ at a resolution of $1.0, ~M_{\odot}$ and \mlt/ is fixed at $2.1$. Here, the observed temperature constraints adjusted to account for the theoretical uncertainty in \mlt/.
All models shown adopt $\eta=0.8$ and $Z=0.02$.
The observed seismic constraints from \citet{Kiss2006} are indicated with blue horizontal lines. 
}
\label{fig:synth_periods}
\end{figure}

%%%%%%%%%%%%%%%%%%%%%%%%%%%%%%%%%%%%%%%%%%

\begin{figure} 
\centering
\includegraphics[width=\columnwidth]{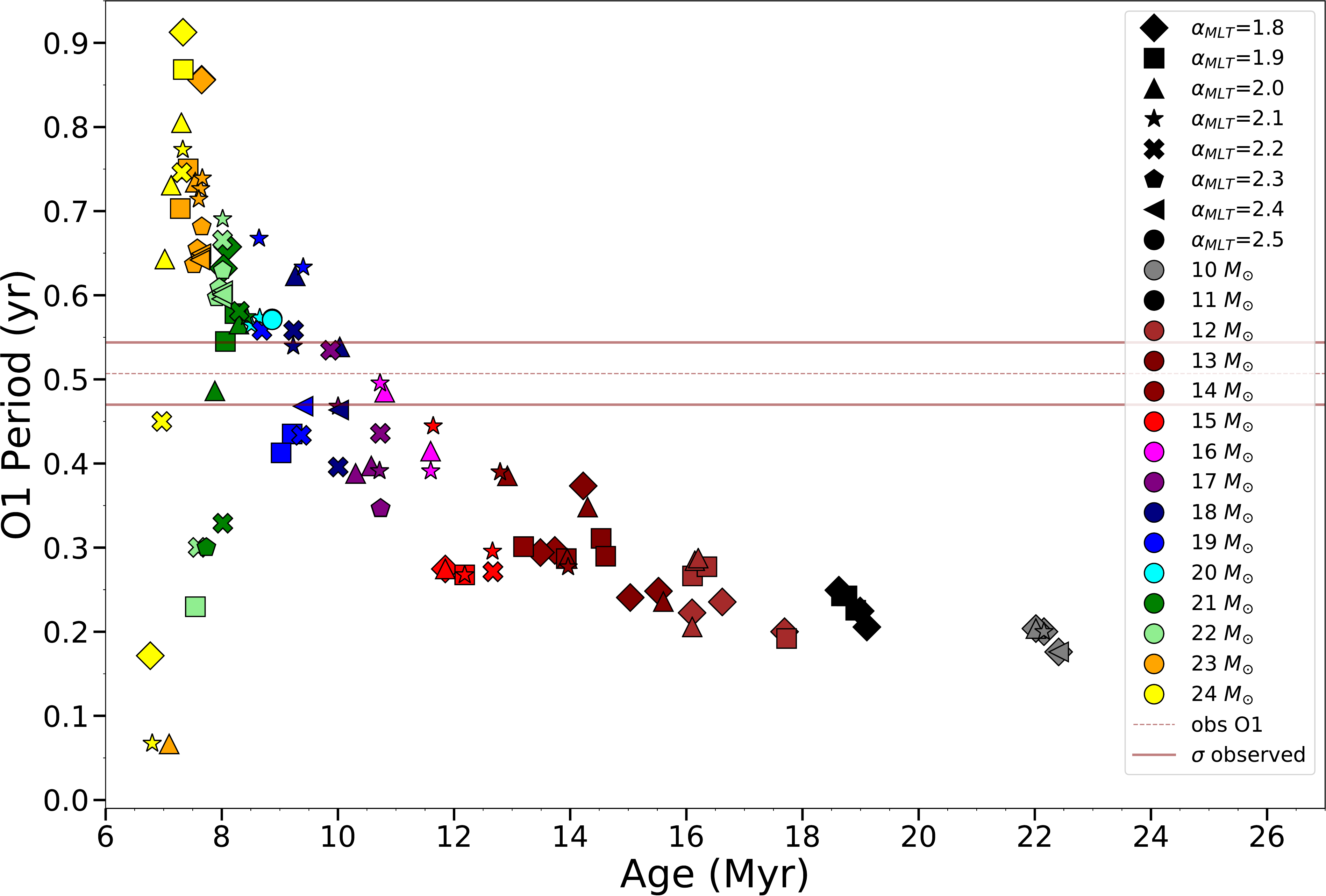}
\includegraphics[width=\columnwidth]{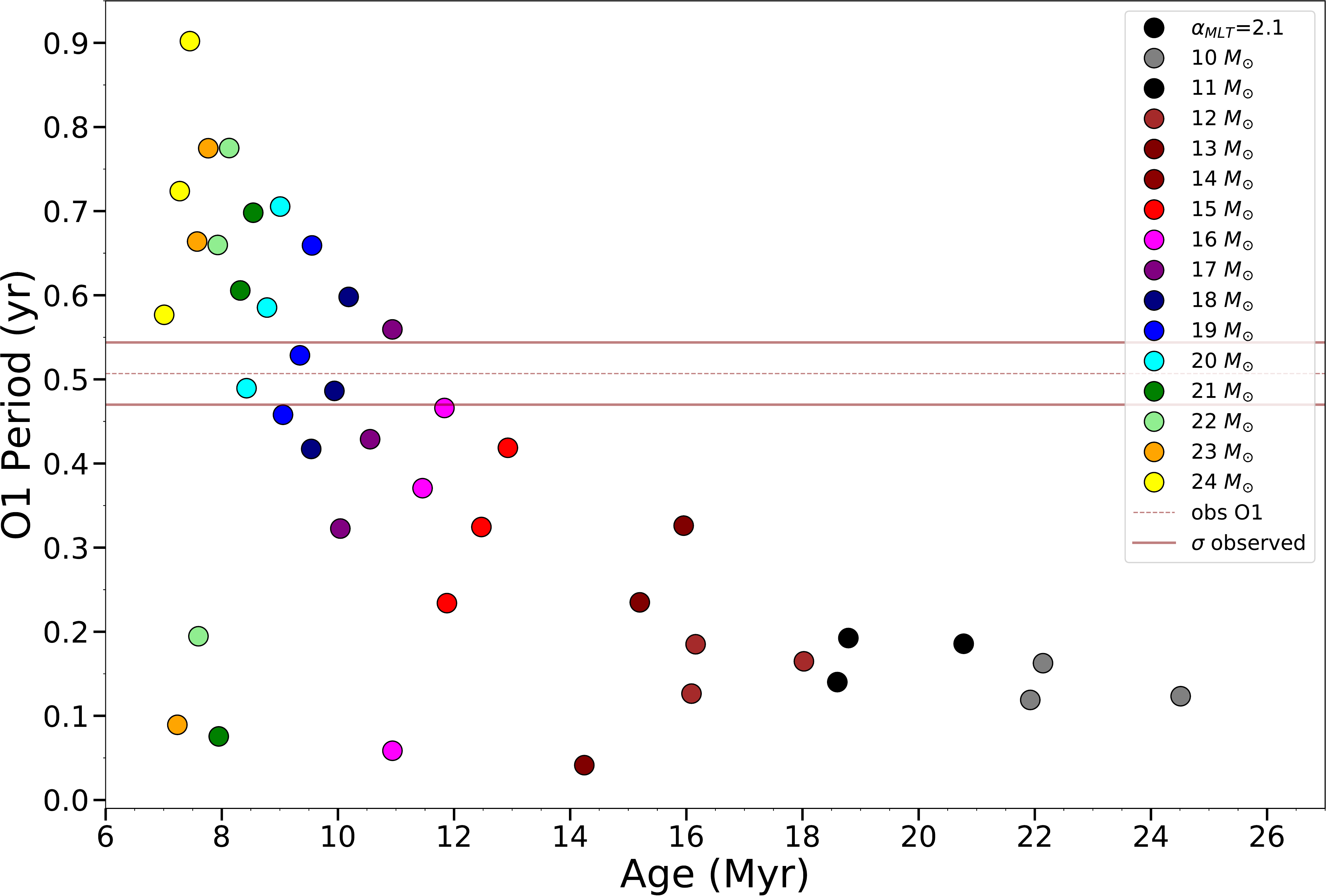}
\caption{ 
Same as Figure \ref{fig:synth_periods}, but for the first overtone, O1. Uncertainties have been scaled by the O1/FM ratio, yielding $\pm13.5$ days, shown with red horizontal lines. }
\label{fig:synth_periods_O1}
\end{figure}

%%%%%%%%%%%%%%%%%%%%%%%%%%%%%%%%%%%%%

%% fundamental mode
\begin{figure} 
\centering
\includegraphics[width=\columnwidth]{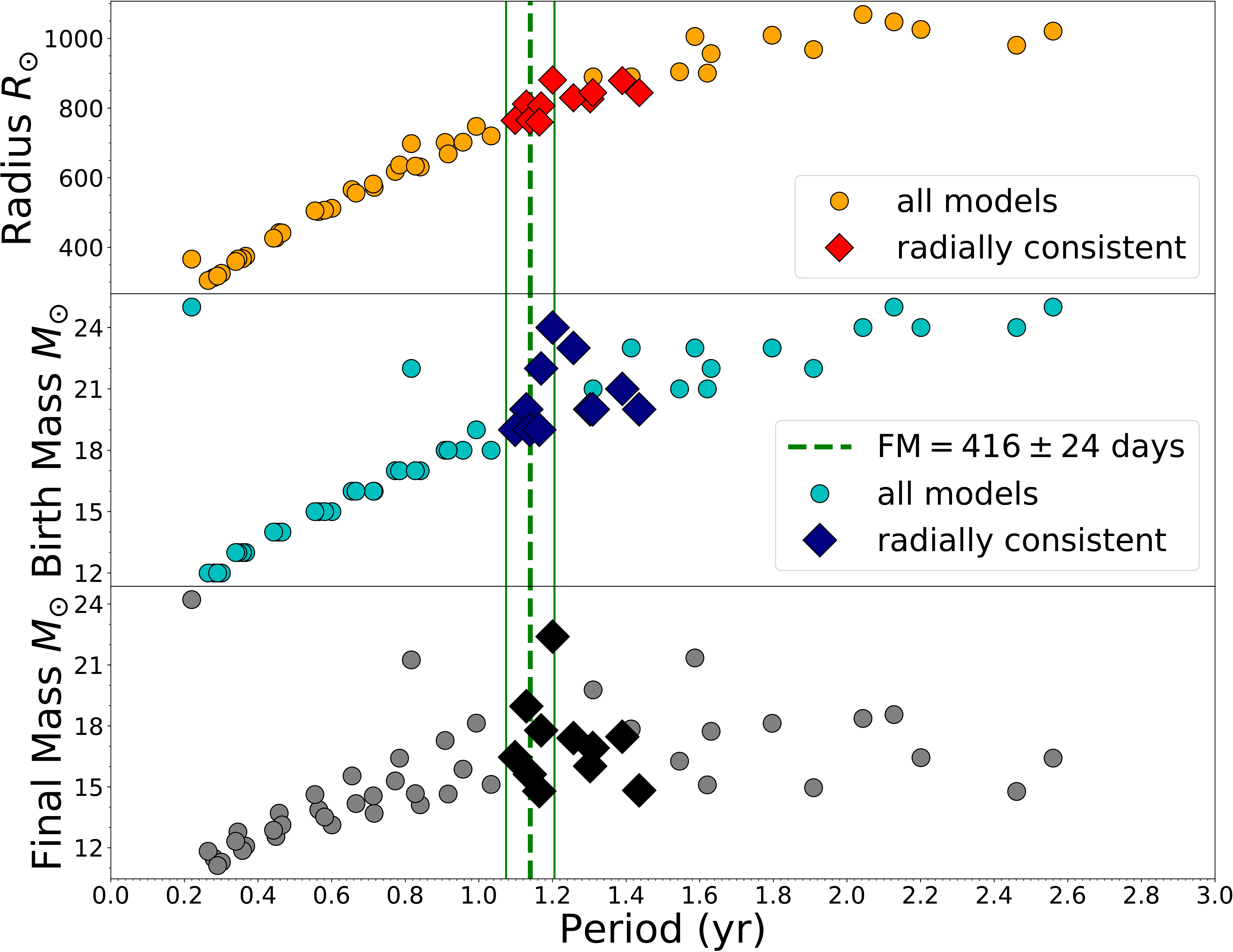}
\vskip6pt
{\footnotesize Graph 1: Fundamental mode; variable mixing length}
\vskip6pt
\includegraphics[width=\columnwidth]{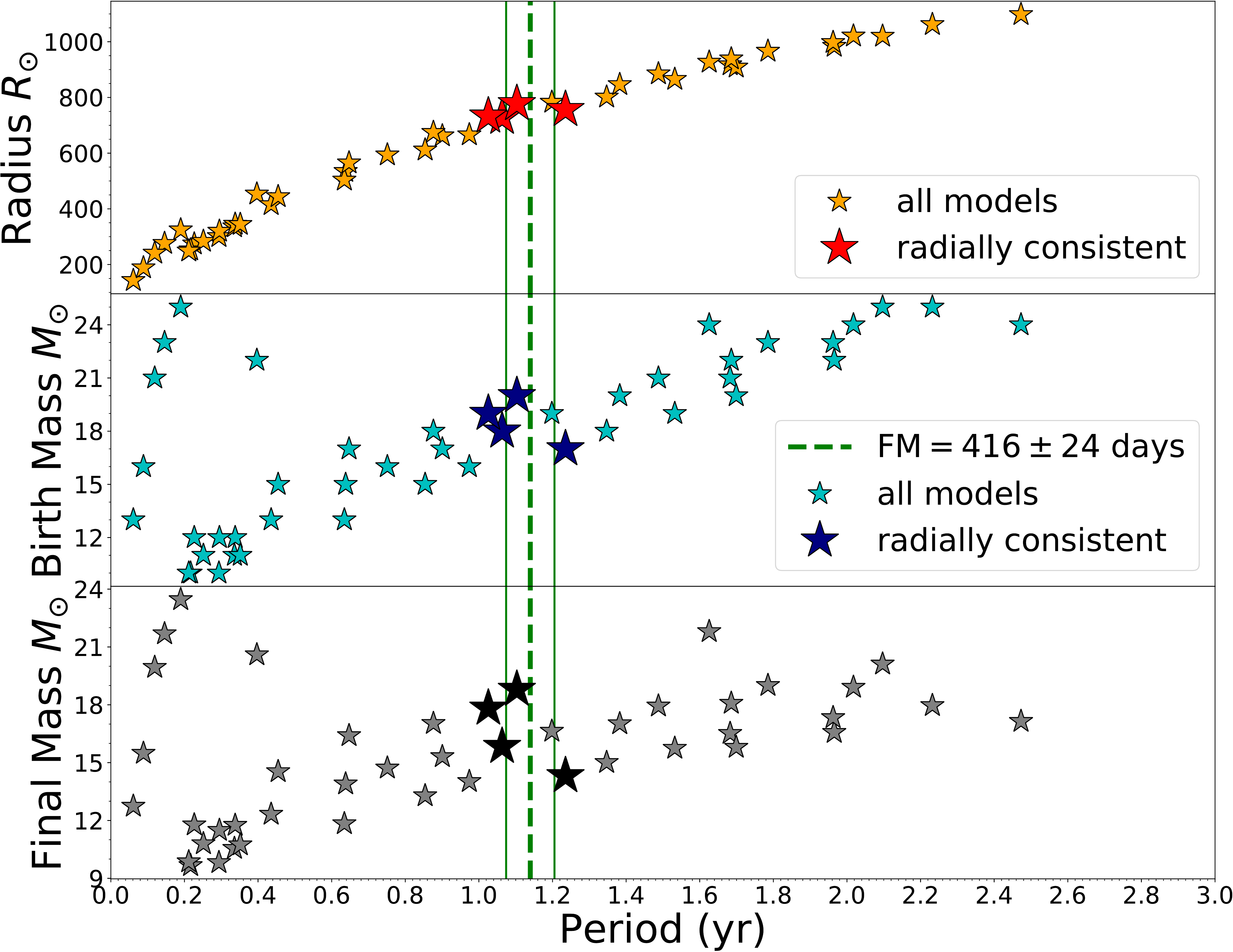}
{\footnotesize Graph 2: Fundamental mode; fixed mixing length}
\caption{Additional parameters, including present-day radius and mass, are shown as a function of period. Period constraints are shown as green, vertical lines. Models in the middle and lower panels of each graph are emphasized if they are also compatible with the radial bounds set by the intersection in the corresponding radius vs period graph.
(UPPER) A random sample of models spanning initial mass, mixing length, mass loss parameter, and degree of helium exhaustion, not pre-selected for agreement with any temperature constraints.
(LOWER) All models with $\mlt/=2.1$, $\eta=0.8$, and terminal conditions determined by agreement with effective temperature, including theoretical uncertainties. 
}
\label{fig:radial_band}
\end{figure}

%%%%%%%%%%%%%%%%%%%%%%%%%%%%%%%%%%%%%%

%% copy, for first overtone
\begin{figure} 
\centering
\includegraphics[width=\columnwidth]{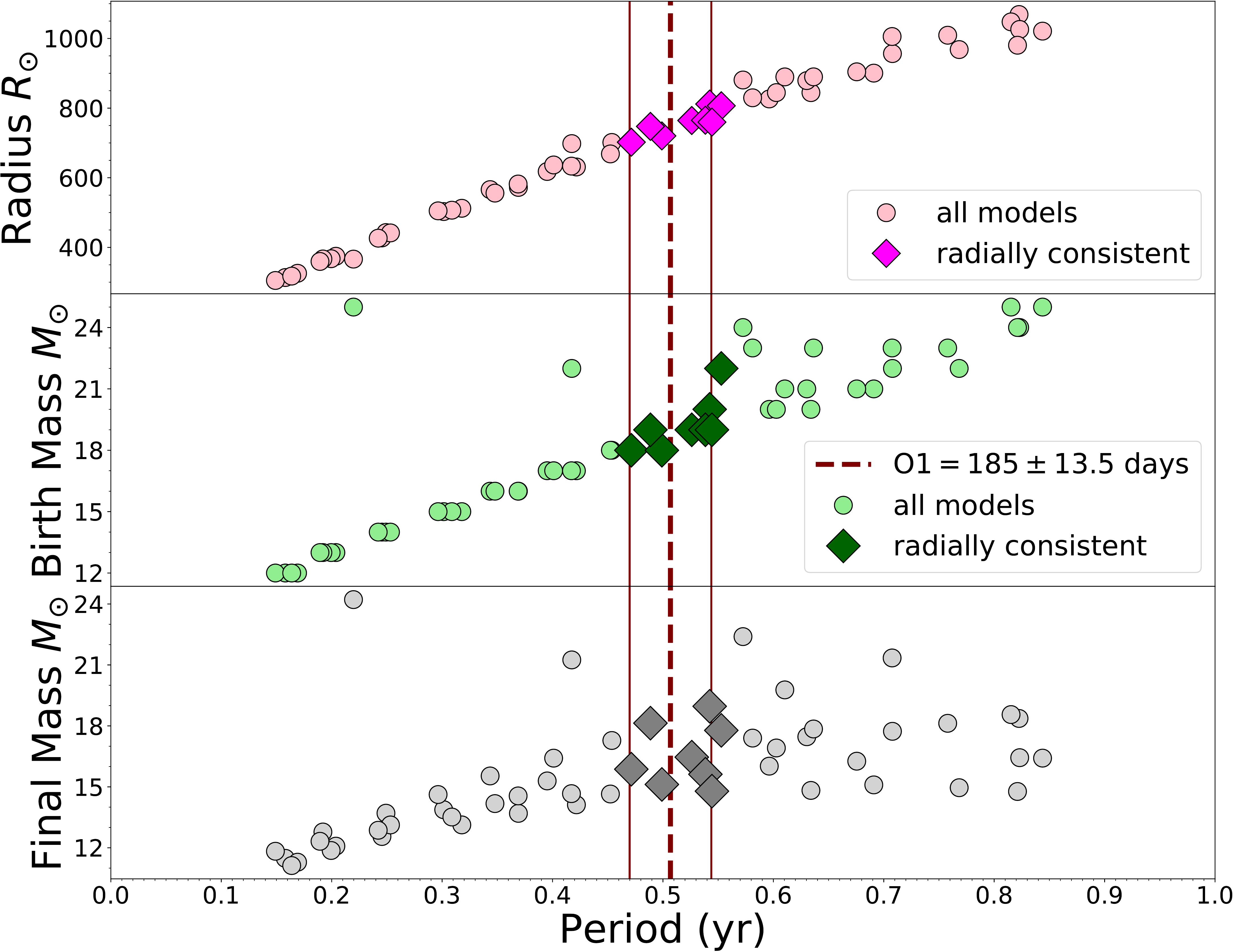}
\vskip6pt
{\footnotesize Graph 3: First overtone; variable mixing length }
\vskip6pt
\includegraphics[width=\columnwidth]{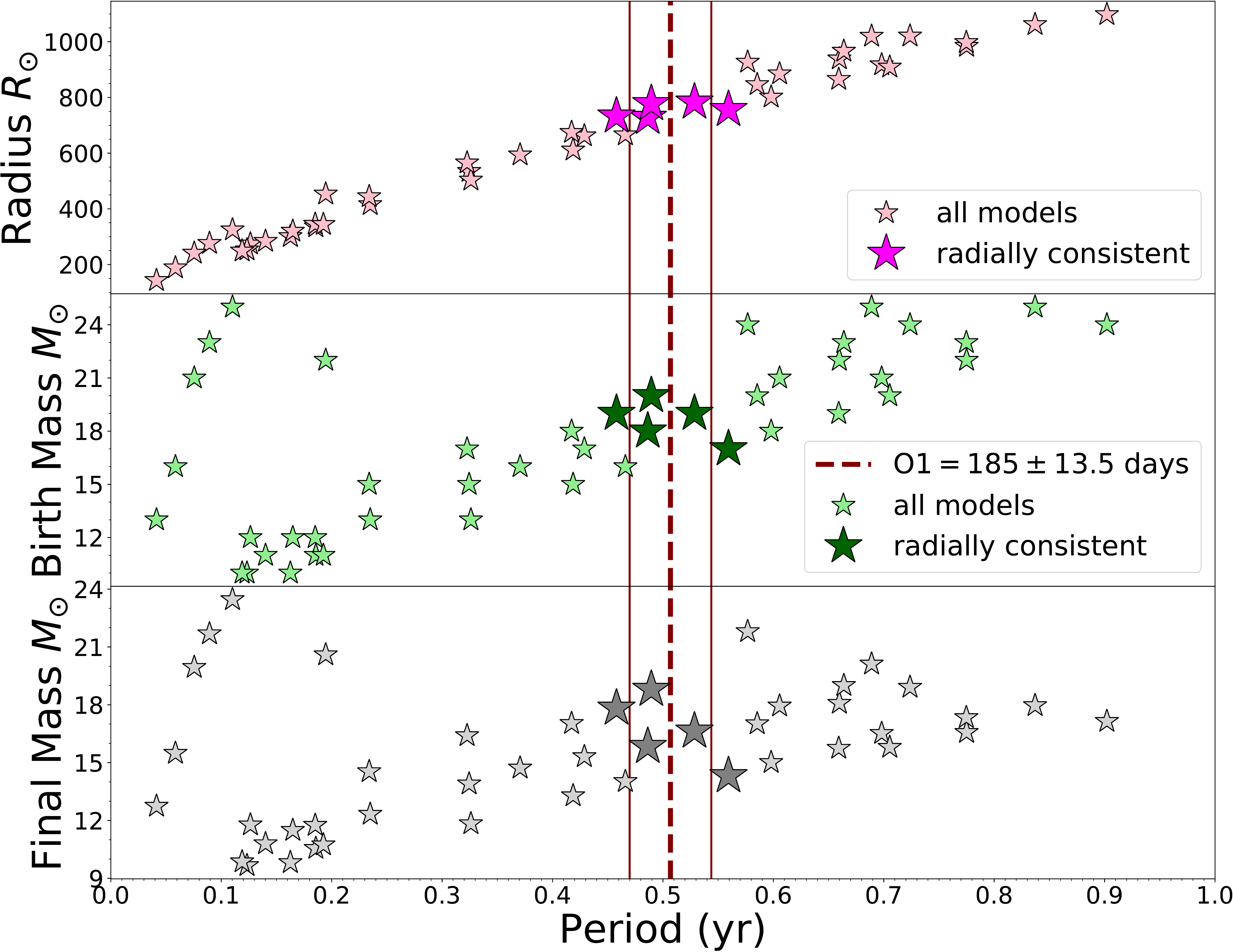}
{\footnotesize Graph 4; First overtone, fixed mixing length}
\caption{Same as Figure \ref{fig:radial_band}, but derived according to seismic agreement with the first overtone, O1.
}
\label{fig:radial_band_O1}
\end{figure}

%%%%%%%%%%%%%%%%%%%%%%%%%%%%%%%%%%%%

Used in conjunction with classical parameters, synthetic frequencies are an extremely powerful tool for discriminating among possible models of a star. The case of $\alpha$ Ori is no exception.

\subsection{Linear perturbations }
\label{subsection:linear_perturbations}
We use GYRE to solve the linearized pulsation equations for high-resolution structural models produced during the RSG phase \citep{GYRE}. The GYRE program is based on a Magnus Multiple Shooting (MMS) scheme and provides both adiabatic and non-adiabatic calculations. We consider only adiabatic results in this analysis. Figures \ref{fig:synth_periods} through \ref{fig:radial_band_O1} show results from these calculations.

As a track that intersects the observational requirements will typically do so at multiple evolutionary timesteps, we can produce several pulsation profiles per track. Where the models are compatible, we generate synthetic frequency spectra at short intervals.
In our frequency modeling, we do not restrict to a search for FMs ($n_{pg}=1$\footnote{The lowest radial order for pressure modes, as defined by GYRE}, $l=0, m= 0$) alone; rather, we note the modes and values of any frequencies in the vicinity of Betelgeuse's two dominant periodicities. However, it is immediately clear from our calculations that periods constituting the primary decaying sequences in both panels of Figure \ref{fig:synth_periods} are fundamental mode periods. 

To account for the theoretical uncertainty on $T_{\text{eff}}$, we consider two metrics by which an evolutionary track is compatible with the observations.
In the upper panel of Figure \ref{fig:synth_periods}, we show the periods of adiabatic $p$-modes versus termination age for a collection of models with a range of initial masses and mixing lengths; here, the structural models used in the seismic analysis have effective temperatures strictly within $3600 \pm 25$ \citep{Levesque2020}.

In the lower panel of Figure \ref{fig:synth_periods}, all models use $\mlt/=2.1$, but are checked for consistency against the extended, theoretical temperature constraints described in Figure \ref{fig:setoftracks}. 
In both panels of Figure \ref{fig:synth_periods}, all masses refer to the initial mass of the model and the $416 \pm 24$ day periodicity is denoted by a blue horizontal band. 

In the upper panel of Figure \ref{fig:synth_periods}, the period--age sequence is tighter and more well-defined, but the results between the upper and lower panels are largely consistent. We note that sub-sequences comprising stars of particular mass are more apparent in the fixed \mlt/ case. This visualization more clearly suggests that, at least for masses in the $15$--$20 ~M_{\odot}$ range, there will necessarily be some point along the helium-burning branch during which the star will pass through the appropriate frequency band. However, the temporal window during which this occurs is quite small in the context of evolutionary timescales--on the order of $0.5$--$1.0$ Myr. The requirement that this time frame align with a particular observed temperature ends up being quite restrictive. 

Collectively, these results suggest a median, model-derived mass range of $16$--$21~M_{\odot}$ (with some outliers as high as $24M_{\odot}$), at a resolution of $1~M_{\odot}$. This is broadly consistent with other modellers' results, though our results are more accommodating at the lower-mass end.

We repeat this analysis using as a period constraint the probable overtone observed in our photometry (Section \ref{sec:freqs}) and confirmed in our synthetic frequency spectra: an O1 mode oscillating with a period of $185$d. 
In lieu of an independent value, we scale the uncertainty of the fundamental mode to derive an uncertainty on the overtone of $13.5$d, yielding $P_1 = 185 \pm 13.5$. 
This is shown in the red, horizontal bars of Figure \ref{fig:synth_periods_O1}. Though suggested by our photometric analysis, confirmation of the detection of this mode and its classification required supporting evidence from theoretical spectra. This argument is detailed further in Section \ref{sec:o1}.

Figure \ref{fig:radial_band} shows other fundamental parameters as a function of period. The FM and its uncertainties are defined by green, vertical bars in all panels.
An analog using the O1 period and its uncertainties, defined by dark red vertical lines, is shown in Figure \ref{fig:radial_band_O1}. 

Models in the upper panels of Figures \ref{fig:synth_periods} and \ref{fig:synth_periods_O1} span the full range of masses and mixing lengths considered in our grid and additionally vary in the prescribed mass loss coefficient ($\eta = 0.2$--$1.0$), but they are not pre-restricted by agreement with temperature constraints. Instead, these evolutionary tracks are terminated at arbitrary intervals along the helium-burning branch, with spacing set by the degree of helium exhaustion in the core. This is done to produce a more well-populated sequence that incorporates additional sources of uncertainty in the modeling assumptions. Despite this added theoretical noise, the range of possible radii across all models remains heavily restricted by the observational period constraints.  

All models in the lower panels of Figures \ref{fig:synth_periods} and \ref{fig:synth_periods_O1} intersect the theoretically extended temperature uncertainties (which essentially sets their termination ages) and adopt $\mlt/=2.1$ and $\eta=0.8$.

In the uppermost panel of Graph 1 in Figure \ref{fig:radial_band} and Graph 3 in Figure \ref{fig:radial_band_O1}, we show radius as a function of FM and O1 period, respectively. Though there is some scatter in the synthetic data, the radial span of the period--compatible models is very narrow in both cases, especially compared to the range of radial estimates collated in \citet{Dolan2016}. We recall from earlier discussion that literature estimates of Betelgeuse's radius range from $\sim500 R_{\odot}$ to nearly $1300~R_{\odot}$, whereas the results presented here suggest little possibility outside $700-900~R_{\odot}$.
%
% the 3sigma limits are wider than this range up here
% yeah this entire thing is a fuck up, working on it
%
If we interpret the FM period measurements as hard limits, our results suggest a radius for $\alpha$ Ori of $764~R_{\odot}$, with $1\sigma$ uncertainties of roughly $30~R_{\odot}$ and non-symmetric limits of
$R_{\text{max}} = 880~R_{\odot}$ and $R_{\text{min}} = 702 ~R_{\odot}$. Figure \ref{fig:radial_band_O1} suggests consistent but even tighter limits of approximately $700-800R_{\odot}$
We thus report a $3\sigma$, model-derived radius of $764 ^{+116}_{-62} ~R_{\odot}$.
% WARNING: DISCUSS ADDITIONAL CONSTRAINTS FROM O1 Figure \ref{fig:radial_band_O1}

%
In the middle and lower panels of each graph, we show the models' initial masses and terminal masses, respectively, as a function of period. In these plots, we emphasize those models that also have radii within our $3\sigma$ uncertainty bounds with larger, darker markers. Considering all possible observational constraints, we report model-derived estimates for the initial and present-day masses of Betelgeuse as approximately $18$--$21~M_{\odot}$ and $16.5$--$19~M_{\odot}$, respectively. 
Though best mass ranges differ slightly among them, these represent median values collated from the four variants of this test (Figures \ref{fig:radial_band} and \ref{fig:radial_band_O1}), encompassing two interpretations of the mixing length--temperature degeneracy for each of {$P_0, P_1$}. The lower rows of Table \ref{table:obs} summarize these findings. 
Taking into account the likelihood of a previous merger event, which would significantly complicate any inferences about the state of Betelgeuse at birth, it is our present-day mass estimates that are most pertinent.

%%%% REORDERED FIGS 7-9 SO THAT THEY CAN BE REFERENCED IN 4.2

%%%%%%%%%%%%%%%%%%%%%%%%%%%%%%%%%%%%%%%%%%

%% period ratio fig
\begin{figure}
    \centering
    \includegraphics[width=\columnwidth]{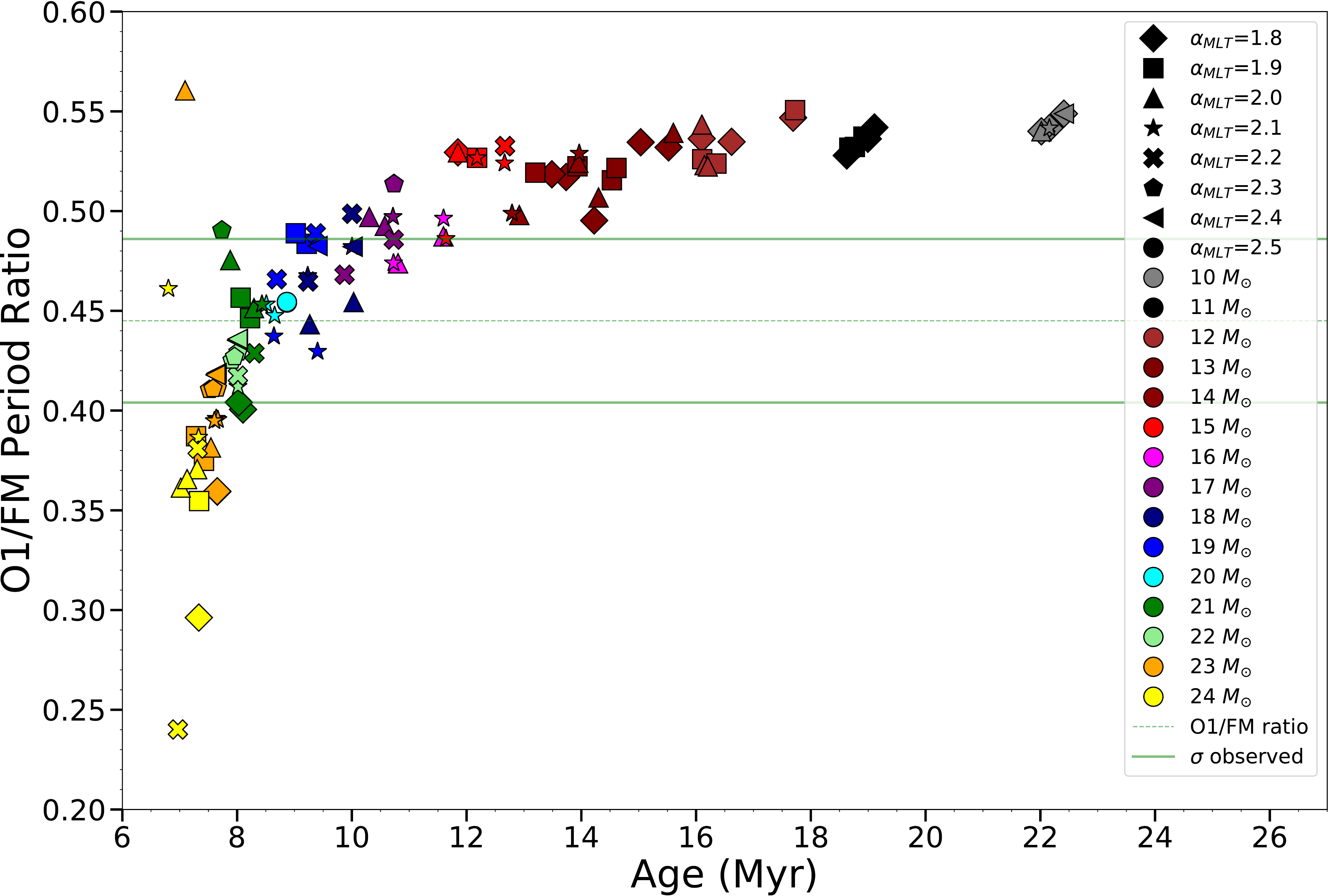}
    \caption{
   	The ratio of the first overtone to fundamental mode periods computed with GYRE is shown
    for the same models and temperature definition used in the upper panel of Fig.~\ref{fig:synth_periods} (variable \mlt/). The range indicated by horizontal, green lines is $P_1/P_0= 0.44 \pm 0.04$. }
    \label{fig:periodratio}
\end{figure}

%%%%%%%%%%%%%%%%%%%%%%%%%%%%%%%%%%%%%%
\subsection{The first overtone}
\label{sec:o1}
In Sect.~\ref{sec:freqs} we presented observational evidence of a new frequency component, $f_1$, that corresponds to a periodicity of {$P_1=185$}~d. 
While strong aliasing caused by annual gaps in the data makes this detection somewhat uncertain, the ratio of this mode to the fundamental suggests that it is the first overtone.
%period ratio with the fundamental mode suggests that this frequency is the first overtone.
%
%However, massive red giant/supergiant seismic models in the literature are scarce,
Because the literature on seismic models of RSGs is sparse (with most works focusing instead on the low- to medium mass regime; c.f. \citealt{trabucchi2019}), we supplement the photometric analysis with theoretical results from GYRE. To test whether the seismic models agree with the observed period ratio we inspect GYRE's prediction for the frequency of the $P_1$ ($n_p=2$, $l=0, m= 0$) mode. With the uncertainties propagated from each period, the relevant range is $P_1/P_0 = 0.445\pm 0.041$. 

We find that GYRE's prediction for $P_1$ is strongly consistent with $185$ d and that the 
%The GYRE models tell us that the 
observed period ratio gives mass and age predictions that are self-consistent with the mass and age ranges constrained by the fundamental mode, as demonstrated in Figure \ref{fig:periodratio}. When mixing length is varied and temperature is fixed, the age and initial mass ranges inferred from the period ratio are 7--11 Myr and 16--23~$~M_{\odot}$, respectively.

If we consider the progression of the models, we find that the period ratio drops in even younger, higher-mass models, with $P_1/P_0$ as small as 0.30--0.36 at around 5 Myr. Older, lower-mass models reach the 0.50-0.55 range, in agreement with period ratios observed in semiregular and Mira variables \citep{kiss1999,TUMi}.
We thus conclude that the signal detected in our photometric data is indeed consistent with the first overtone, making Betelgeuse a double-mode star.

Observing multiple modes in a pulsating star can provide stringent constraints on its physical parameters: this is the main principle behind asteroseismology \citep{ConnyReview}. 
The Fourier analysis in Sect.~\ref{sec:freqs} only provided us with a formal error for the single detected frequency peak, but we expect the overtone to have a limited mode lifetime, just like the fundamental mode.  
As Figures \ref{fig:synth_periods_O1} and \ref{fig:radial_band_O1} show, and as is discussed in Section \ref{subsection:linear_perturbations}, the first overtone prefers largely the same models as the fundamental mode.

%%%%%%%%%%%%%%%%%%%%%%%%%%%%%%%%%%%%%%%%%%
\subsection{Seismic parallax and luminosity}

\begin{figure}
    \centering
    \includegraphics[width=\columnwidth]{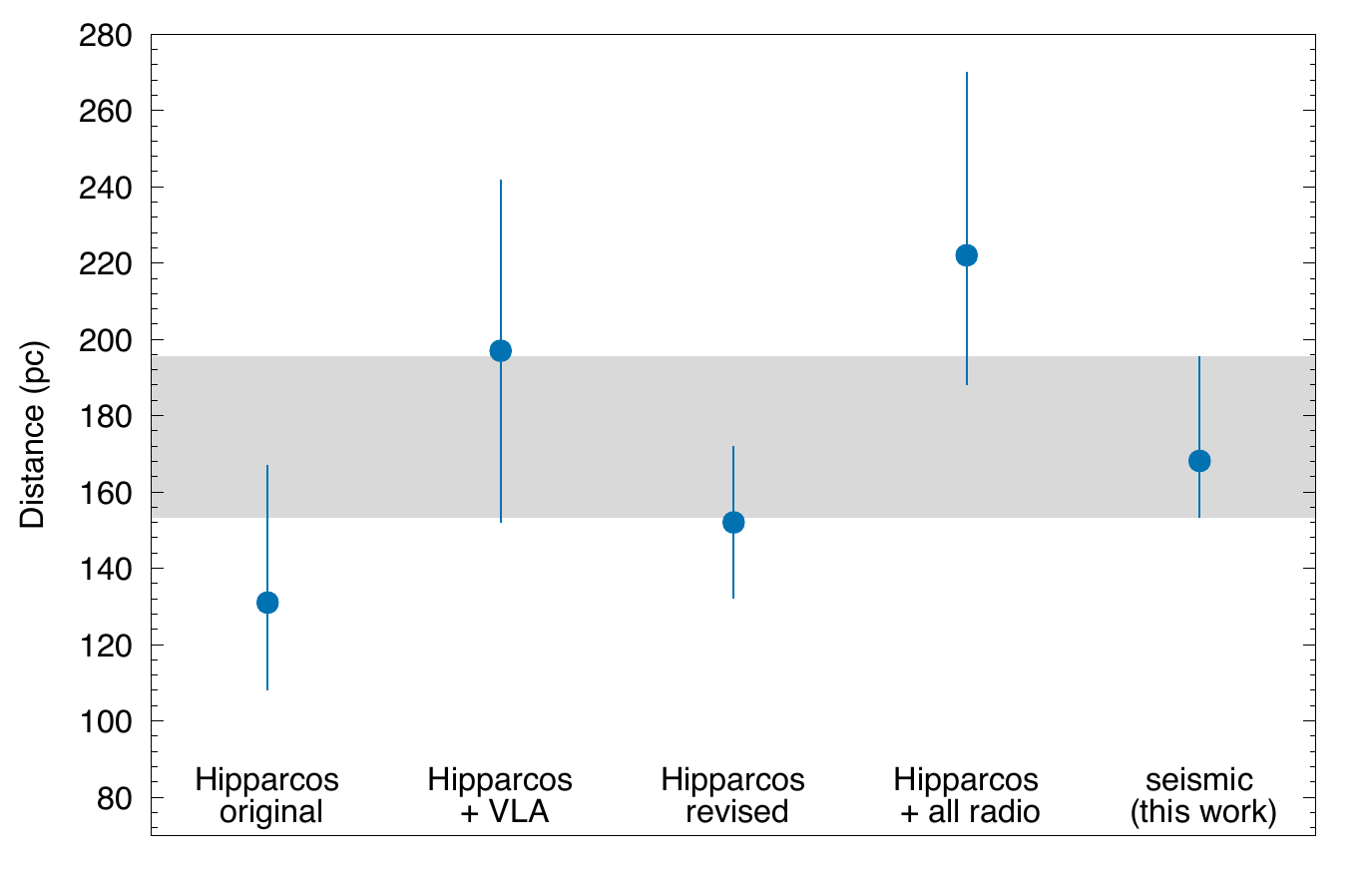}
    \caption{A summary of recent literature distances to $\alpha$~Ori. The gray bar represents the seismically-derived values reported in this work.}
    \label{fig:distances}
\end{figure}

With the radius of $\alpha$ Ori heavily constrained by the seismic models, we can calculate the distance to the star based on the measured angular diameter. Using an angular diameter of $42.28\pm0.43$~mas and 
the $3\sigma$ uncertainty range of the seismic radius estimate, $764^{+116}_{-62}R_{\odot}$, we calculate $168^{+27}_{-15}$~pc for the distance and $\pi=6.06^{+0.58}_{-0.85}$~mas for the parallax. 
Our values are in agreement with the parallaxes derived entirely or in large part from the \textit{Hipparcos} measurements (see \citealt{vanLeeuwen2007} and \citealt{harper2008}), and place $\alpha$~Ori nearer to us. It is, however, somewhat in tension with the more recent results based on radio observations, with disagreement at the $1$--$2\sigma$ level \citep{Harper2017}. Figure \ref{fig:distances} shows our results in context.

This discrepancy could stem from various observational or theoretical shortcomings. 
One possibility is that the period shift is caused by large-amplitude, non-linear pulsation. Stellar structure adjusts dynamically to the changes caused by coherent pulsation, which may cause a shift in the eigenfrequencies of the structure. Therefore even if the physical parameters of a linear seismic model agree with those of the star, the calculated and observed periods may not. 
In the case of $p$-modes, the radius relates to the pulsation period as $R \sim P^{2/3}$ for a given mass. From this alone, we estimate that the linear period of Betelgeuse should be $500\pm40$~d if its radius is 887~R$_\odot$, as adopted by \citet{Dolan2016}. This means that a $\sim20\%$ non-linear decrease would be required to reproduce the observed $416\pm24$~d pulsation period.
Given that we cannot yet evolve RSG hydrodynamic models to full-amplitude pulsation, we cannot infer the period shift directly.
For comparison, studies show that pulsating Mira models produce period shifts of up to --23\% and +15\%, which is of the appropriate order \citep[see, e.g.,][]{lebzelter-wood-2005,IrelandScholzWood2011}.
However, non-linear period shifts scale with the pulsation amplitude. As such, the relatively low-amplitude, short-lifetime mode seen in $\alpha$~Ori makes such a large shift implausible. Thus, non-linear pulsation could be, at best, only partially responsible for the discrepancy we observe.
%}

%%%%%%%%%%%%%%%%%%%%%%%%

%%%%%%%%%%%%%%%%%%%%%%%

Another possibility is that the true Rosseland angular diameter of the star is smaller than the 41.8~mas value adopted in the present analysis, and thus the diameter of the photosphere could be as small as 36~mas. 
However, this would suggest that none of the direct imaging and interferometric observations, including the multi-band models created by \citet{perrin2004} and \citet{Montarges14}, were capturing the photosphere itself, but, rather, only two separate layers in the extended atmosphere. We consider this very unlikely.

On the other hand, consistency between our parallax estimate and the \textit{Hipparcos} value could indicate that the astrometric cosmic noise has been underestimated in previous studies, or that it is correlated over the timespan of the astrometric observations.

\citet{Harper2017} illustrates that the addition of a large cosmic noise term to the radio data would be required to bring the combined \textit{Hipparcos}$+$radio parallax value close to 6.0~mas. 
Nevertheless, this could be an indication that various effects, such as photocenter displacements and non-axisymmetric stellar disk shapes caused by convective hot spots, have been previously underestimated. 
A resolved image obtained from ALMA shows a large hot spot towards the disk limb with a temperature contrast of $\Delta T\approx 1000$\,K, which coincides with the rotational axis but provides no context for the temporal evolution of the photocenter displacements \citep{ogorman2017}. \citet{Kervella2018} proposed that this hot spot corresponds to a rogue convection cell that might also be magnetically connected to ongoing mass ejection from the polar region of the star. Presence of one or more persistent spots could mean that the cosmic noise level of the star is higher than reported by \citet{Harper2017}, and/or that it needs to be modeled as a correlated noise source. 
Confirming or ruling out this possibility would require either a sustained, multi-year interferometric observational campaign or
the development of accurate 3D physical simulations of supergiant stellar atmospheres that can handle the evolution of hot spots.
Finding new ways for \textit{Gaia}  to process heavily saturated stars and thus allowing us to obtain better parallaxes and possibly astrometry of individual hot spots is yet another avenue forward \citep{Sahlmann2018}. However, any of these endeavors would require substantial investment from the respective experts.

Finally, with a seismic parallax in hand, we can estimate a tighter luminosity range based on the Stefan-Boltzmann law. Adopting the strict $T_{\rm eff}$ uncertainty range reported by \citet{Levesque2020},
we report the luminosity constraints of $\alpha$~Ori to be $\log_{10} L = 4.94^{+0.06}_{-0.04}$, where \textit{L} is in units of $L_\odot$. The effective temperature range permitted by taking into account theoretical uncertainties extends this range to $\log_{10} L = 4.94^{+0.10}_{-0.06}$. 
%
%removed L_Sun unit as logL is in dex, i.e. unitless. - LM
% good, logL was causing political tension
%
We note that if we were to superimpose this range on Figure \ref{fig:setoftracks}, it would intersect the RSB at the appropriate temperatures for tracks with initial masses between $16$--$21 ~M_{\odot}$. This range is self-consistent with our other means of estimating mass; however, as mass is a derived quantity, we do not use it to restrict the domain of our seismic grid.

%%%%%%%%%%%%%%%%%%%%%%%%%%%%%%%%%%%%%%%%%%%%%%%%%%%%%%%%%%%%%%%%%%%%%%%%%%
\section{Hydrodynamic Analysis}
\label{section:hydro}
The third component of our modeling relies on MESA's implicit hydrodynamics solver, which we use to explore the non-linear oscillatory behavior of the models' envelopes on decadal timescales.
% ShingChi: Since at the end we only explored linear modes before the code crash
% OK -MJ
Though the term ``non-linear'' can accurately be used to describe any use of the Euler equation, which is non-linear by nature, or to describe any models that make full use of the stellar structure and evolution equations as opposed to perturbation analysis, we apply this term only to particular hydrodynamic behavior.
We use the term ``linear'' in the hydrodynamic context to describe any oscillation that does not excite other modes or cause the development of shocks.

\subsection{Method}
We use MESA (version 8118---\citealt{MESAIII}) to solve the stellar structure and evolution equations by  means of an implicit hydrodynamics scheme in the Lagrangian formalism. Critically, this means that velocity becomes a dynamical variable similar to density, luminosity, and other physical quantities.

To capture the shock propagation, artificial viscosity $q$ is included in MESA's hydrodynamical scheme \citep{Richtmyer1967}. However, in our calculations, the global motion of the H-envelope remains subsonic ($\sim 0.5$ km s$^{-1}$) until the end of the simulation. Since this is lower than the typical speed of sound in the atmosphere ($\sim 10$ km s$^{-1}$), capturing shocks is not integral to our work, and hence $q$ acts only as a safeguard to prevent numerical instabilities.

Furthermore, the code uses the energy-conserving,  time-discretization scheme of \citet{Grott2005}, which ensures that the models at two consecutive timesteps are consistent with each other. We describe the precise configuration for our simulations in more detail in Appendix \ref{section:appendix}, and provide the inlists necessary for reproduction\footnote{We will publicly release the inlist files on Zenodo at the conclusion of the refereeing process.}.

It is important to note that MESA's implicit hydrodynamics scheme does not include the equations of stellar pulsation directly; rather, MESA has a dedicated module, RSP (Radial Stellar Pulsations; \citealt{Smolec2016}), for calculating non-linear pulsating models. However, RSP is not suitable for stars with the luminosity-to-mass ratios ($L/M$) of giants such as $\alpha$ Ori \citep{MESAV}. 
Previous work has shown that the pulsation modes of stars with $L/M$ ratios similar to that of Betelgeuse have high enough growth rates to induce shocks even if the implicit solver is employed \citep{Heger1997, MESAII, Smolec2016, YoonCantiello2010, Goldberg2020}. 

\subsection{Period--Radius estimates from hydrodynamic runs}
% Our hydrodynamic method is introduced in full detail in Section \ref{section:hydro}, where we
While the main goal of incorporating hydrodynamic simulations into our analysis is to study a canonical model of Betelgeuse's envelope rigorously,
non-tailored hydrodynamic simulations also provide a second method of calculating short-order pulsation modes, and thus a means of independently verifying the linear calculations. From a small grid of cursory hydrodynamic runs of varying mass, we can estimate theoretical pulsation cycle lengths directly. 

We first conduct an exploratory investigation of the hydrodynamic evolution for a small subset of the models in our grid, restricting to those with initial masses between $17$--$23\,M_{\odot}$. We require that the timestep does not exceed some fixed, small value---typically 5000--10,000 seconds---and compute the temporal evolution for several decades.

\begin{figure} 
\centering
\includegraphics[width=\linewidth]{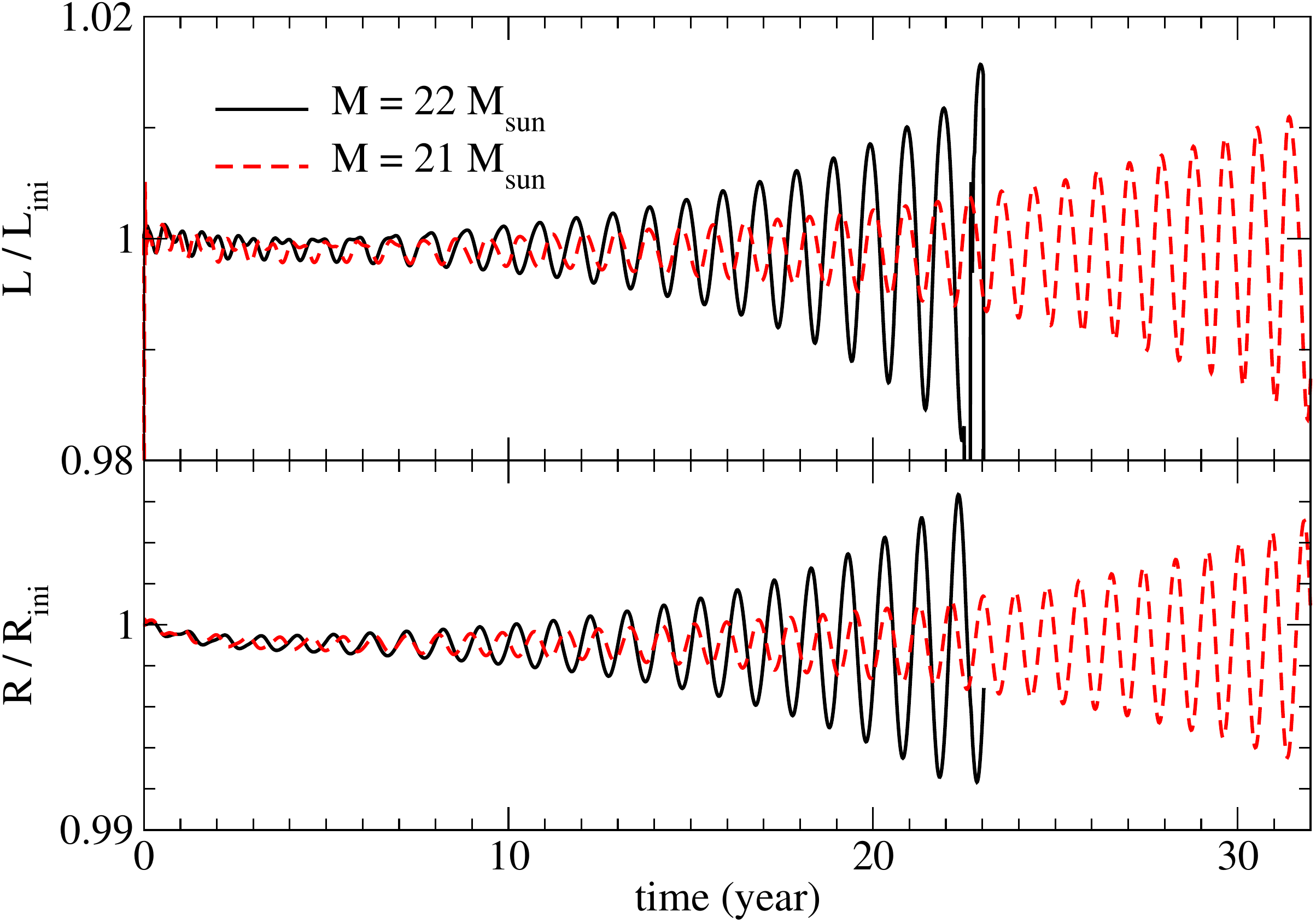}
\caption{
We show the short-timescale, hydrodynamic evolution for two MESA models with observationally consistent features at the termination of their evolution in terms of normalized luminosity (UPPER) and normalized radius (LOWER) versus time. From these simulations, we can extract cycle lengths as a secondary means of estimating oscillation periods. 
}
\label{fig:hydro_grid}
\end{figure}

Figure \ref{fig:hydro_grid} demonstrates the oscillatory behavior of two
hydrodynamic models with slightly different initial masses. 
If not handled correctly, the hydrodynamic models will rapidly expand from their evolutionary initial conditions---in most cases, to nearly double our radial limits---before stable pulsations emerge. This is caused by a discrepancy between the luminosity of the inner boundary of the simulated stellar envelope and the actual stellar luminosity. Thus, over time, the star deposits its energy near the surface, making the star expand.
This can be mitigated by applying relaxation procedures to the initial hydrostatic model 
(\citealt{LSP-wood2004, NichollsWoodCioni2009, IrelandScholzWood2011, Hideyuki2015}).

\begin{figure*} 
%% Figure 9
\centering
\includegraphics[width=\linewidth]{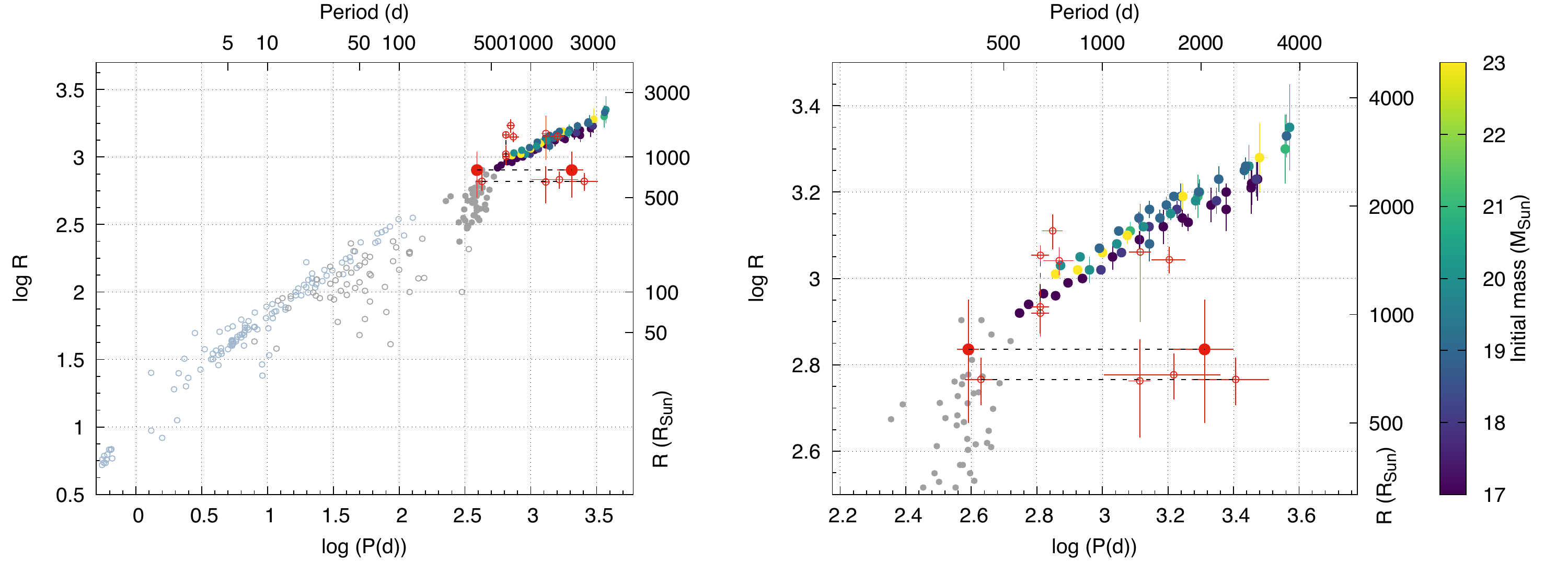}
\caption{The left panel shows the sequence of stellar radii against pulsation periods, extending from RR Lyrae and Cepheid stars to Miras and RSGs.
The right panel highlights the region containing observations of other variable red giants and measurements extracted from the hydrodynamic simulations, demonstrating that Betelgeuse's $416$ day, rather than $2050$ day, periodicity lines up better with the modeled sequence.
In both panels, gray dots correspond to observations of lower-mass pulsators. Variable red giants in particular are shown in open red circles, with Betelgeuse's two modes represented by closed red circles. Points derived from models are colored, with the colorbar indicating their mass. }
\label{fig:hydro_period_logR}
\end{figure*}

However, it is still possible to derive the pulsation periods and average radii of these models based upon selected cycles before shocks and/or numerical failure occur, thus providing a cursory but independent validation of the pulsation periods computed with GYRE. 

The modeled data are produced by estimating the instantaneous period and radius values from the hydrodynamical models using a combination of quadratic and a sine functions fit to short segments of the radial evolution. In this way, we can extract multiple theoretical $R,P$ estimates from one hydrodynamic model. 
We compare these to a set of direct and inferred observations of pulsation periods and radii of variable stars. 
The bulk of these data come from the collection of \citet{szatmary2004},\footnote{\url{http://astro.u-szeged.hu/P-R_relation/pr_poster.html}} which contains a variety of variable stars including RR Lyrae, Cepheids, and Miras. 
In addition, we collate period and radius estimates for a number of other supergiants from available literature: CE~Tau, TV~Gem, $\alpha$~Sco, V766~Gem, AH~Sco, UY~Sct, V602~Car, VY~CMa and KW~Sgr \citep{wasatonic1998,levesque2005,Kiss2006,ohnaka2013,wasatonic2015,wittkowski2017}. 
Of these, pulsation and LSP period measurements are available for $\alpha$~Ori and TV~Gem, and KW~Sgr has two distinct radii published---these have dashed lines connecting their measurements.

Figure \ref{fig:hydro_period_logR} shows a set of synthetic periods and radii derived from the hydrodynamic grid. Also shown in Figure \ref{fig:hydro_period_logR} are 
(1) the $P$-$R$ sequence constructed from observations of pulsators across a wide mass range (gray dots);
(2) the observed FM and LSP periodicities for the small number of red giants listed above (red, open circles); and 
(3) the 416 and $2050$ day periodicities of Betelgeuse (red, closed circles). Masses of the synthetic stars are indicated via the color bar. 

It is well-known that acoustic modes scale with the average density of the star, which itself largely depends on the radius. We should therefore expect a clear correlation between radius and period, as seen both here and in the linear seismic analysis (Figure \ref{fig:radial_band}).

As is clear in the left panel of Figure \ref{fig:hydro_period_logR},
there is a well-defined $P,R$ sequence spanning RR Lyrae up to synthetic supergiants.
The periods and radii extracted from the hydrodynamic models extend the established sequence of pulsating stars to higher radii in a systematic and continuous way, indicating that our models experience $p$-mode pulsation
until our simulations are terminated or the numerics break down. A track of given mass can produce multiple data points by sampling at different times, and hence different degrees of radial expansion, during the hydrodynamic calculation.

In turn, the right panel hints at certain mode classifications for some of the observations. The periods for a number of stars with measured radii fall cleanly on the model sequence, and some fall above: the latter could suggest either pulsations in an overtone or that their radii have been overestimated. Given the complicated circumstellar environment surrounding many supergiants, and our own findings on the radius of $\alpha$ Ori, the latter is a plausible explanation. Finally, the LSP signals are clearly separate from the model sequence, confirming once again that the $2050$ day periodicity is not driven by acoustic variations.

Even before more careful modeling of the hydrodynamic evolution, it is evident that pulsation periods emerging naturally in the simulations are of the same order as Betelgeuse's 416 day periodicity. The linear and hydrodynamic seismic analyses both demonstrate that this is $\alpha$ Ori's fundamental mode, a fact which, when combined with other classical observations, places particularly strong constraints on the star's radius.

%%%%%%%%%%%%%%%%%%%%%%%%%%%%%%%%%%%%%%%%%%%%%%%%%%%%%%%%%%%%%%%%%%%%%%%%%%%%%%%%%%%%%%%
\subsection{Possibility of self-excitation due to non-linear effects}
\label{subsection:modeexcitation}

As stars evolve across the HRD, they may undergo mode transitions when a new pulsation mode becomes unstable. At this point, the star can switch to the new mode---a phenomenon observed directly in RR Lyrae stars---or transition to a multimode pulsator. 

Mode growth rates have various definitions. In the linear framework, they usually represent the natural timescale of changes in the pulsation energy of the star \citep{pulsatingstarsbook}. Growth rates are sometimes calculated directly from the change in amplitude between successive cycles, but one must keep in mind that in non-linear calculations, amplitudes do not grow indefinitely. Hence, non-linear growth rates only agree with the linear values initially, eventually fading back to zero \citep{YoonCantiello2010}. Normalized growth rates are thus scaled with the pulsation frequencies. 
In the case of, e.g., OGLE--BLG--RRLYR--12245, this mode transition lasted for hundreds of cycles, as is consistent with the small growth rates of the modes \citep{ogle2014}.
But, in contrast to classical pulsators, semiregular stars have very large growth rates that can lead to strong mode interactions, some of which may even become chaotic \citep{buchler1996}.
As such, it is theoretically possible that Betelgeuse has recently experienced a rapid mode transition, or a rapid increase in amplitude of an overtone mode already present, and that the superposition of the resulting modes created the unusually low brightness minimum seen in November of 2019. It is thus worth investigating whether such a situation can be simulated; however, modeling multimode pulsation in the non-linear regime is notoriously difficult; at present, this is only reliably reproducible for stars with much lower $L/M$ ratios \citep{kollath2002, Smolec2016}. 
It is thus beyond the scope of the current paper to investigate such a situation, though this scenario is one we hope to address in a subsequent investigation.

%%%%%%%%%%%%%%%%%%%%%%%%%%%%%%%%%%%%%%%%%%%%%%%%%%%%%%%%%%%%%%%%%%%%%%%%%%%%%%%%%%%

\subsection{Analysis of Canonical Hydrodynamic Model}
\label{sec:canonical_hydro}
We consider the evolution of a canonical hydrodynamic model whose initial and terminal evolutionary conditions are consistent (as best as possible) with the parameters reported in Section \ref{section:seismic}. 
We construct a star with initial mass 21 $M_{\odot}$ and terminal mass of 19.54 $M_{\odot}$ during core He-burning. In order to force the stabilized radius to be consistent with our reported values, we must inflate the mixing length parameter to $\mlt/ =3.0$. However, we still require that our hydrodynamic model intersect the theoretical temperature uncertainties described above, $3600 \pm 200$ K, during its oscillations.

Following the general methodology outlined in \citet{Goldberg2020}, we cease tracing the evolution of the innermost $6.0 M_{\odot}$, representing the core, at the conclusion of the classical evolutionary run. 
At this point (often called ``core removal'' or, more accurately, ``core freezing''), a typical timestep in the code is still 100 - 1000 years. As a numerical test, we run an evolutionary (non-hydrodynamic) model of a $20M_{\odot}$ star and find that after He exhaustion, it takes approximately 50,000 years for the luminosity to change by 0.05, whereas our hydrodynamic models show luminosity variations of similar scale over the course of 40 years. Hence, the core can be considered unchanged to good approximation over the duration of the hydrodynamical simulations. 
It is thus valid to adopt constant inner boundary conditions set by $r_{\text{cut}}$ such that $M(r_{\text{cut}})=6.0 M_{\odot}, L = L(r_{\text{cut}})$ throughout our calculations of the short-timescale, dynamical evolution of the envelope.
The value of $6.0 M_{\odot}$ is chosen so that $1.0 ~M_{\odot}$ of the outer He-layer remains along with the entire H-envelope. The He layer, which sets the base of the hydrodynamical simulation,
forms a core-envelope structure with the H-envelope, and the higher density core ensures
that the oscillation of the envelope does not interact directly with the mass gap.

In order to maintain a stable configuration after ceasing to evolve the core, we allow the model to settle into a hydrostatic approximation before turning on the hydrodynamic solver.
To capture the short timescale motion to adequate resolution, we limit the timestep of the hydrodynamical evolution to a maximum of $10^4$ s. 
A larger timestep of only $\sim 10^5$ s can already result in the erasure of modes with a sub-annual period; this is due to the implicit nature of the hydrodynamical solver.
We note that an implicit hydrodynamic scheme is necessary in order to follow the global motion of the star because the relative distances among mass shells near the surface are small. In terms of the Courant timescale, it is $\sim 10^6$ larger than that required by explicit time discretization ($\sim 0.5$ s). Thus, in order to track the motion of the surface with sufficient temporal resolution, implicit hydrodynamics must be employed.

Our canonical model is evolved for a total of $\sim$ 10,000 steps from the
initiation of hydrodynamics until the point at which stellar expansion begins to 
disturb the pulsation frequency. We find that beyond $\sim 30$ years, the motion in the star becomes large enough to interact with the convective layer, causing the timestep to drop as low as $1 \times 10^2$ s before the code is unable to evolve the model forward in time. At this point, we stop the simulation.

\subsubsection{Global Features}
\begin{figure*}
    \centering
    \includegraphics[width=18cm]{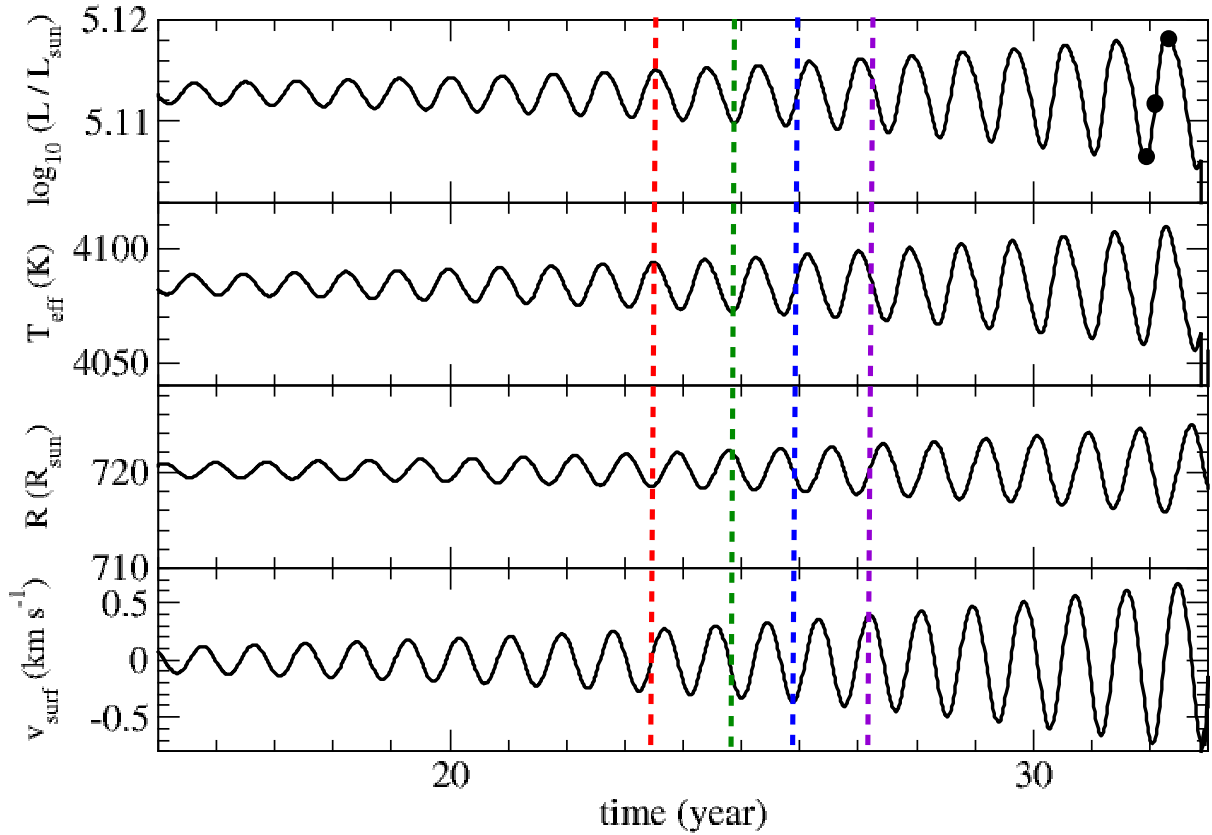}
    \caption{(UPPER) The temporal evolution of the luminosity (L), effective temperature ($T_{\rm eff}$), radius ($R$) and surface velocity ($v_{\rm surf}$) for the characteristic model. The red, green, blue and purple vertical dashed lines are added for comparing the four quantities at the same time slices.
    The black dots are the time moments where the stellar profiles are plotted in Figure \ref{fig:profile2}.}
    \label{fig:history}
\end{figure*}

We first discuss the temporal evolution of the critical observables in the canonical model.
In the four panels of Figure \ref{fig:history}, we plot the luminosity, effective temperature, radius and surface velocity of the star. 

The system enters into a state of pulsation with steady but growing cycles a few years after the hydrodynamic solver is switched on. 
Early in the evolution, quasi-annual oscillatory behavior is present in the luminosity and effective temperature, and
the stellar radius exhibits a consistent periodic motion on top of a steady exponential expansion. In this work, we will consider the stellar pulsation only when the motion remains linear; 
we note that once the behavior becomes non-linear, the timestep becomes too small to follow the pulsation effectively. Moreover, non-linear pulsations greatly disturb the profile of the stellar envelope, particularly in terms of opacity and free electron fraction, which makes direct comparison difficult. 

In Figure \ref{fig:history}, four vertical, dotted lines indicate moments at which we compare the instantaneous values of the four quantities.
The red and green lines correspond to timesteps where the luminosity is at a local maximum
and minimum, respectively. 
The blue and purple lines correspond to timesteps where the surface velocity is at a local minimum and maximum, respectively.

When the star reaches its brightest point in the pulsation cycle, the effective temperature also reaches its maximum. Concurrently, the star is in its most contracted state (radial minimum) and displays a nearly zero surface velocity.
This is consistent with the behavior of a classical harmonic oscillator where the displacement is largest during one cycle. 
Conversely, the luminosity and effective temperature are minimal when the star is most radially extended, and when the star is maximal in surface velocity, the luminosity, effective temperature, and radius are near their average values. 
We then observe that as the star continues to expand, a wobble in its motion emerges. As indicated in the hydrodynamic evolutionary tracks of Figure \ref{fig:hydro_grid}, 
%the star will eventually approach a hydrodynamical instability at the end of its helium-burning phase. 
the star will eventually approach a state in which it is capable of developing a hydrodynamical instability in the outer envelope---the exact pulsation properties are themselves a function of evolutionary stage and mass, as shown in \citet{YoonCantiello2010}.
The evolutionary time at which we launch our hydrodynamic simulations is chosen so as best to reproduce the constraints on the classical parameters found in the previous analysis. We adopt a uniform starting point at which the He mass fraction in the core is $10^{-4}$. Though this is somewhat late in the helium burning phase and thus does not correspond to the evolutionary phase statistically preferred by our classical models, this starting condition lends itself to more stable hydrodynamic models. We find that when the simulations are initiated at earlier or later burning phases, it is difficult to produce stable, suitable stellar models that fit the radius and luminosity constraints derived in previous sections.
As our main priority is to match these parameters, we do not further investigate the effects of evolutionary starting condition. We require only that our hydrodynamical initial conditions yield models that reproduce the radius, luminosity, and mass of the evolutionarily preferred models, as these are the features relevant for studying pressure-driven pulsation in the envelope.

The expansion of the outer radius gradually affects the $P$--$R$ relation, as the sound speed travel time increases with increasing distance. Analysis of the radius is additionally complicated by (1) how the outermost boundary of the star is defined and (2) radiation pressure outside the photosphere. A rigorous treatment of radiative transport is necessary in the photosphere regime, and so we stop the simulation to analyze the motion only when the radius is beneath this threshold. 

The effective temperature $T_{\rm eff}$ and luminosity $L$ behave similarly to radius in the simulations. 
In the former case, this is because MESA calculates the effective temperature directly from 
the luminosity and radius via $T_{\rm eff}^4 = (L / 4 \pi \sigma_B R^2)$,
where $\sigma_B$ is the Stefan--Boltzmann constant. Given the slow change of the radius,
$T_{\rm eff}$ primarily mirrors the fast-evolving $L$.

Regarding the evolution of luminosity, we note that from year 15 onward, the star exhibits 
periodic motion in its brightness. The early motion is highly regular: as in the preliminary grid of hydrodynamic models (see Figure \ref{fig:hydro_grid}), we observe a 
quasi-annual rise and fall---the correct timescale for the FM.
Near the end of the simulation, the large oscillation begins to trigger non-oscillatory motion in all quantities, and this is responsible for the rapid drop in luminosity.

\begin{figure*}
    \centering
    \includegraphics[width=18cm]{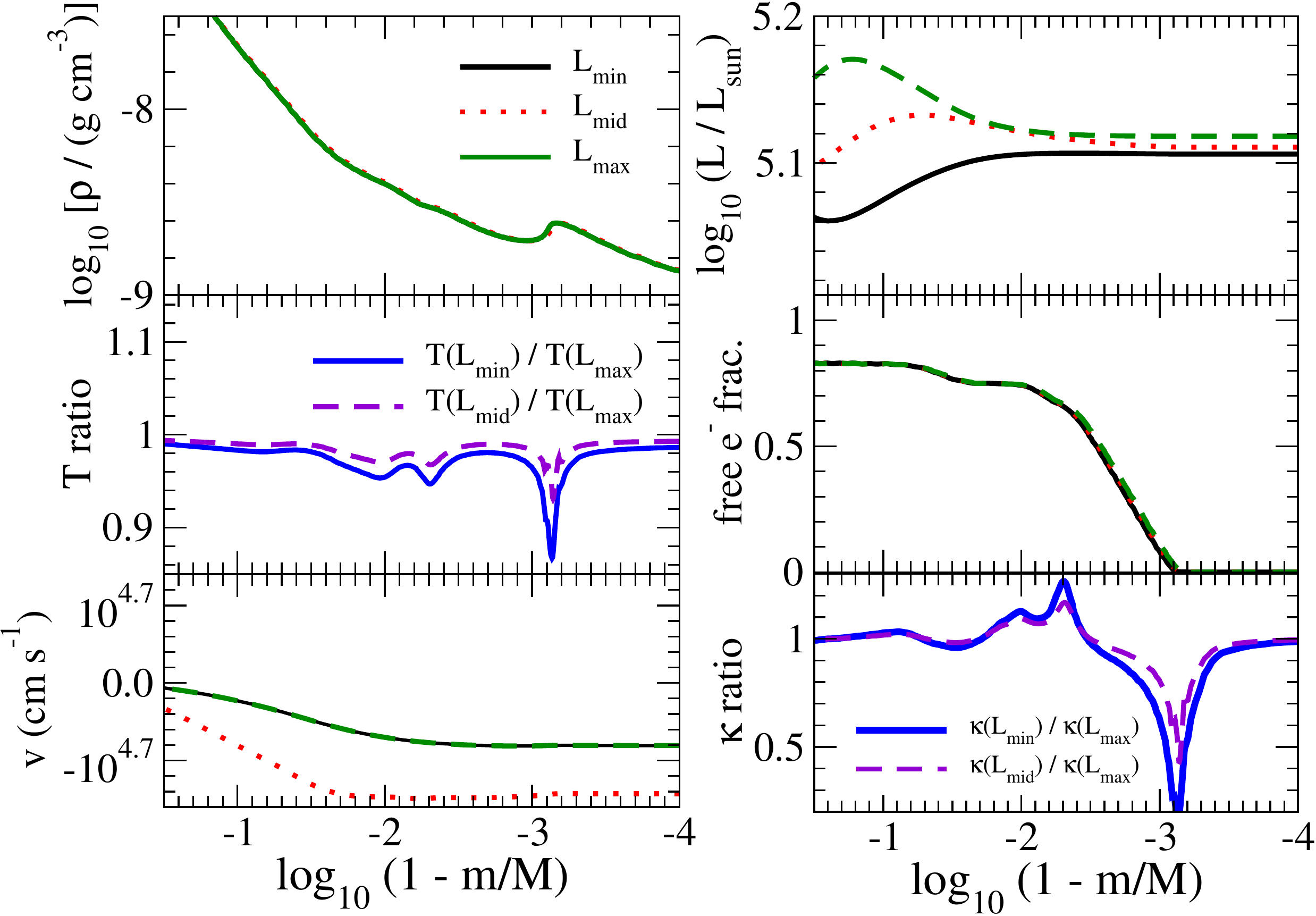}
    \caption{Analysis of model star's structure at three points selected during the pulsation, indicated by black markers in Figure \ref{fig:history}.
    LEFT: The density (top panel), temperature (middle panel) 
    and velocity (lower panel) profiles for the moments at the luminosity minimum (black solid line),
    midpoint (red dotted line), and maximum (green dashed line).
    RIGHT:  Same as the left panel, but for the luminosity, 
    opacity and free $e^-$ fraction. 
    In both panels, $m$ represents the enclosed mass, $M(r)$.}
    \label{fig:profile2}
\end{figure*}

In Figure \ref{fig:profile2}, we plot the structural profiles of six quantities at points indicated by black circles in Figure \ref{fig:history}.
The global features of the density and temperature profiles show that outermost 10\% of the stellar mass has a relatively low density, sitting between $10^{-9}$--$10^{-7}$ g cm$^{-3}$, while the temperature lies between $10^{3.5}$--$10^{5}$ K. A small density bump appears 
at $\log_{10}(1-q) = \log_{10}(1-M(r)/M) = -3$, which is accompanied by a sharp fall in temperature. This occurs in order to maintain hydrostatic equilibrium. 
We note that a density inversion can occur only when convective mixing is inefficient. 

It should be noted that a density inversion, in general, cannot exist when there is convection: the mixing will smooth out the density difference. 
The efficiency of this process depends on the ratio of the timescale of photon diffusion to the dynamical time, as discussed in \citet{Jiang2015}. A slow photon diffusion time, as applicable to the stellar interior, implies efficient convection. Hence the density discontinuity is an evanescent phenomenon. However, near the surface where the photon diffusion timescale is much shorter---similar to what we have observed---inefficient convective mixing cannot remove the density inversion. As described in \cite{Jiang2015}, this density inversion can form and be destroyed repeatedly in the optically thin regions. Numerical schemes, such as the inclusion of a photon ``porosity'' by modifying the photon acceleration terms, can suppress this phenomenon in optically thick regions \citep{Paczynski1969}.

When convection is less efficient and has a timescale comparable with the dynamical timescale, the density and pressure gradients can change signs, creating a Rayleigh-Taylor instability.
Through mixing, the excess density can gradually reduce via diffusion. However, modeling this phenomenon would require a detailed, time-dependent convective scheme, which is not included in this work. For our case, the density inversion plays a less important role in the luminosity evolution, given that this quantity remains steady from $\log_{10} (1-q) = -1$ upward (see subsequent discussion).

The free electron fraction shows that up to $\log_{10} (1-q) = -3$, the matter is partially ionized. Beyond that, the low temperature causes the nuclei to recombine with the free electrons. The opacity profile is richer; rather than falling monotonically like the free electron fraction, we see two major opacity bumps near $\log_{10}(1-q) = -1.2$ and $-2.5$. 
These correspond to the partial ionization zones of H-HeI and HeII, respectively \citep{Cox1973,Kiriakidis1992}.

The velocity and the luminosity vary dynamically during the pulsation. When the stellar luminosity is mid-phase, the whole envelope is contracting with a constant velocity of $\sim 0.7$ km s$^{-1}$. 
The whole star contracts more slowly when it is close to its luminosity maximum or minimum. Meanwhile, the luminosity profiles show that when the star is at its local minimum, the luminosity near the interior of the envelope is lower. The opposite applies during its local maximum. 
We note further that the motion vanishes approaching the core-envelope interface. This suggests that the observed pulsation is a collective motion of the envelope without interaction with the core.

\begin{figure}
    \centering
    \includegraphics[width=\columnwidth]{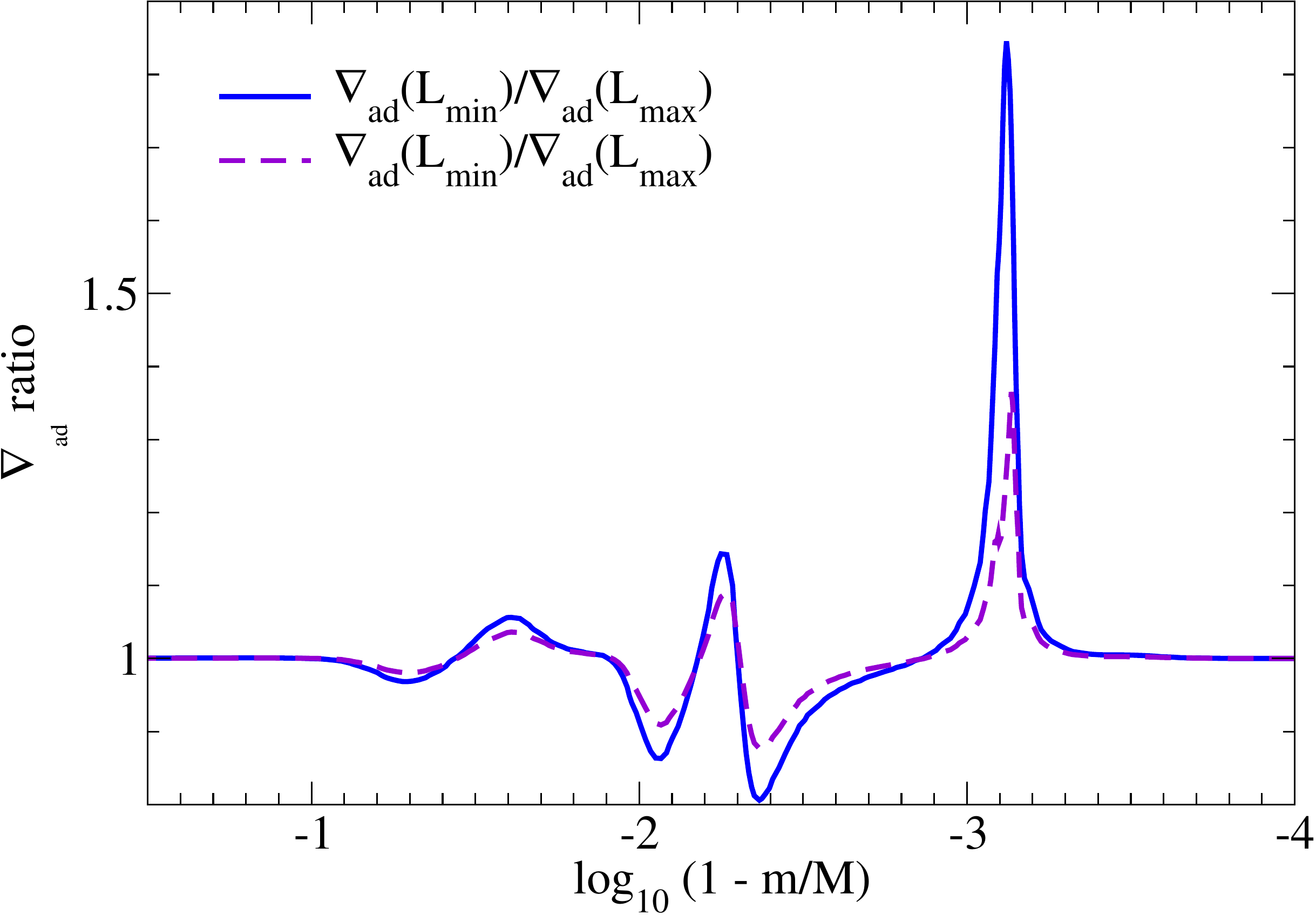}
    \caption{Similar to Figure \ref{fig:profile2}, but 
    for the ratio of the adiabatic temperature gradient for the profiles at luminosity minimum (blue solid line) and luminosity median (purple dashed line), with respect to that at luminosity maximum.}
    \label{fig:nabla}
\end{figure}

To further quantify changes in the stellar profile during pulsation, we plot in Figure \ref{fig:nabla} the ratio of the adiabatic temperature gradients for the profiles at luminosity minimum and medium with respect to luminosity maximum. 
In Figure \ref{fig:nabla}, we observe a sinusoidal-like variation of the profile with about 7 nodes, located at $\log_{10} q \approx -1, -1.5, -1.9, -2.2, -2.3, -2.9$ and $-3.5$. The star at luminosity maximum shows the highest temperature gradient near the surface at $\log_{10} q \approx -2.2$ and $-3.1$. These correspond to the minima in the temperature ratio profile and the extrema in the opacity ratio profiles.

These trends are indicative of the $\kappa$-mechanism, and thus explain why the pulsation gradually grows over many periods. In particular, during contraction, the higher opacity prevents the heat from being stored in the deeper layers, which in turn prevents unstable energy extraction by ionization.

From this collection of profiles, we can deduce that the $\sim1$ year variation is driven by the collective expansion and contraction of the recombined hydrogen layer. 
%The small motions in the layer interior to this interface imply a shorter transition time, as the propagation time required depends on the sound travelling time between the interface and the surface. 
%
We thus conclude that it is this $\kappa$-related interaction driving the fundamental pulsation mode in Betelgeuse.
% nice

\subsubsection{Literature Comparison}
\begin{figure}
    \centering
    \includegraphics[width=\columnwidth]{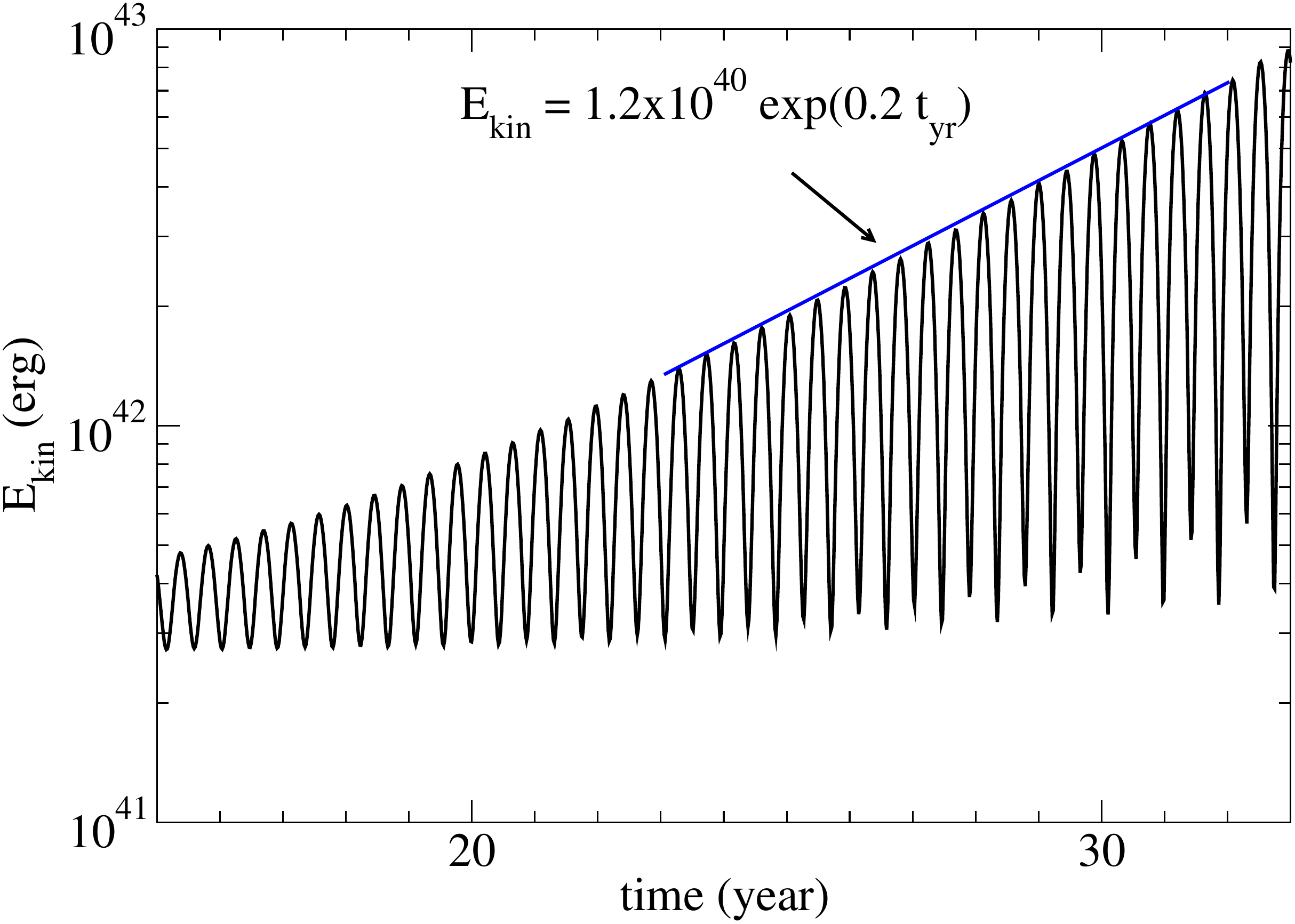}
    \caption{Kinetic energy as a function of time for the hydrodynamical model. The blue line is an exponential fit of the form $1.2 \times 10^{40} \exp(0.2 t)$, with time in years.}
    \label{fig:ke}
\end{figure}

To examine the growth of the oscillation further, we plot the total kinetic energy of the system against time in Figure \ref{fig:ke}. 
The kinetic energy, which is dominated mostly by the motion of the atmosphere, is much smaller than both the total energy and the total gravitational energy, which are, on average, $-6 \times 10^{49}$ and $-1 \times 10^{51}$ ergs, respectively. 
These are several orders of magnitude larger than the kinetic energy, which is $10^{42\text{--} 43}$ erg, in agreement with earlier results (see \citealt{cox1980book}). We note that it is the outermost $q = 0.1 - 1$ that contributes to the atmospheric behavior, including the pulsation. This corresponds to $\sim 2 \times 10^{47}$ erg when this matter is moving at a speed comparable to the escape velocity.

We provide an exponential fit in blue on Figure \ref{fig:ke} to characterize the rate of growth of the kinetic energy. A function of $E_{\rm kin} = A \exp(bt_{\rm year})$ with parameters $A = 1.2 \times 10^{40}$ erg and $b = 0.2$ provide a good fit to the hydrodynamic component of the simulation. 
Naively, this suggests that when the oscillation begins to grow, we should expect that the outermost layers of material will be expelled by the pulsations within a time of $\sim 83$ years. 
In reality, however, we see this rate level off---see, for example, \citet{cox1966}, who showed this in some early pulsation models. 
This is because viscous and turbulent dissipation limit the maximum amplitude of the star in the non-linear regime. Historically, the level of dissipation in 1D pulsation models has been tuned to match the observed amplitudes of RR Lyrae and Cepheid stars and to reproduce double-mode pulsation in the models. This dissipation was first applied via the ``artificial viscosity'' term and later through the eddy viscosity and other $\alpha$ parameters of time-dependent, turbulent convection \citep[see, e.g.,][]{buchler1990,Takeuti1998,Kollath1998,smolec2008}.
%} 

We note that the models in Fig. \ref{fig:mass_compare} stop at about $\pm2\%$ luminosity variation or less, whereas Betelgeuse itself varies by $\pm10$--$30\%$ in $V$ and about $\pm20$--$30\%$ in near-IR (the latter being more closely representative of Betelgeuse's bolometric variation). 
As such, there is plenty time remaining for the kinetic energy and amplitude to grow and eventually saturate at that level, but this is beyond what can be achieved with hydrodynamics today before encountering a numerical runaway episode.

We note that when the oscillation becomes strong, heat deposition effects near the surface, where there is a sharp density gradient, become important. The extra heat can change the opacity of the matter by increasing the ionization fraction, resulting in stronger amplification of the pulsation and in turn accelerating the predicted timescale from the first pulsation until mass ejection. 
Concurrently, however, shock dissipation can convert the kinetic energy of the envelope into heat, while mass loss can also efficiently remove kinetic energy from the star. Heat deposition, shock dissipation, and mass loss interact in complicated ways to form the full dynamical picture of the envelope in the later phases.

Similar ejections of the outer layers in models of luminous Cepheid models, however, are known to be related to numerics rather than physics; see \citet{Smolec2016}. Regardless, we do not follow the code until this phase because the timestep becomes prohibitively small ($\sim 100$ s).
In particular, numerical difficulties arise in the Newton-Raphson iterations, during which the code fails to resolve the formation of convection zones around the shock front. Due to shock compression, the extra heating also invalidates the equilibrium assumptions of the mixing length theory \citep{ErikaBohmVitense1953}. 
To prevent nonphysical convective behavior from developing near the surface, we set a ceiling for the relative convective velocity with respect to local velocity and the speed of sound in the simulations (see also the Appendix for related setting in MESA).
To limit the steepness of shockwaves and distribute them over multiple zones,
explicit pulsation codes like RSP include either artificial viscosity or eddy viscosity terms (or both), but this is only effective up to certain $L/M$ ratios \citep{stellingwerf1975,Smolec2016}. 

Also at this stage, non-linear effects become dominant, causing sub-annual features to appear gradually on top of the linear pulsation. As observations of Betelgeuse do not show periodicities on sub-annual timescales, we do not consider this phase of pulsation further, though a study of non-linear pulsation with the dynamical coupling of opacity and ionization will be interesting future work.

By comparing the general features of our hydrodynamical model with the pulsation patterns of Betelgeuse, it becomes clear that the quasi-annual variation is indeed caused solely by the contraction and expansion of the star.
It is interesting to note that in this linear oscillation phase, we do not see any evidence of longer timescale variations, such as the 6-year and 35-year periodicities. 
In fact, the hydrodynamic simulations never reproduce any of the observed variations besides the currently presented quasi-annual pulsation, even when the initial mass is varied. This implies that these periodicities are driven by some mechanism outside the scope of what 1D hydrodynamic simulations can reproduce, i.e., not the $\kappa$-mechanism. 
It would be interesting to conduct further dynamical studies on how the star relaxes when the opacity effects becomes important; 
however, work in this domain will require an algorithm to suppress the development of the $\kappa$-mechanism so that the pulsation can be sustained without triggering excessive mass loss. 

There are similar works in the literature that focus on the pulsational features of massive stellar envelopes using the stellar evolution code described in \cite{Langer1988}. 
In particular, \citet{Heger1997} present the dynamical evolution of massive stars from 10--20 $M_{\odot}$ and analyze their linear stability.
In Figure \ref{fig:phase_diagram}, we plot the phase diagram of 
our canonical model's
$\log_{10} (L/L_{\odot})$  against $\log_{10} T_{\rm eff}$
during the hydrodynamic evolution as a means of comparing directly with Figure 5 in \citet{Heger1997}.
In their work, the 11 $M_{\odot}$ Red Supergiant model is followed for about 75 periods of oscillation, whereas ours capture the first 45 periods. Beyond the 45$^{\text{th}}$, our models show numerical instability where the expansion and compression interact with the convective mixing zone, which largely suppresses the timestep and creates non-linear behaviour. 

\citet{Heger1997} show an approximately circular trajectory that spirals outwards from 
%
%$\log_{10} T_{\rm eff} (K) \approx 3.52$ and $\log_{10} L = 4.90/L_{\odot}$, 
%
$\log_{10} T_{\rm eff} (K) \approx 3.52$ and $\log_{10} (L/L_{\odot}) = 4.90$, 
whereas our model shows an elliptical trajectory, vacillating between high $L$ and high $T_{\rm eff}$ on one side and low $L$ and low $T_{\rm eff}$ on the other. 
The outward spiraling in our work and theirs demonstrates that both stars are undergoing dynamical instability with a growing amplitude. 
As expected, \citet{Heger1997}'s model has a lower period because it is a lower-mass model. This implies a more compact envelope, which allows all 75 periods of oscillations to happen within 30 years---this is only half the time of our model.

Both models show a clockwise trajectory. Since the radius, temperature, and luminosity are related by the blackbody radiation formula $L = 4 \pi \sigma_B R^2 T_{\rm eff}^4$,
this means that when the stellar models resume their initial luminosities, the models achieve a higher maximum $T_{\rm eff}$ (i.e., smaller $R$) and a lower minimum $T_{\rm eff}$ (i.e., larger $R$). 
These features suggest that  $T_{\rm eff}$, $L$ and $R$ achieve their local extrema simultaneously in \citet{Heger1997}'s model, but in our case, this relationship is slightly lagged.
As shown in Figure \ref{fig:history}, our model approaches its local extrema with a non-zero velocity; thus, the stellar radius, which affects $T_{\rm eff}$, reaches its local maximum and minimum later than $L$. Our calculations therefore reproduce the phase lag between the luminosity and the velocity that has been observed in many other, smaller pulsators before \citep{Castor1968,szabo2007}.

In \cite{YoonCantiello2010}, the hydrodynamical features of a 20 $M_{\odot}$ star with a luminosity of 
%
%$\log_{10} = 5.05~L/L_{\odot}$ 
%
$\log_{10} (L/L_{\odot}) = 5.05$ and temperature of $T_{\rm eff}  = 3198$ K are analyzed. 
Their stellar parameters are similar to ours, where our hydrodynamical model assumes a 21 $M_{\odot}$ star with a slightly lower initial hydrostatic luminosity at 
%
%$\log_{10} L = 5.01~L/L_{\odot}$
%
$\log_{10} (L / L_{\odot}) = 5.01$ and $T_{\rm eff} = 4000$ K. 
They model about 50 years of the stellar pulsation; Figure 2 in their work shows the surface velocity and is comparable to Figure \ref{fig:history} in this work.
Approximately 20 pulsation cycles are followed in their work, where a higher period of $\sim 1000$ days is observed. Compared to our $\sim 400$ day period, this indicates that their envelope is more relaxed and expanded. 
Both \citet{YoonCantiello2010} and our work show a consistent growth of the surface velocity. It takes about 5 cycles for the surface velocity to reach a ten-fold of amplification, while our model takes much longer---almost 20 cycles. 
This suggests that the $\kappa$-mechanism is less efficient in our model, where the star exhibits behavior closer to adiabatic oscillations than driven oscillations. From the growth of kinetic energy of the system, we can estimate that it takes a further $\sim 10$ years for the pulsation to grow to a surface velocity comparable with \citet{YoonCantiello2010}. This would correspond to another $11$--$15$ cycles in our case. Despite this, the robust exponential growth of the pulsation energy (see Figure \ref{fig:ke}) in both works implies that the pulsation could remove the outermost layers of the H-envelope from the star. However, this is not consistent with observational evidence; the RSGs we have observed pulsate with limited amplitude for several decades.
The mass loss rate of the Betelgeuse is observed to be roughly $\sim 10^{-6}~M_{\odot}$ yr$^{-1}$ \citep{Dolan2016}. 
We may speculate that driven pulsation is partially responsible for this driven mass loss, but the pulsational growth rate is limited by the constant energy dissipation through shock heating and mass ejection. It is also true that the typical pulsation amplitude is not as large as would be expected for a driven wind \citep{fox-wood-1982,Wood2000,LSP-wood2004}.

Whether or not mass loss can be driven depends on the degree of saturation in the pulsation of the surface layers \citep{Cox1966b}.
When the mode is permitted to develop, this process can be influential in the formation of circumstellar matter in Type-IIn supernovae for massive stars close to $\sim 20~M_{\odot}$ \citep[e.g.,][]{Smith2017book}. 
However, given the regulated oscillation amplitude observed in a number of RSGs empirically, additional mechanisms not modeled in this work must become dominant in regulating the growth of these oscillation patterns.

The most recent analysis of this kind can be found in \cite{Goldberg2020}. The pulsation of a red supergiant with 16.3 $M_{\odot}$ is computed using MESA with the GYRE extension (version 11701). In contrast to the approaches discussed above, their hydrodynamic models involve an initial perturbation to the density distribution to trigger direct pulsation of not only the fundamental mode, but also the first overtone. 
They obtain a star of 
%$\log_{10} L = 5.2~L/L_{\odot}$
$\log_{10} (L / L_{\odot}) = 5.2$ and $880 R_{\odot}$, which is about 10\% larger than our model. As a result, their pulsation shows a fundamental mode with a longer period---about 600 days---and first overtone of $\sim 300$ days.
We reiterate that since we do not perturb the density profile at the onset of hydrodynamic evolution, and we assume that all pulsation is triggered by numerical perturbations, it is consistent that we do not see higher order excited modes alongside the fundamental mode.
However, the strength of the quasi-annual variation of Betelgeuse but absence of clear, shorter-timescale periods suggests that we should not be concerned by a lack of overtone activity in our hydrodynamic modeling. 
The results in  \citet{Goldberg2020} demonstrate that an overtone in our model would give rise to significant sub-annual motion, which is not observed in Betelgeuse.
Furthermore, the growth rate depends in part on the $L/M$ ratio, and this factor may not be high enough in our models to excite an overtone.

\citet{Goldberg2020}'s density perturbation approach can also cause the star to pulsate with a much larger amplitude initially, which is not achieved in our work nor in the two analyses presented above (\citealt{Heger1997, YoonCantiello2010}). Therefore, it is unclear if their work shows a similar exponential growth as in the literature and in our model. 

\subsection{Impact of Initial Mass on Pulsation}
Thus far, we have presented one model with an initial mass of 21 $M_{\odot}$. Now, we consider a series of hydrodynamic models of different initial mass and discuss how the
progenitor mass affects the pulsation pattern. 

We repeat the simulations by varying the progenitor mass, while fixing the mixing length parameter (see Section \ref{section:appendix} for details on the exact configuration) so that we can compare consistently among models.
In Table \ref{table:hydro_model}, we tabulate the global parameters and pulsation statistics of these models. The data present the following trends: when the progenitor mass increases, the present day mass $M_{\rm fin}$, helium core mass $M_{\rm He}$, the radius at the end of He-burning $R$, and its corresponding luminosity $L$ all increase.  There is a weak decreasing trend for the effective temperature $T_{\rm eff}$. Meanwhile, the time required for the non-linear pulsation to emerge decreases. We note a severe drop between models of 20.2 and 20.5 $M_{\odot}$ during which the associated number of pulsations also decreases sharply. Below $M = 19~M_{\odot}$, the oscillation does not amplify significantly within 300 years, at which point we terminate the simulation.

We note in particular two entries in Table \ref{table:hydro_model} showing simulations with initial (evolutionary) masses of $20 M_{\odot}$: one with \mlt/=3.0 and one with \mlt/=2.5. In the latter case, the non-linear excitation time drops considerably. This data point disrupts an otherwise monotonically decreasing trend in pulse number with increasing present-day mass (above $19M_{\odot}$), demonstrating that the relationship between pulse number, $t_{\rm run}$, and present-day mass is not totally straightforward.
It is well-known, however, that below a certain level of precision, the impact on global parameters caused by changes in the mixing length are indistinguishable from variations in mass and metallicity in models of low to intermediate mass stars with convective envelopes \citep{Joyce2018a}.
Further, the late stage evolution of high mass stars is especially sensitive to convective parameters; in practice, \mlt/ is often tuned arbitrarily until the model converges or behaves as desired.
We include the last row of Table \ref{table:hydro_model} to highlight this degeneracy and caution against over-interpretation. 
%}

\begin{table}[]
    \centering
    \caption{The global properties and pulsation statistics of the hydrodynamical models studied in this work. $M$, $M_{\rm fin}$ and $M_{\rm He}$ are the initial, final and He-core masses of the star in units of $M_{\odot}$, respectively. $R$, $\log_{10} L$ and $T_{\rm eff}$ are the initial radius in units of $R_{\odot}$, luminosity in units of $L_{\odot}$ 
    and the effective temperature in units of K at the beginning of the hydrodynamical phase, respectively. $\alpha$ is the mixing length parameter. $t_{\rm run}$ is the time the pulsation of the star becomes non-linear, where we stop the simulations, in years. ``Pulse" is the number of pulsation cycles experienced by the star before the onset of non-linear pulsations. No number is available when $t_{\rm run}$ is larger than 300 years. 
	All hydrodynamic models are launched from the evolutionary point at which the helium mass fraction in the core is $10^{-4}$. Though this does not correspond to the evolutionary phase statistically preferred by our classical models, this starting condition lends itself to more stable hydrodynamic models and allows us to explore the appropriate radius, luminosity, and mass regimes. }
    \begin{tabular}{ccccccccc} %c
    \hline
    \hline
    $M$ & $\alpha$ & $M_{\rm fin}$ & $M_{\rm He}$ & $R$ & $\log_{10}L$ & $T_{\rm eff}$ & $t_{\rm run}$ & Pulse \\%& stage\\ 
    \hline
    18   & 3 & 17.12 & 5.57 & 550 & 4.92 & 4160 & >300 & N/A  \\%& end of He-burn\\
    19   & 3 & 17.90 & 6.06 & 624 & 5.00 & 4115 & >300 & N/A\\% & end of He-burn\\ 
    20   & 3 & 18.80 & 6.47 & 655 & 5.04 & 4117 & 166.2 & $\sim 230$ \\%& end of He-burn\\
    20.2 & 3 & 18.95 & 6.60 & 659 & 5.05 & 4120 & 141.8 & $\sim 200$ \\%& end of He-burn \\ 
    20.5 & 3 & 19.17 & 6.75 & 707 & 5.10 & 4081 & 31.5 & 43 \\%& end of He-burn \\
    21   & 3 & 19.54 & 7.00 & 721 & 5.11 & 4083 & 27.0 & 40 \\%& end of He-burn \\
    22   & 3 & 20.30 & 7.46 & 787 & 5.18 & 4053 & 21.0 & 24 \\%& end of He-burn \\
    23   & 3 & 20.92 & 8.04 & 875 & 5.25 & 4008 & 16.7 & 18 \\%& end of He-burn \\
\hline
    19   & 2.5 & 17.78 & 6.07 & 724 & 5.01 & 3832 & 102.8 & 122 \\%& end of He-burn \\
    20   & 2.5 & 18.57 & 6.59 & 794 & 5.08 & 3801 & 41.1 & 37 \\%& end of He-burn \\
    \hline
    \end{tabular}
    \label{table:hydro_model}
\end{table}

\begin{figure*}
    \centering
    \includegraphics[width=18cm]{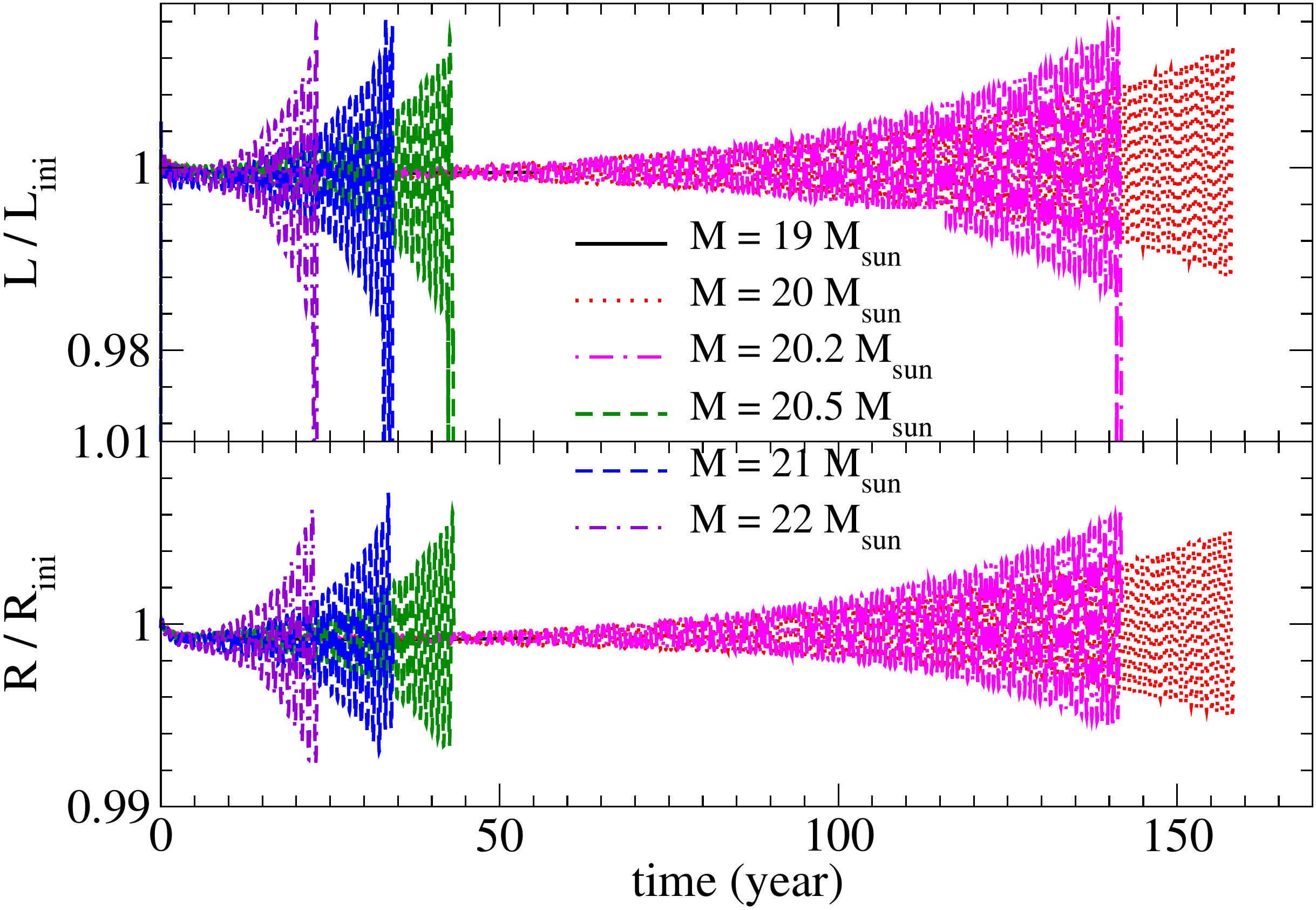}    
    \caption{The temporal evolution of the luminosity and stellar radius, scaled by their initial values, is shown for models with different progenitor masses studied in this work. Time 0 marks the transition from hydrostatic stellar evolutionary calculations to the hydrodynamical prescription. See also Appendix \ref{section:appendix} for the exact numerical treatment.}
    \label{fig:mass_compare}
\end{figure*}

In Figure \ref{fig:mass_compare}, we present the time evolution of the pulsation pattern for models with the progenitor mass from 19 to 22 $M_{\odot}$. We choose these masses as their timescales are more relevant to that of Betelgeuse.
Before non-linearity disturbs the simulation, all models behave similarly in both luminosity and radius.

Despite the fact that the excitation time apparently depends on the progenitor mass and mixing length parameter, 
the the means by which the star becomes excited---i.e. the pulsation driving mechanism itself---is less sensitive to these choices.
The dynamical pulsation always concludes with a significant drop in the stellar luminosity, and the peak luminosity and maximum radius are similar among all models near the end of the simulation.

\begin{figure*}
    \centering
    \includegraphics[width=\columnwidth]{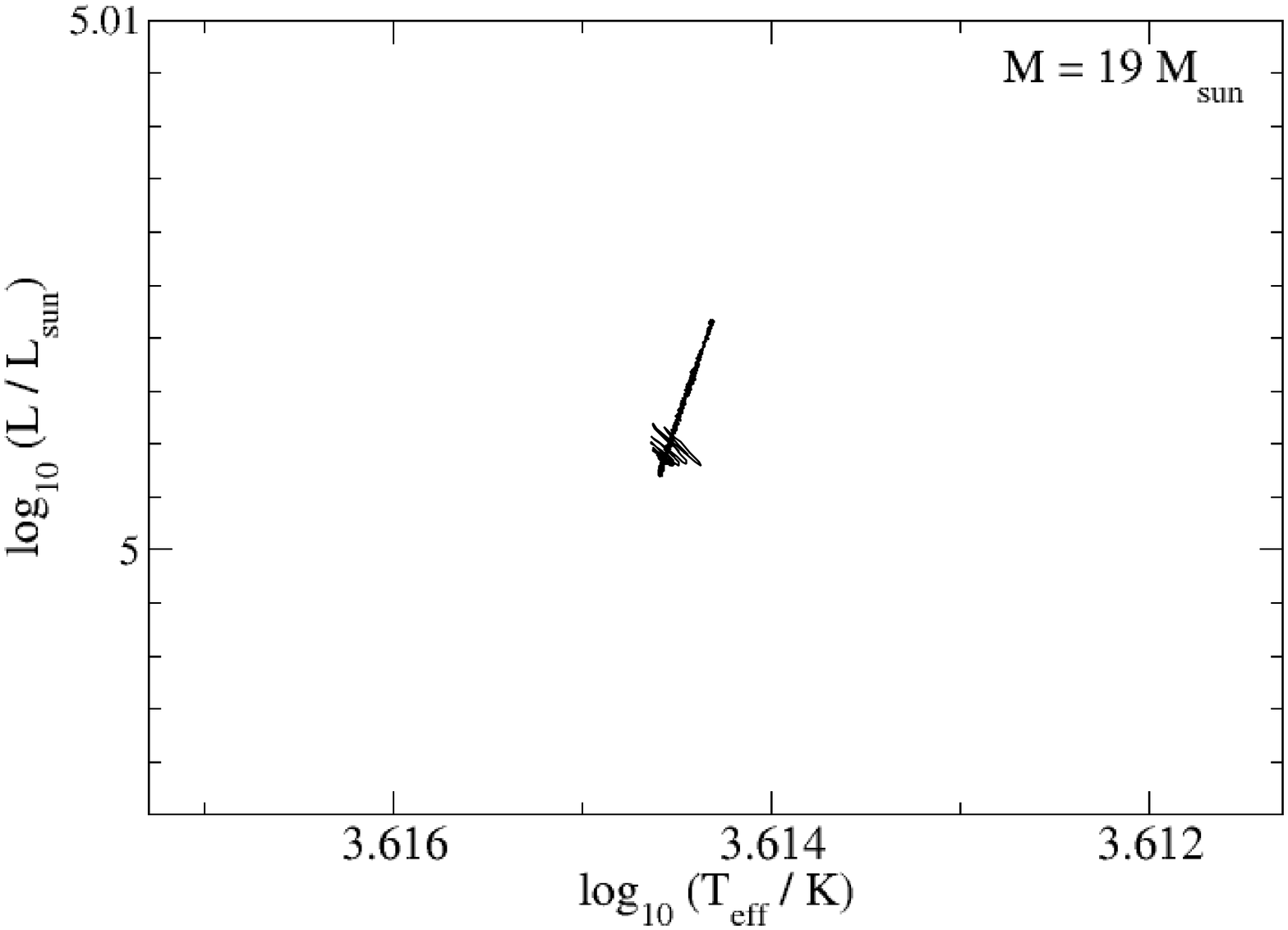}
    \includegraphics[width=\columnwidth]{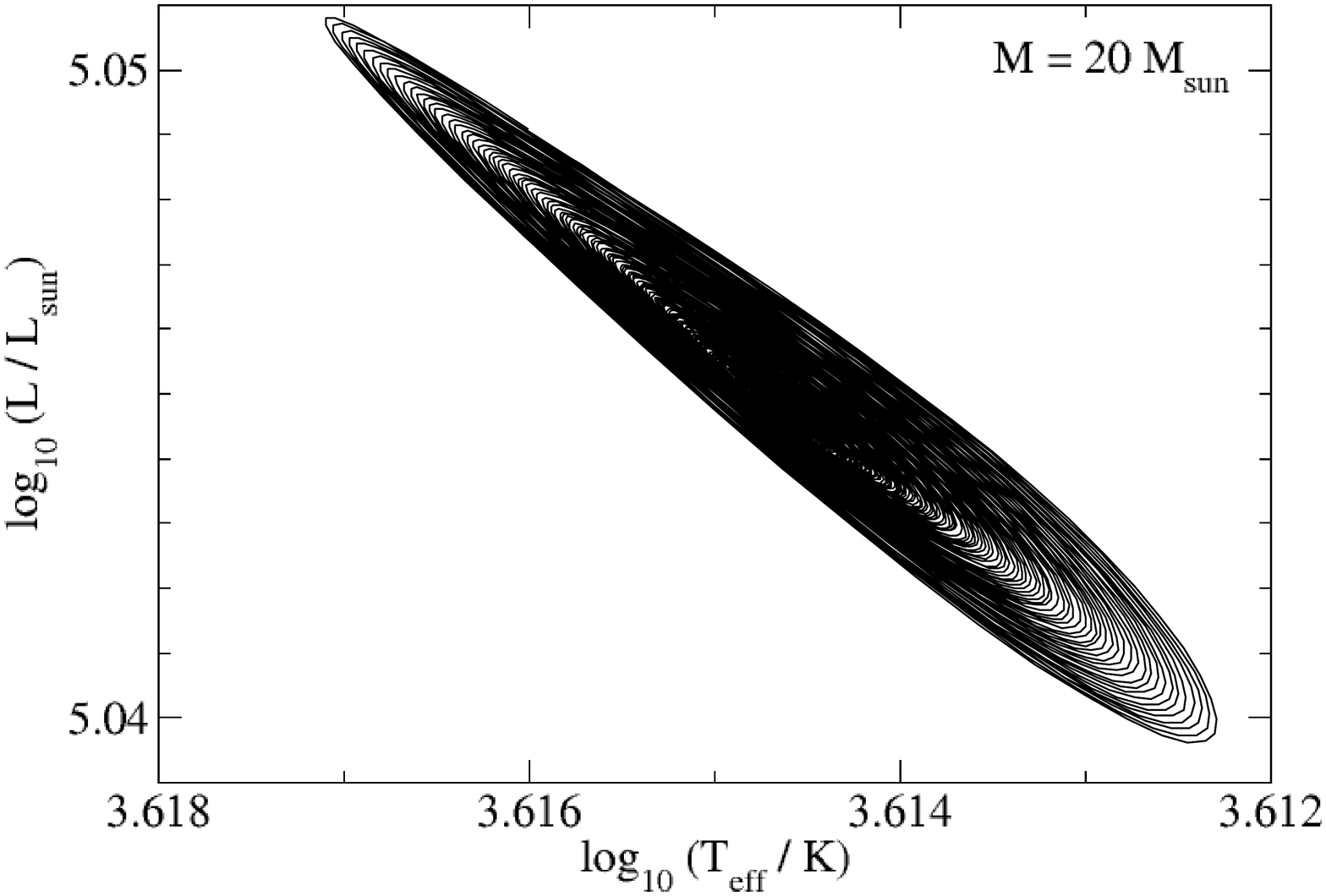}
    \includegraphics[width=\columnwidth]{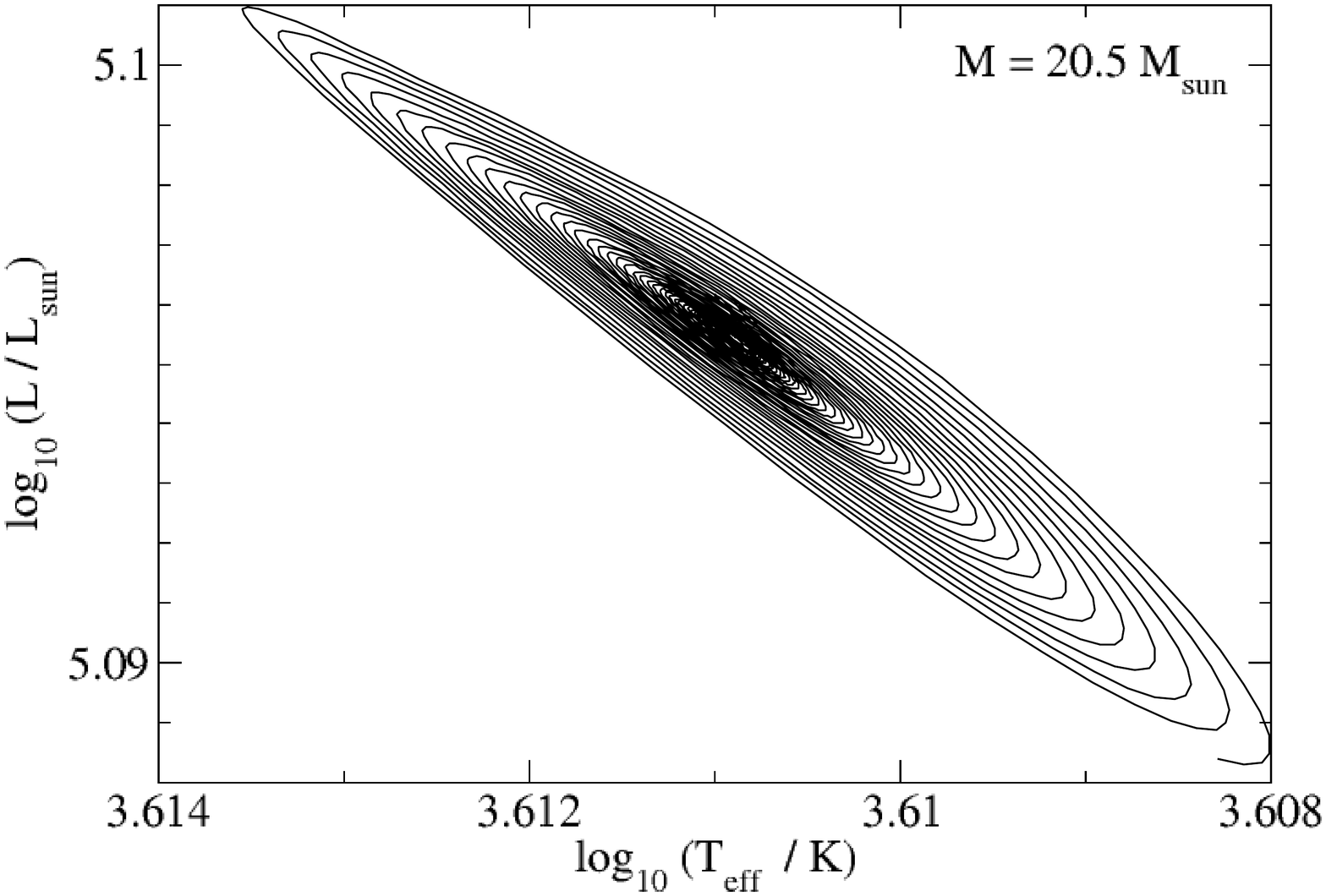}
    \includegraphics[width=\columnwidth]{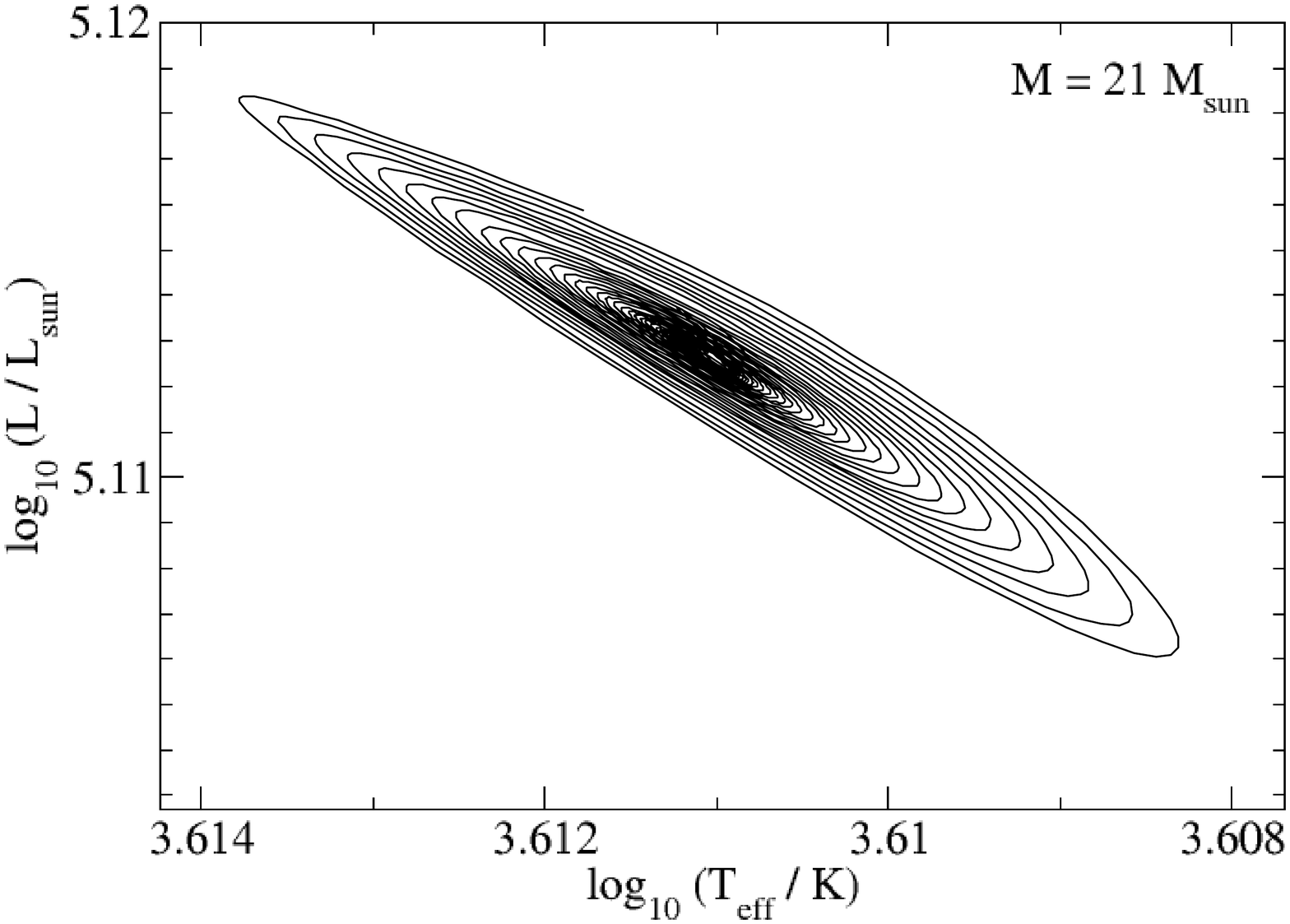}
    \includegraphics[width=\columnwidth]{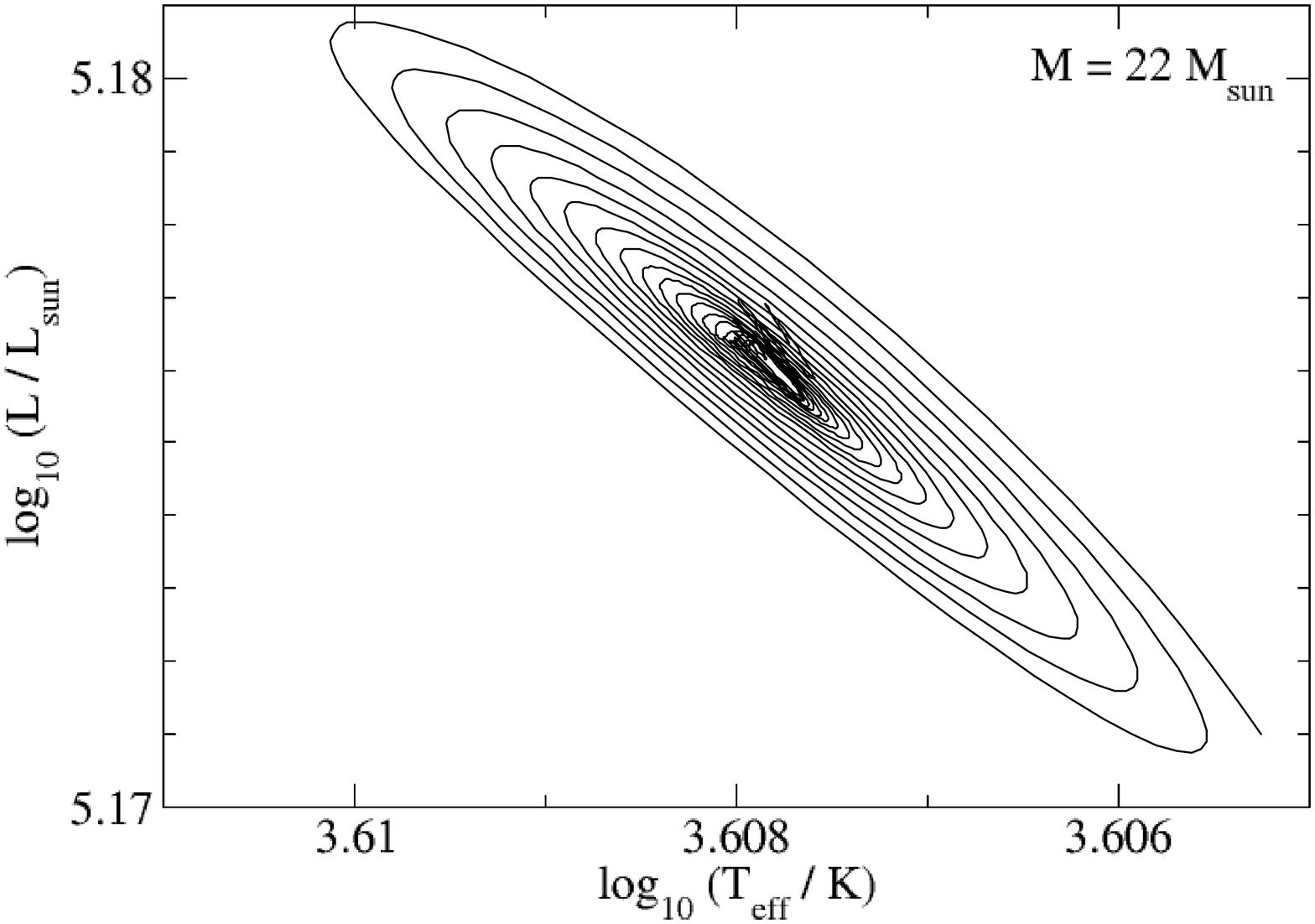}
    \includegraphics[width=\columnwidth]{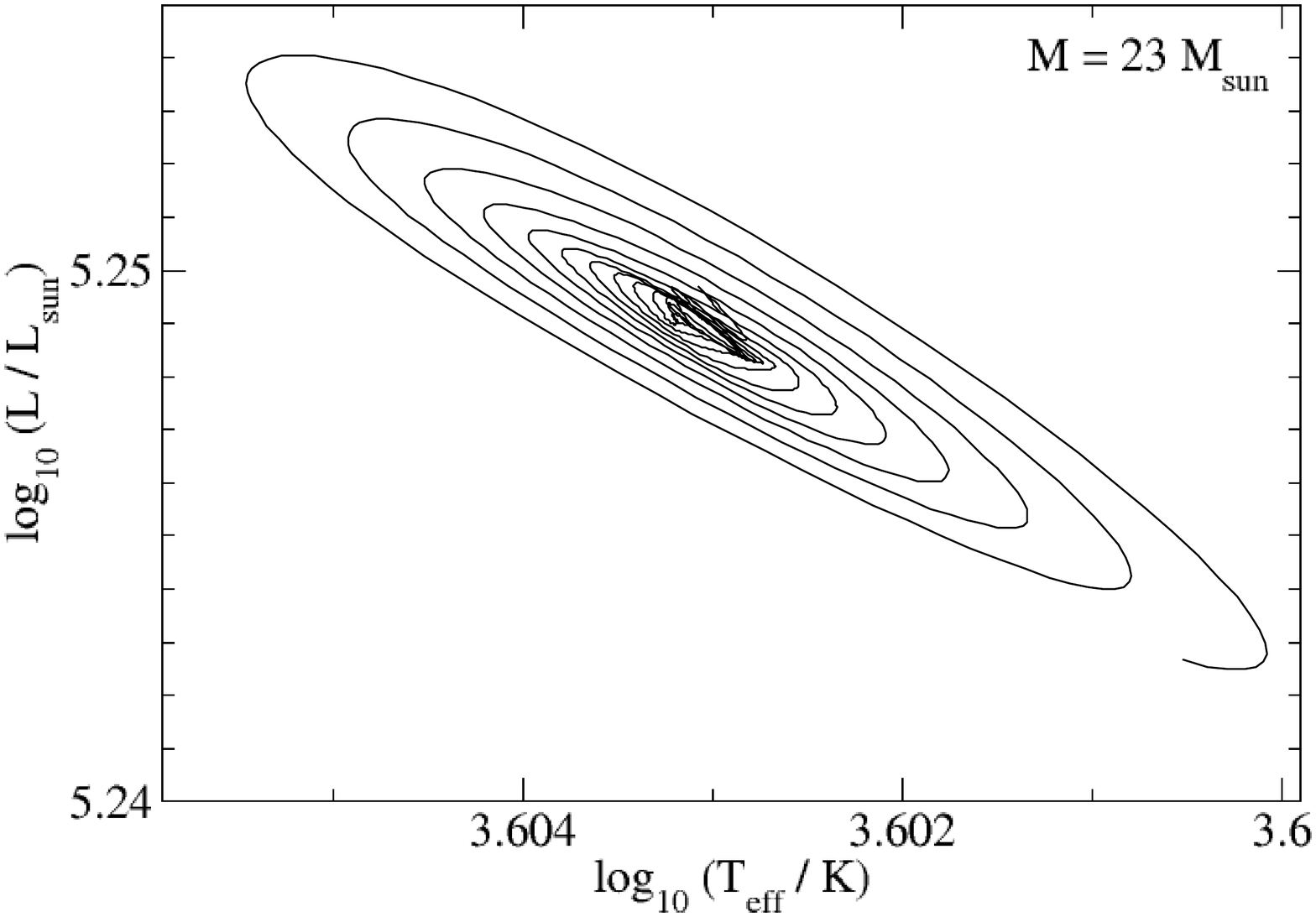}
    \caption{The phase diagrams for the model with a progenitor mass 19 (top left), 20 (top right), 20.5 (middle left), 21 (middle right), 22 (bottom left), 23 (bottom right) $M_{\odot}$ respectively. The trajectory is cut when the non-linearity begins to disturb the elliptical pattern in each figure.}
    \label{fig:phase_diagram}
\end{figure*}

To further outline the similarity, we plot in Figure \ref{fig:phase_diagram} the phase diagram of representative models from $19$--$23 ~M_{\odot}$. 
Clear similarity can be seen for models above $19~M_{\odot}$. In particular, for $M = 20~M_{\odot}$, the model has a highly extended $t_{\rm run}$ of $\sim 166$ year. All models have an elliptical structure, which is actually a clockwise outward-going spiral. They show once again that all stars evolve toward a high $L$ and a high $T_{\rm eff}$ state simultaneously, or the converse. 
This suggests that the driving mechanism in all of these models is qualitatively the same, too. A higher progenitor mass gives rise to a sparser trajectory; however, we notice that for $M = 19~M_{\odot}$, there is no regularity in the trajectory. This suggests that the $\kappa$-mechanism fails to stimulate residual numerical noise into periodic motions.

To further characterize the runaway time of the $M = 20~M_{\odot}$ model, we compare the amount of time needed for the star to develop non-linear pulsation ($t_{\rm run}$) after using the hydrodynamical prescription. For progenitor masses above 20.5 $M_{\odot}$, $t_{\rm run}$ decreases slowly with time. As shown in Figure \ref{fig:mass_compare}, nonlinear activation timescales for masses of this range are between $15$--$40$ years. However, below 20.5 $M_{\odot}$, there is a sudden jump in $t_{\rm run}$, and the star requires more than roughly $\geqslant 150$  in order for non-linearity to become significant. 
The sudden jump could signify some qualitative changes in the stellar profile, namely that the $\kappa$-mechanism becomes much less effective in amplifying the acoustic wave inside the star; a detailed comparison to and analysis of the means of formation for the $\kappa$-mechanism will be an interesting future project, but is beyond the scope of the present work. 
Crucially, this mass-sensitive timescale bifurcation suggests that the time required for the star to develop non-linear pulsation could be a highly discerning attribute among models of Betelgeuse.

As an order of magnitude estimation, the typical luminosity of our model star is $10^5 ~L_{\odot}$. The amount of energy dissipated is then $\sim 10^{46}$ erg per year, but the kinetic energy is only on the order of $10^{41\text{--}43}$ erg.
This is because radiation acts as a damping force through photon emission, and
without a consistent driving force for the pulsation, the oscillation would quickly dissipate.

In the previous text, we have shown that there are multiple periodicities in Betelgeuse's lightcurve. These include a quasi-annual mode, a 6-year mode, a 30-year modulation, and, potentially, an overtone mode with 185 day period. In the hydrodynamic models, we recover only the 416 day period. These results are largely self-consistent, as the 416 d mode is driven by the $\kappa$ mechanism, the LSP is not, and the 30 year modulation is most likely 
caused by rotation, which is not an internally driven form of variability.   
In the case of the overtone, however, we must address the question of how multiple modes may appear in the first place. 

One possibility is by non-linear mode excitation, as touched upon in Section \ref{subsection:modeexcitation}. Through large amplitude oscillations, the outer layers can accumulate sufficient energy and momentum to compress matter beneath the stellar surface. This results in compression heating, which in turn raises the local temperature. This may impact the convective structure in the near-surface regions, thus presenting an additional source of energy that alters the net energy flow inside the star. 
Capturing this scenario numerically is particularly challenging because it involves modeling the dynamics of mixing behaviour in the convection zone. 
Meanwhile, the standard mixing length theory adopted in our work assumes the convective mixing is in equilibrium \citep{ErikaBohmVitense1953}. Modeling this phase properly would require a more sophisticated approach to time-dependent mixing and a robust solving mechanism. 
We reiterate that the development of the overtone mode can also be sensitive to the numerical setting. In particular, the implicit nature of the hydrodynamics tends to suppress acoustic waves naturally, regardless of the use of artificial viscosity. Therefore, the current non-detection can be attributed to numerics alone. Future exploration using, for example, an explicit hydrodynamical scheme would shed light on this matter but is beyond the scope of this study.

We observe that in our hydrodynamical grid, models with $M > 21~M_{\odot}$ develop non-linear pulsation much more quickly than the lower mass models, as shown in Figure \ref{fig:mass_compare}.
%which occurs once per a few decades. 
We emphasize that our results only suggest that the star is developing non-linear pulsation and/or near-surface shocks; how strong the final shock is remains unclear. Because we do not follow our hydrodynamical simulations far enough to investigate rigorously the various energy dissipation channels discussed in section \ref{sec:canonical_hydro}, it remains to be studied whether $\alpha$ Orionis's strong pulsations eventually reach a new equilibrium or behave in some other way.

Another possible excitation mechanism is wave-driven pulsation, as described in  \citet{Shiode2014S, Fuller2017, Fuller2018}. 
This mechanism proposes that a convective wave in the star can partially penetrate through the evanescent regions\footnote{Zones dominated by thermal radiation.} and approach the stellar surface. 
Although wave-driven pulsation was described in the context of very late phases of stellar evolution in those works (i.e., Neon--Oxygen burning, rather than He),
the theory suggests that as long as convection is activated, energy can be transferred from the interior convection zone to regions near the surface, where it can then excite surface motion.
However, depending on the convective luminosity, such a mechanism would provide a heavily condensed energy deposition near the surface, in turn triggering enormous losses in mass of $0.01$--$1~M_{\odot}$ yr$^{-1}$. Mass loss of this order is not observed in Betelgeuse.

\section{Conclusions}
\label{section:conclusions}
We have presented a detailed observational and theoretical analysis of $\alpha$ Orionis, including the presentation of new photometry and three different types of numerical predictions from classical evolutionary, linear seismic, and hydrodynamic simulations. 
Our critical results are summarized as follows.

%%%%%%%%%%%%%%%%%%%%%%%%%%%%%%%%%%%%%%%%%%%%%

We present a new set of processed, space-based photometric data from the SMEI instrument, filling a gap in precise, publicly available photometry during the late 2000s. 
These data reveal variation on monthly timescales, which is likely the signature of convective cell turnover. In combination with longitudinal data collected by the AAVSO, the photometry confirms the presence of several key periodicities and contextualizes the recent dimming behavior of $\alpha$ Orionis in the long-term. 

We determine that the fundamental mode and the LSP are longer according to the SMEI and ground-based \textit{V}-band photometric data than according to the long-term visual results of \citet{Kiss2006}: {$P_0=416\pm24$} d and {$P_{\text{LSP}}=2365\pm10$} d in our work versus $P_0=388 \pm 30$ d and {$P_{\text{LSP}}\sim2050$} d in theirs. We conclude that the semiregular variability of the star---except the primary dimming event of 2019-2020---can be explained by phase changes in a short-lifetime pulsation mode and the photometric effects of giant convective cells. We also detect a new component with $185\pm13.5$ d period. We identify this as the first overtone, thus classifying $\alpha$~Ori a double-mode pulsator.

We conduct a grid-based analysis of evolutionary tracks to estimate the fundamental, model-derived parameters of $\alpha$ Ori. Supported by previous studies, we take special account of the theoretical uncertainty imparted by an ad hoc choice of the mixing length parameter, \mlt/, and reconsider the uncertainties on Betelgeuse's effective temperature accordingly \citep{Joyce2018a, Joyce2018b, Levesque2020}. We perform a probabilistic age prior analysis and find good agreement between our estimates of Betelgeuse's current evolutionary stage (RSB core helium burning) and present-day mass range ($16.5$--$19 ~M_{\odot}$) with previous modeling initiatives \citep{Neilson2011, Dolan2016, BetelgeuseProjectI, BetelgeuseProjectII}. Our seismic analysis prefers a median initial mass range of {$18-21 M_{\odot}$}.
However, we find that the observed, present-day rotational velocity of $\alpha$ Ori cannot be reproduced using single-star evolution; a merger or some other source of spin-up is required, in agreement with \citet{BetelgeuseProjectI, Chatzopoulos2020}.
The likelihood of a previous interaction is also supported by our kinematic argument in Section \ref{section:constraints}.

Linear seismic analysis with GYRE heavily constrains the radius of Betelgeuse, for which we report a value of $764 ^{+116}_{-62} R_{\odot}$. Combining this result with existing angular diameter and temperature data, we obtain a parallax value of $\pi=5.95^{+0.58}_{-0.85}$\,mas for $\alpha$ Orionis based on seismic constraints, resulting in a precise and independent distance estimate of $168^{+27}_{-15}$~pc. Our results are consistent with reprocessed \textit{Hipparcos} measurements but in disagreement with recent radio parallax observations \citep{vanLeeuwen2007, Harper2017}, highlighting the difficulty of estimating cosmic noise when deriving the geometric parallax of this star. To the best of our knowledge, this is the first time that a seismic parallax has been obtained for Betelgeuse.

Deeper analysis of emergent periodicities in both hydrostatic seismic and hydrodynamic models, in conjunction with existing observational data on variable stars across the mass spectrum, unambiguously demonstrate that the $416$~d period derived in this work is due to pulsation in the fundamental $p$-mode.

Finally, using hydrodynamic models with six different masses, we investigate the physics of these oscillations. All hydrodynamic models in the prescribed mass range manifest similar quasi-annual behavior as the fundamental mode, in agreement with similar studies. Our hydrodynamical simulations thus confirm that the 416 day pulsation is driven by the $\kappa$-mechanism.

We find that stars with an initial mass below $\sim20~M_{\odot}$ take much longer for the pulsation to excite other oscillation modes; in particular, a 19 $M_{\odot}$ model can take as long as 150 years to build up to non-linearity. The similarity among models suggests that the exact parameters of the model play a less important role in reproducing the fundamental mode of the star.
Importantly, if non-linear excitation is assumed to be correlated to the $\kappa$-mechanism's triggering of overtone modes, and if the observed mass loss in Betelgeuse is not pulsationally driven, our hydrodynamic simulations constrain against progenitor masses above $\sim 20~M_{\odot}$. On the contrary, if the large amplitude pulsation fails to reach an equilibrium and instead triggers shock waves and consecutive mass loss, it would strongly
suggest that Betelgeuse has a mass greater than $19 M_{\odot}$. As we do not evolve our simulations far enough to characterize the late-stage pulsational behavior, we cannot infer mass constraints definitively from these simulations.

It is unclear whether the excited fundamental mode can be modulated by other radiative mechanisms or lead to observable mass loss. If mass loss can be triggered, the short runaway time from the appearance of the first wave until mass ejection suggests that the star can lose a considerable amount of its H-envelope during its post-main-sequence evolution. 
If the observed mass loss in Betelgeuse can be connected to the instability observed in this work, we could potentially make additional inferences about the initial mass of Betelgeuse based on the timescale of non-linear excitation.

The sudden bifurcation in excitation time as a function of mass in our hydrodynamical models provides some constraint on Betelgeuse's upcoming, pre-supernova evolution. For models with an initial mass above $\sim 20~M_{\odot}$ (present-day mass $18.8~M_\odot$), the $\kappa$-mechanism driven pulsation and the mass loss it incites could partially remove the H-envelope prior to the final explosion. 
This would give rise to 
%the transition of
a Type-IIp, Type-IIL and then Type-IIn supernova. 
Meanwhile, for models with initial masses below this break-off point, the very long excitation time of the $\kappa$-mechanism means that the star would retain most of its H-envelope. In this case, an alternative mass loss channel would be required for the formation of a circumstellar medium.

Conclusively determining which of these two possible evolutionary channels $\alpha$ Ori will take would require disentangling the degeneracy between mass and mixing length in the simulations, but our work here suggests that a predictive investigation in this vein is possible.

\section*{Acknowledgements}
M.J.\ was supported the Research School of Astronomy and Astrophysics at the Australian National University and funding from Australian Research
Council grant No.\ DP150100250. 
M.J.\ was likewise supported by Ken'ichi Nomoto and invitation to the Kavli Institute for Theoretical Physics at the Institute for the Mathematics and Physics of the Universe (IPMU) at the University of Tokyo in January of 2020. Collaboration with Chiaki Kobayashi was made possible in part through the Stromlo Distinguished Visitors Program.

M.J.\ wishes to thank Peter Wood and Matteo Cantiello for helpful discussion regarding construction and interpretation of hydrodynamic simulations. 
M.J.\ further acknowledges Richard Townsend for management of the GYRE forums and the rest of the MESA developers for their support and expert guidance.

This work was also supported by World Premier International Research Center Initiative (WPI), and JSPS KAKENHI Grant Numbers JP17K05382 and JP20K04024.
S.C.L.\ thanks the MESA development community for making the code open-source.
S.C.L.\ acknowledges support by funding HST-AR-15021.001-A and 80NSSC18K101.

L.M.\ was supported by the Premium Postdoctoral Research Program of the Hungarian Academy of Sciences. The research leading to these results received funding from the LP2014-17 and LP2018-7/2019 Lend\"ulet grants of the Hungarian Academy of Sciences and the KH\_18 130405 grant of the Hungarian National Research, Development and Innovation Office (NKFIH). L.M.\ wishes to thank Bernard Jackson for discussions about the SMEI photometry. 

C.K. acknowledges funding from the UK Science and Technology Facility Council (STFC) through grant ST/M000958/1 \& ST/R000905/1, and the Stromlo Distinguished Visitorship at the ANU.

We acknowledge with thanks the variable star observations from the AAVSO International Database contributed by observers worldwide and used in this research. This research has made use of the SIMBAD database, operated at CDS, Strasbourg, France, and NASA's Astrophysics Data System Bibliographic Services.

\facilities{AAVSO (\url{http://aavso.org}), SMEI \citep{hick2007}}
\software{MESA \citep{MESAIV}, GYRE \citep{GYRE}}, Period04 \citep{period04}, Python: numpy, matplotlib, Bokeh \citep{numpy,matplotlib};  gnuplot

\appendix

\section{MESA Configurations}
\label{section:appendix}
In this section, we detail the configuration profile for the evolutionary and hydrodynamical portions of the simulations.

The evolutionary phase inherits settings from the \verb|massive_star_defaults| inlist. Additionally, we set the ``Dutch'' mass loss prescription with a parameter 0.8, namely:
%}
\begin{verbatim}
hot_wind_scheme = 'Dutch'
Dutch_scaling_factor = 0.8 
hot_wind_full_on_T = 1d0
hot_wind_full_off_T = 0d0
\end{verbatim}

In order to construct a star that maintains the proper radius for hydrodynamic evolution, we must adjust the mixing length parameter: 
\begin{verbatim}
mixing_length_alpha = 3
MLT_opion = 'Henyey' 
\end{verbatim}
We notice that a larger mixing length parameter results in a smaller radius at the end of the He-burning. The mass of the star is selected such that the luminosity 
is within the expected range ($\sim 4.8 $--$5.1$) and a radius between $700$--$800 R_{\odot}$ for consistency with the seismic parameters.

A requirement of our configuration is that the star should exhibit an observable amount of pulsation within a reasonable amount of time ($\sim 100$ years). A small progenitor mass results in a very long quiescent time. Meanwhile, a higher mass can trigger observable pulsation quickly, but its luminosity and radius can be too high. As a result, for the hydrodynamics, we pick the high mass end $M = 21 ~M_{\odot}$ with a large mixing length parameter $\alpha = 3$. This is slightly higher than what is used in the evolutionary calculations ($\alpha \leq 2.5$), but our model gives the correct radius at 720 $R_{\odot}$ and a luminosity $\sim 10^{5.1} ~L_{\odot}$. The final mass (present-day mass) is 19.5 $M_{\odot}$ and the helium core is $7.00 M_{\odot}$

As we require that the stellar profile transition smoothly from the evolutionary phase to the hydrodynamical phase, we use identical settings in the dynamical phase. 
\begin{verbatim}
T_mix_limit = 0
min_T_for_acceleration_limited_conv_velocity = 0
okay_to_reduce_gradT_excess = .false.
\end{verbatim}

In the hydrodynamical phase, we patch extra settings onto this configuration such that the hydrostatic equilibrium constructed in the previous phase can be well maintained.
However, one qualitative change is included, where the mass loss is suspended.
\begin{verbatim}
Dutch_scaling_factor = 0.0d0    
\end{verbatim}
This is a reasonable approximation given that we are simulating a short period of time: $\sim 100$ years.

To trigger the hydrodynamics, we use the standard settings as provided by the \verb|test_suite| test case \verb|ccsn| in MESA version 8118. This includes
\begin{verbatim}
use_ODE_var_eqn_pairing = .true.
use_dvdt_form_of_momentum_eqn = .true.
use_dPrad_dm_form_of_T_gradient_eqn = .true.
use_dedt_form_of_energy_eqn = .true.
use_momentum_outer_BC = .true.        
use_ODE_form_of_density_eqn = .true.     
\end{verbatim}
These settings have been used in our previous work modeling the dynamical pulsation in pulsation pair-instability supernovae. See \cite{Leung2019PPISN1,Leung2020PPISN2}
for the application of these setting to the more massive star counterpart. 

Furthermore, to ensure the code captures the early oscillation when the simulation has begun, we impose a maximum evolutionary timestep of $10^5$ s.
\begin{verbatim}
max_timestep = 100000
\end{verbatim}
We also remove the temperature limitation in which the hydrodynamics is solved. This means the Euler equations are solved throughout the star, without assuming the envelope is in hydrostatic equilibrium:
\begin{verbatim}
velocity_logT_lower_bound = 0
\end{verbatim}

To prevent supersonic convection from occurring in the simulation and invalidating the assumptions of the mixing length theory, we impose a cap on the convective speed via
\begin{verbatim}
mlt_accel_g_theta = 1      
max_v_div_cs_for_convection = 1.0d-1
max_conv_vel_div_csound = 1.0d0,
\end{verbatim}
ensuring that the convective behavior remains physical.

At last, we turn on the artificial viscosity so that all potential shocks can be resolved by the simulation. This happens, in particular, near the surface where the density gradient is the highest.
\begin{verbatim}
use_artificial_viscosity = .true.
shock_spread_linear = 0
shock_spread_quadratic = 2d-2
\end{verbatim}
We find that a higher artificial viscosity parameter can result in the code crashing earlier in the simulation, whereas a value too small can result in too strong of a shock when the global pulsation amplitude is still weak.

A simulation of $\sim 30$ years requires approximately 10000 timesteps. 

%\newpage
\bibliographystyle{apj}
\bibliography{Betelgeuse_submitted.bib} % if your bibtex file is called example.bib

\label{lastpage}
\end{document}